\documentclass[published]{JHEP3}

\JHEP{ }

\usepackage{epsf}
\usepackage{latexsym}
\usepackage{epsfig,multicol}

\def\yzero{\smash{\hbox{$y\kern-4pt\raise1pt\hbox{${}^\circ$}$}}}

\def\a{\alpha}
\def\b{\beta}
\def\g{\gamma}
\def\d{\delta}
\def\beq{\begin{equation}}
\def\eeq{\end{equation}}
\def\beqa{\begin{eqnarray}}
\def\eeqa{\end{eqnarray}}
\def\Om{\Omega}

\def\th{\theta}
\def\vt{\vartheta}

\def\-{\hphantom{-}}

\def\s2{\frac{1}{\sqrt2}}

\def\oh{\frac{1}{2}}
\def\beq{\begin{equation}}
\def\eeq{\end{equation}}
\def\beqa{\begin{eqnarray}}
\def\eeqa{\end{eqnarray}}

\def\IF{\relax{\rm I\kern-.18em F}}
\def\II{\relax{\rm I\kern-.18em I}}
\def\IP{\relax{\rm I\kern-.18em P}}
\def\IC{\relax\hbox{\kern.25em$\inbar\kern-.3em{\rm C}$}}
\def\IR{\relax{\rm I\kern-.18em R}}

\def\cn{{\cal N}}

\def\Dsl{\,\raise.15ex\hbox{/}\mkern-13.5mu D} 
\def\IZ{Z\kern-.4em  Z}

\def\t{\times}
\def\eps{\epsilon}

\def\car{{\cal R}}
\def\b{\beta}
\def\q{\vec{q}}
\def\B{{\bf B}}
\def\lam{\lambda}

\newbox\pippobox

\title{Intersecting Brane Models of Particle Physics and the Higgs Mechanism}

\author{D.~Cremades, L.~E.~Ib\'a\~nez and  F.~Marchesano \\
 	Departamento de F\'{\i}sica Te\'orica C-XI
	and Instituto de F\'{\i}sica Te\'orica  C-XVI,\\
	Universidad Aut\'onoma de Madrid,
	Cantoblanco, 28049 Madrid, Spain.
}
\received{\today}               
\revised{\today}
\accepted{}               

\preprint{\hepth{0203160}}
\preprint{FTUAM-02/8 IFT-UAM/CSIC-02-7}

\abstract{We analyze  a recently constructed class of
D-brane theories with the fermion spectrum of the SM 
at the intersection of D6-branes wrapping a compact
toroidal space. We show how the SM Higgs mechanism appears 
as a brane recombination effect in which the branes giving rise
to $U(2)_L\times U(1)$ recombine into a single brane related
to $U(1)_{em}$. We also show how one can construct D6-brane
models which respect some supersymmetry at every
intersection. These are {\it quasi-supersymmetric} models
of the type introduced in hep-th/0201205 which may be 
depicted in terms of {\it  SUSY-quivers} and may stabilize
the hierarchy between the weak scale and a fundamental scale 
of order 10-100 TeV present in low string scale models.
Several explicit D6-brane models with three generation 
of quarks and leptons and different SUSY-quiver structure
are presented. One can prove on general grounds
that if one wants to build 
a (factorizable) D6-brane configuration with the SM 
gauge group and $N = 1$ SUSY (or quasi-SUSY), also a massless
$(B-L)$ generator must be initially present in any  model.
If in addition we insist on left- and right-handed fermions
respecting the {\it same } $N=1$ SUSY, the brane configurations 
are forced to have intersections giving rise to Higgs multiplets,
providing for a rationale for the very existence of
the SM Higgs sector.}


\begin{document}

\section{Introduction}

Chirality is probably the deepest property 
of the standard model (SM) of particle physics.
It thus seems that, in trying
to build a string theory description of the SM,
the first thing we have to obtain is massless chiral fermions.
There is a number of known ways in order to achieve chirality
in string theory models. From the point of view of
explicit D-brane model building there are however only
a few known ways. One of the simplest is to locate 
D-branes (e.g., D3-branes) at some 
(e.g., orbifold or conifold) singularity
in transverse space \cite{dm}. Explicit models 
with branes at singularities have been built \cite{aiqu,evenmore} 
 with three generations and massless spectrum 
close to that of the SM or some simple left-right
symmetric extension (see also \cite{bjl,bailin,gerardo}).

Another alternative way in order to get massless 
chiral fermions in D-brane models is intersecting branes \cite{bdl}.
Under certain conditions,  D-branes intersecting at angles 
give rise to massless chiral fermions localized at the 
intersections. Explicit D-brane models with realistic 
three-generation particle spectrum lying at intersecting branes 
have been constructed in the last couple of years 
\cite{bgkl,afiru,afiru2,bkl,imr,bklo,csu,cim1,pheno,blumy} 
(see also \cite{cosmo}). 
One of the nice features of this type of constructions  
is that family replication appears  naturally as a consequence of the 
fact that in a compact space branes may intersect a multiple 
number of times. Another  
attractive property of this scenario is that  quarks  
and leptons corresponding to different generations  
are localized at different points in the transverse 
 compact space. It has been suggested \cite{afiru2,imr} that this may 
provide a geometrical understanding of the observed 
hierarchy of quark and lepton masses.

In ref.\cite{imr}  
a particularly interesting  class of models was found  
in which the massless chiral   
fermion spectrum is identical to that 
of the SM of particle physics. 
They  were obtained from 
sets of intersecting D6-brane wrapping an 
(orientifolded) six-torus \cite{bgkl}. These models are  
non-supersymmetric but it has recently been found \cite{cim1} that  
analogous models with one $\cn = 1$ supersymmetry  
at each brane intersection may be obtained by 
appropriately varying the geometry (complex  
structure) of the torus. In these models each 
 intersection respects  in general a {\it different} 
$\cn = 1$ supersymmetry so that the model 
is globally non-supersymmetric but in some sense 
locally supersymmetric. These type of models  
are called in \cite{cim1} {\it quasi-supersymmetric} (Q-SUSY) 
and have the property that quantum  corrections to scalar  
masses appear only at two-loops. 
The different $\cn = 1$ structure of each intersection in this class 
of models may be represented in terms of {\it SUSY quivers}. 
 This Q-SUSY 
property may be of  
 phenomenological interest in order to  stabilize 
the hierarchy between the weak scale and a fundamental scale 
of order 10-100 TeV present in low string scale models.

The present article has two main purposes: 
 
{\it i)} We  
construct explicit D6-brane Q-SUSY models with  
three quark-lepton generations and study some of their 
properties. All of these models have four stacks of 
D6-branes : the {\it baryonic}, {\it leptonic},  
{\it left} and {\it right} stacks  
and either the SM group or a slight generalization  
with an extra $U(1)$. 
We show that there are just four classes of quivers  
(see fig. \ref{quivers}) with 
four stacks of branes and no $SU(3)$ anomalies yielding 
realistic models. We call them  
the triangle, linear, square and rombic quivers.  
Each of these classes of quivers differ on the  
$\cn = 1$ SUSY preserved by the quark, lepton and Higgs sector. 
In some models ({\it triangle quiver}) all  
of them preserve the same $\cn = 1$ SUSY whereas in others 
e.g., left-handed fermions preserve one supersymmetry 
and right-handed fermions a different one ({\it linear quiver}).  
The ({\it square quiver}) class corresponds to 
Q-SUSY versions of the models of \cite{imr} 
and have the SM fermion spectrum. 
They have the property that left-handed quarks, 
right-handed quarks, left-handed leptons and 
right-handed leptons have each a different $\cn = 1$ SUSY 
inside a $\cn = 4$ living in the bulk. 
Finally, models constructed from the ({\it rombic quiver}) have 
 the property that 
all quarks and the Higgs set respect the same $\cn = 1$ supersymmetry 
whereas the leptonic sector does not. The 
example provided has  a left-right 
symmetric gauge group. 
Whereas in the linear and square quiver models the Higgs sector is  
non-SUSY, it is so in the other two, providing  
explicit examples of models with weak scale stabilization. 
 
There are a couple of results which look fairly general: 
 
\begin{itemize} 
 
\item 
Imposing  
supersymmetry at all the intersections implies necessarily the presence 
of a very definite extra $U(1)$, the $U(1)_{B-L}$  
familiar from left-right symmetric models.  
 
\item 
If in addition we insist on some left- and right-handed  
fermions to respect the {\it same} $\cn = 1$ SUSY 
(as happens e.g. in the triangular and rombic quivers), 
the brane configurations are forced to have  
intersections giving rise to SM Higgs multiplets 
\footnote{This is in contrast with other popular 
embeddings of the SM like SUSY-GUT, $CY_3$ or Horava-Witten 
heterotic compactifications etc., in which the presence of 
light SM Higgs multiplets is  
somewhat ad-hoc and misterious.}, providing for a 
rationale for the very existence of the SM Higgs sector.  
 
\end{itemize}

We also  show how slight variations of torus  moduli give rise to 
Fayet-Iliopoulos (FI) terms for the $U(1)$'s in the theory. 
These FI-terms are only present for the subset of  
$U(1)$'s which become massive by combining with antisymmetric 
$B_2$ fields of the closed string sector of the theory. 
Since those $U(1)$'s are massive, the scalar D-term masses 
look like explicit soft SUSY-breaking masses for the 
sparticles charged under those $U(1)$'s. 
We finally analyze the structure of gauge coupling constants 
in this class of models. 
For these intersecting brane 
models gauge couplings do not unify at the string scale. 
Rather they are inversely proportional to the volume  
wrapped by each brane. In the case of Q-SUSY models  
those volumes get a particularly simple expression in terms 
of the torus moduli, and we give explicit formulae for them.

{\it ii) } We describe how the SM Higgs mechanism 
in intersecting brane world models has a nice geometrical interpretation  
as a brane recombination process, in which the branes  
giving rise to $U(2)_L\times U(1)$ recombine into a single 
brane related to $U(1)_{em}$. In the SM Higgs 
mechanism the rank of the gauge group is reduced.  
The stringy counterpart of this rank-reduction 
is brane recombination \footnote{Note that the SM Higgs mechanism 
cannot be described by the familiar process in which two 
parallel branes separate. Brane separation does not 
lower the rank and corresponds to adjoint Higgsing, which is 
not what the SM Higgs mechanism requires.}. 
The chiral fermion spectrum  in these models is 
determined by the intersection numbers $I_{ab}$  
which are topological in character. As branes recombine, 
$|I_{ab}|$ decreases,  signaling that some chiral 
fermions become massive. This is the stringy  
version of the fermions getting masses from Yukawa 
couplings after the Higgs mechanism takes place. 
We exemplify this brane recombination interpretation 
of the Higgs mechanism  in some of the specific examples 
introduced in the paper, although the general physics  
applies to any intersecting brane model.

As we said, having some $\cn = 1$ SUSY at all brane intersections 
implies generically the presence of a $B-L$ massless generator 
in the spectrum. We discuss how this symmetry may be Higgssed 
away in terms of a brane recombination process 
in which the {\it leptonic} and {\it right} branes  
recombine into a single brane. That recombination process  
gives masses to the right-handed neutrinos and we argue that,   
after the SM Higgs/recombination mechanism  takes place, 
the left-handed 
neutrinos get at some level Majorana masses.

The structure of this paper is as follows. In the next two sections we  
give short introductions to both the D6-brane toroidal models 
introduced in \cite{imr} and to the concept of 
quasi-supersymmetric models of ref.\cite{cim1}. 
In Section 4 we construct several three generation 
models with SM gauge group (or some simple extension) 
and with some quasi-supersymmetry.  
We analyze SUSY-breaking effects from FI-terms for some of those 
models in Section 5 whereas we provide formulae for the  
gauge coupling constants in Section 6.  
In Section 7 we interpret the SM Higgs mechanism in terms 
of recombination of intersecting branes.  
Some final comments and general conclusions  
are presented in Section 8.  A couple of appendices  
concerning the presence of extra $U(1)$'s  
in Q-SUSY models and models with D6-branes  
wrapping  non-factorizable cycles on the tori 
are provided.

\section{The Standard Model at intersecting branes revisited} 
 
In this section we summarize the construction of the Standard Model  
fermionic spectrum arising from intersecting brane worlds, as presented  
in \cite{imr}. We refer the reader to this paper for details regarding  
this construction. 
 
We will consider Type IIA string theory compactified on a  
factorized 6-torus $T^2 \times T^2 \times T^2$. In this setup, 
we introduce sets of D6-branes  
with their 7-dimensional volume containing four-dimensional Minkowski 
space and wrapping 3-cycles $[\Pi]$ of $T^6$ \cite{bgkl,afiru,bkl} 
(for related constructions see also \cite{inter1}).  
We will further assume that these 3-cycles can be factorized as  
three 1-cycles, each of them wrapping on a different $T^2$. 
We denote by $(n_a^i,m_a^i)$, $i = 1,2,3$ the wrapping numbers of each  
$D6_a$-brane, on the $i^{th}$ torus, $n_a^i$ ($m_a^i$) being the number 
of times the brane is wrapping around the basis vector $e_1^i$ ($e_2^i$) 
defining the lattice of the $i^{th}$ torus, as depicted in figure  
\ref{compact2}. 

\EPSFIGURE{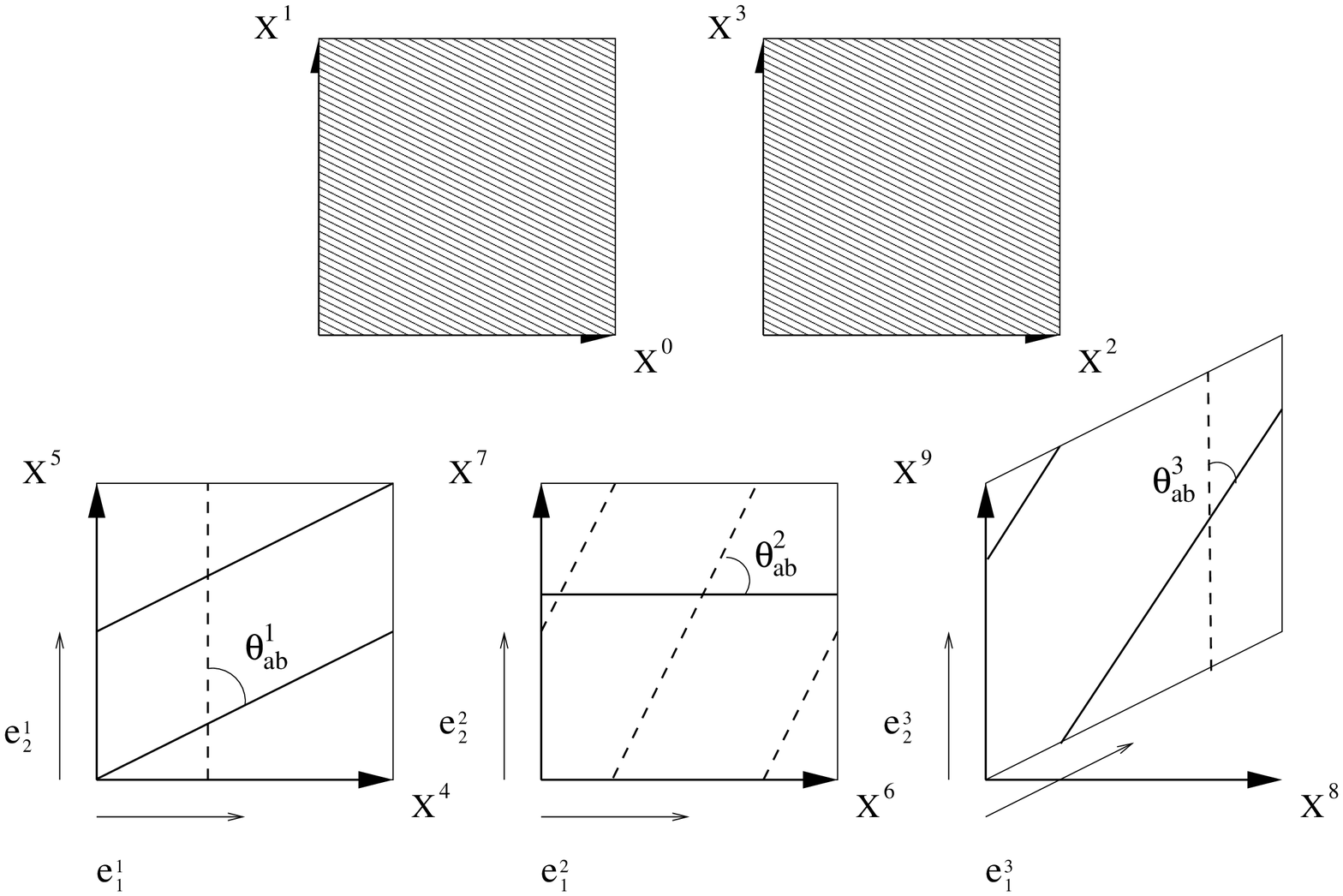, width=6in}
{\label{compact2}
Intersecting brane world setup. We consider two D6-branes  
containing four non-compact dimensions, to be identified with $M_4$, 
and wrapping factorizable 3-cycles of $T^2 \t T^2 \t T^2$.  
In this specific example the wrapping numbers are $(2,1)(1,0)(1,1)$ 
(solid line) and $(0,1)(1,2)(0,1)$ (dashed line). Notice that the 
wrapping numbers of the third torus can be also expressed as fractional 
cycles, as we have a tilted complex structure (see text). 
In this fractional language the {\it dashed} brane remains as before, 
whereas the {\em solid} brane wrapping numbers become 
$(2,1)(1,0)(1,\frac 32)$ (see Appendix I). Note that the net 
intersection number in this example is $2\times2\times1 = 4$. } 

Actually, we will study an orientifolded version of the theory 
obtained by modding out by $\Omega \car$ \cite{bgkl}, where $\Omega$ is the  
worldsheet parity operator and $\car = \car_{(4)} \car_{(6)} \car_{(8)}$  
is a reflection operator with respect to the real axis of each internal  
complex dimension $Z_j = X_{2j+2} + iX_{2j+3}$, $j = 1,2,3$  
(that is, $\car_{(4)}: Z_1 \mapsto \bar Z_1$, etc.).  
Since we must have $\Om \car$ symmetry in our models, for each  
$D6_a$ brane we introduce we must also add its mirror image  
$\Om \car D6_a$ or $D6_{a^*}$, whose geometrical locus will be determined  
by a reflection on the real axes $X_4, X_6, X_8$. Being identified,  
both branes will give rise to the same unitary gauge group.  
This symmetry under $\Om \car$ does also imply that we are allowed to  
consider only some specific choices of $T^2$. We may either have  
rectangular tori (as the first two tori in fig. \ref{compact2})  
or tilted tori (as the third torus in fig. \ref{compact2}).  
When we have a tilted torus, we can easily describe our  
configurations in terms of fractional wrapping numbers,  
where the $m$'s may take $\IZ/2$ values (see \cite{bkl,imr}  
and Appendix I for more detailed discussions). 
 
The number of times two branes $D6_a$ and $D6_b$ intersect  
in $T^6$ is a topological quantity known as the intersection number. 
When dealing with factorizable branes it can be easily expressed as 
\beq 
I_{ab}\ =\ 
(n_a^1m_b^1-m_a^1n_b^1)(n_a^2m_b^2-m_a^2n_b^2)(n_a^3m_b^3-m_a^3n_b^3).   
\label{internumber} 
\eeq 

Open strings stretching around the intersections give rise to 
chiral fermions in bifundamental representations $(N_a,{\bar N}_b)$ 
or $(N_a,{N}_b)$ under the gauge group of the two branes $ 
U(N_a)\times U(N_b)$. 
Thus, these configurations  yield $I_{ab}$ copies of the same 
bifundamental representation. In \cite{imr} it was shown how one  
can build a configuration giving just three generations of a  
$SU(3)\times SU(2)\times U(1)_Y$ gauge group.  
To achieve this construction one must consider four stacks  
of D6-branes whose multiplicities and associated gauge group is 
shown in table \ref{SMbranes}. 
\TABLE{\renewcommand{\arraystretch}{1.7}
\begin{tabular}{|c|c|c|c|} 
\hline 
Label & Multiplicity & Gauge Group & Name \\ 
\hline 
\hline 
stack $a$ & $N_a = 3$ & $SU(3) \times U(1)_a$ & Baryonic brane\\ 
\hline 
stack $b$ & $N_b = 2$ & $SU(2) \times U(1)_b$ & Left brane\\ 
\hline 
stack $c$ & $N_c = 1$ & $U(1)_c$ & Right brane\\ 
\hline 
stack $d$ & $N_d = 1$ & $U(1)_d$ & Leptonic brane \\ 
\hline 
\end{tabular} 
\label{SMbranes}
\caption{\small Brane content yielding the SM spectrum.}}
Now, given this brane content, it is relatively easy to achieve 
the desired SM spectrum as arising from massless fermions  
living at brane intersections. It is enough to select  
branes' wrapping numbers $(n_a^i,m_a^i)$ in such a way that 
the intersection numbers $I_{ij}$, $i,j = a,b,c,d$ are given by \cite{imr} 
\beq 
\begin{array}{lcl} 
I_{ab}\ =   \ 1, & & I_{ab*}\ =\ 2, \\ 
I_{ac}\ =   \ -3, & & I_{ac*}\ =\ -3, \\ 
I_{bd}\ =   \ -3,  & & I_{bd*}\ =\ 0, \\ 
I_{cd}\ =   \ 3, & & I_{cd*}\ =\ -3, 
\end{array} 
\label{intersec2} 
\eeq 
all other intersections vanishing. 
Here a negative number denotes that the 
corresponding fermions should have opposite chirality to those 
with positive intersection number. 
The massless fermionic spectrum arising from (\ref{intersec2}) is 
shown in table \ref{tabpssm}, as well as the charges with respect to the 
four $U(1)$'s. Note in this respect that eventually only  
one $U(1)$ (the hypercharge) remains massless, the other three become massive 
by swallowing certain closed RR fields, as discussed below. 
\TABLE{\renewcommand{\arraystretch}{1.25}
\begin{tabular}{|c|c|c|c|c|c|c|c|} 
\hline Intersection & 
 Matter fields  &   &  $Q_a$  & $Q_b $ & $Q_c $ & $Q_d$  & Y \\ 
\hline\hline (ab) & $Q_L$ &  $(3,2)$ & 1  & -1 & 0 & 0 & 1/6 \\ 
\hline (ab*) & $q_L$   &  $2( 3,2)$ &  1  & 1  & 0  & 0  & 1/6 \\ 
\hline (ac) & $U_R$   &  $3( {\bar 3},1)$ &  -1  & 0  & 1  & 0 & -2/3 \\  
\hline (ac*) & $D_R$   &  $3( {\bar 3},1)$ &  -1  & 0  & -1  & 0 & 1/3 \\ 
\hline (bd) & $L$    &  $3(1,2)$ &  0   & -1   & 0  & 1 & -1/2  \\ 
\hline (cd) & $N_R$   &  $3(1,1)$ &  0  & 0  & 1  & -1  & 0  \\ 
\hline (cd*) & $E_R$   &  $3(1,1)$ &  0 & 0 & -1 & -1  & 1 \\ 
\hline 
\end{tabular} 
\label{tabpssm}
\caption{\small Standard model spectrum and $U(1)$ charges. The hypercharge 
generator is defined as $Q_Y = \frac 16 Q_a - \frac 12 Q_c - \frac 12 Q_d$.}}
In \cite{imr}  a general class of solutions was given for the wrapping numbers 
$(n_a^i,m_a^i)$ giving rise to a SM spectrum. These are shown in table  
\ref{solution}. In this table we have several discrete parameters. 
First we consider $\beta^i =1 - b^i$,  
with $b^i = 0, 1/2$ being the T-dual NS B-background 
field discussed in \cite{bkl} (see also \cite{Bachas,bfield,fluxes} 
for previous related discussions). 
As shown there, the addition of this  
background is required in order to get and odd number of quark-lepton 
generations. From the point of view of branes at angles 
$\beta^i = 1$ stands for a rectangular lattice for the $i^{th}$ torus,  
whereas $\beta^i = 1/2$ describes a tilted lattice allowed 
by the $\Om\car$ symmetry. Notice that in the third torus  
one always has $\beta^3 = 1/2$, as in figure \ref{compact2}.   
We also have two phases $\epsilon, \tilde \eps = \pm 1$ and 
the parameter $\rho$ which can only take the values $\rho = 1, 1/3$. 
Furthermore, each of these families of D6-brane configurations depend on 
four integers ($n_a^2,n_b^1,n_c^1$ and $n_d^2$). 
Any of these choices lead exactly to the same massless 
fermion spectrum of table \ref{tabpssm}. 
\TABLE{\renewcommand{\arraystretch}{2.5}
\begin{tabular}{|c||c|c|c|} 
\hline 
 $N_i$  &  $(n_i^1,m_i^1)$  &  $(n_i^2,m_i^2)$   & $(n_i^3,m_i^3)$ \\ 
\hline\hline $N_a=3$ & $(1/\beta ^1,0)$  &  $(n_a^2, \epsilon \beta^2)$ & 
 $(1/\rho, - \tilde \eps/2)$  \\ 
\hline $N_b=2$ &   $(n_b^1,\tilde \eps\epsilon \beta^1)$     
&  $ (1/ \beta^2,0)$  & $(1,-3\rho \tilde\eps /2)$   \\ 
\hline $N_c=1$ & $(n_c^1,3\rho \epsilon \beta^1)$  &    
 $(1/\beta^2,0)$  & $(0,1)$  \\ 
\hline $N_d=1$ &   $(1/\beta^1,0)$  &  $(n_d^2,\epsilon\beta^2/\rho)$  & 
$(1, 3\rho \tilde\eps /2)$ \\
\hline 
\end{tabular} 
\label{solution}
\caption{\small D6-brane wrapping numbers giving rise to a SM spectrum. 
The general solutions are parametrized by two phases 
$\epsilon, \tilde \eps =\pm1$, the NS background 
on the first two tori $\beta^i=1-b^i=1,1/2$, four integers    
$n_a^2,n_b^1,n_c^1,n_d^2$ and a parameter $\rho=1,1/3$.}} 
The wrapping numbers of the complete set of D6-branes have to verify 
the RR tadpole cancellation conditions \cite{bgkl} 
\beq 
\begin{array}{c} 
\sum_a N_a n_a^1n_a^2n_a^3 = 16 \\ 
\sum_a N_a n_a^1m_a^2m_a^3 = 0 \\ 
\sum_a N_a m_a^1n_a^2m_a^3 = 0 \\ 
\sum_a N_a m_a^1m_a^2n_a^3 = 0 
\label{tadpoleO6} 
\end{array} 
\eeq 
which just state that the total RR-charge of the configuration has to vanish. 
The orientifold modding leads to the presence of 8$\b^1\b^2\b^3$
orientifold planes wrapping the cycle $(1/\b^1,0)(1/\b^2,0)(1/\b^3,0)$,
each with net RR charge -2 (compared to the charge of a pair of mirror 
D6-branes), as the first condition shows.

These conditions automatically guarantee 
cancellation of chiral anomalies. In this  class of models the last  
three conditions are automatically verified whereas the first requires: 
\beq 
\frac{3n_a^2}{\rho \beta^1} \ +\ \frac{2n_b^1}{\beta^2} \ +\ 
\frac{n_d^2}{\beta^1} \ = \ 
16 \ . 
\label{tadsm} 
\eeq 
Note however that one can always relax this constraint by 
adding extra D6-branes with no 
intersection with the previous ones and not contributing to the 
rest of the tadpole conditions. In particular, any set of branes 
with wrapping numbers of the form 
\beq 
N_h\, (1/\beta_1,0)( 1/\beta _2,0)(n_h^3,m_h^3)  
\label{hidden} 
\eeq 
does not intersect the branes from table \ref{SMbranes} whereas contributes  
with $N_hn_h^3/(\beta_1\beta_2)$ to the first condition (without affecting the
other three). We will call such D6-brane a {\it hidden brane}. 

As advanced in \cite{cim1}, we can implement this same idea in a more 
general situation. Let us consider an arbitrary configuration of  
D6-branes whose low energy chiral spectrum has neither chiral nor 
mixed anomalies. Since tadpoles imply anomaly cancellation 
but not the other way round, it could happen that the brane content 
giving rise to such anomaly-free spectrum did not satisfy tadpoles. 
We must then `complete' our model by simply adding the necessary 
brane content to cancel RR charges. In particular, we can complete 
our brane configuration by a single brane, let us call it $H$-brane. 
Unlike in (\ref{hidden}), in general this $H$-brane will not have a simple  
expression, but will be wrapping a non-factorizable cycle 
\footnote{See Appendix I for a discussion of non-factorizable cycles.}  
$[\Pi_H] $ of $T^6$. 
The important point to notice is that, 
as long as the low-energy fermion spectrum 
is anomaly-free, this $H$-brane will have no net 
intersection with any brane belonging to our initial  
configuration, just as in our previous example. Thus, it will constitute 
a hidden sector, and its presence will not imply the existence 
of new chiral massless fermions in our low energy spectrum. 
 
Alternatively, and following the recent proposal in \cite{uranga} 
(see also \cite{flujos}), 
we could equally well turn on an explicit NS-NS background  
flux $H_{NS}$ in our configuration. This flux will have an 
associated homology class $[H_{NS}]$, behaving as an 
RR source just as a D6-brane in this same homology class  
would do. Thus, we can consider configurations where tadpoles 
cancel by a combination of D6-brane content and NS-NS flux, 
the presence of the latter not implying new chiral spectrum. 
Furthermore, the presence of these fluxes may relax even more 
the model-building constraints, since the addition 
of $H_{NS}$ can compensate the anomaly generated by an anomalous  
$D = 4$ chiral spectrum. 
The reason for this is that the presence of $H_{NS}$ will induce  
a Wess-Zumino term 
in the low energy Lagrangian, which will cancel the  
$SU(N_a)^3$ and $U(1)_a-SU(N_b)^2$ anomaly developed from 
the naive `fermion content' point of view (see \cite{uranga}).  
Unlike the addition of some $H$-brane to complete an  
anomalous configuration, such background flux will not imply  
new chiral content to our spectrum\footnote{If we added such a 
D6-brane $H$ in order to complete an {\it anomalous } configuration, 
its net intersection number would not vanish for some brane $a$  
contained in it. Clearly, in this case we would 
not be allowed to call this extra brane a {\it hidden brane}.}. 
We will use the addition of both RR sources in some of the explicit 
models built in Section 4. However, in order to  
limit the arbitrariness,  we will only consider the  
possible presence of H-flux induced  Wess-Zumino terms  
for anomalous massive $U(1)$'s so that none of the 
anomalies of the SM gauge groups have to be canceled via 
Wess-Zumino terms.

One of the most interesting 
aspects of this class of theories is the structure of 
Abelian gauge symmetries. The four $U(1)$ symmetries $Q_a$, $Q_b$, $Q_c$ and 
$Q_d$ have clear 
interpretation in terms of global symmetries of the standard model. 
Indeed, $Q_a$ is $3B$, $B$ being the baryon number, and $Q_d$ is nothing 
but lepton number 
\footnote{In all the brane intersection models discussed in 
this article the baryon number symmetry $U(1)_a$ is automatically 
gauged. Although the corresponding generator becomes  
eventually massive, baryon number remains as an accidental  
symmetry in perturbation theory \cite{imr}. Thus the 
proton is perturbatively stable.} 
. Concerning $Q_c$, it is twice $I_R$, the third 
component of right-handed weak isospin familiar from left-right 
symmetric models. Finally $Q_b$ has the properties of a Peccei-Quinn   
symmetry, having mixed $SU(3)$ anomalies. 
Two out of the four $U(1)$'s have triangle 
anomalies which are canceled by a generalized Green-Schwarz mechanism 
\cite{imr}. 
The anomalous symmetries correspond to generators  $Q_b$  
and $3Q_a+Q_d$, the latter being identified with $(9B+L)$. 
For a general brane configuration the  
 anomaly cancellation mechanism goes as follows. 
There are four closed string antisymmetric 
fields $B_2^{I}$, $I = 0,1,2,3$  which couple to the four $U(1)_a$ fields 
associated to each set of $N_a$ D6-branes in the form: 
\beq 
 \int_{M_4}\, \sum_a \,  N_a(  m^1_a\, m^2_a\, m^3_a \ B_2^0   \ 
+\  
\sum_I\, n^J_a\, n^K_a\, m^I_a \  B_2^I \ )\ \wedge F_a  \, \, ,\, 
I\not=J\not=K  
\label{bf} 
\eeq 
On the other hand, the Poincare duals of those antisymmetric fields, 
denoted $C^I$, $I=0,1,2,3$ couple to both Abelian and non-Abelian fields  
$F_b$ as follows: 

\beq 
 \int_{M_4} \, \sum_b \, ( \,  n^1_b\, n^2_b\, n^3_b \, C^0  \ +\  
 \sum_I\, n^I_b\, m^J_b\, m^K_b \, C^I\,  ) \, \wedge F_b\wedge F_b  \ . 
\label{cff} 
\eeq 
The combined effect of these two couplings cancel the residual 
$U(1)_a\times U(N_b)^2$ triangle anomalies by the tree level exchange  
of the four RR fields. At the same time the $B_2^I\wedge F$ couplings 
in (\ref{bf}) give masses to the following four linear combinations 
of Abelian fields  
\beq 
\begin{array}{c} 
B_2^0 \ :\quad\sum_a   \, N_a \,   m^1_a\, m^2_a\, m^3_a  \, F_a, \\ 
B_2^1 \ :\quad\sum_a   \, N_a \,   m^1_a\, n^2_a\, n^3_a  \, F_a, \\  
B_2^2 \ :\quad\sum_a   \, N_a \,   n^1_a\, m^2_a\, n^3_a  \, F_a, \\  
B_2^3 \ :\quad\sum_a   \, N_a \,   n^1_a\, n^2_a\, m^3_a  \, F_a.  
\label{caplillos} 
\end{array} 
\eeq 
Since in the specific models we are constructing at least one of the  
$m_a$'s of each brane vanishes, the first linear combination  
is trivially vanishing and there is always a massless $U(1)$.  
The three linear combinations which are massive are : 
\beq 
\begin{array}{c} 
Q_{b} \\ 3Q_a + Q_d \\ 
Q_m = \left( {3\beta^2 n_a^2} Q_a \, + \, 6\rho \beta^1 n_b^1 Q_{b}  
\, - \, 2\tilde \eps\beta^1n_c^1 Q_c \, -  \, 3\rho \beta^2 n_d^2  Q_d 
\, - \, 2N_h \tilde \eps m_h^3  Q_h \right) 
\end{array} 
\label{massive} 
\eeq 
where for the sake of generality we have included the effect 
of a set of $N_h$ parallel ``hidden'' branes as discussed above.  
The first two massive generators are model independent 
and correspond to the anomalous $U(1)$'s, whereas $Q_m$ is model-dependent 
and anomaly-free. 
The massless $U(1)$'s are: 
 
{\it i)\ $m_h^3\ =\ 0$} 
 
In this case the presence of a hidden brane does not affect the  
form of the massless linear combinations which are given by: 
\beq 
Q_0\ =\ {1\over 6} Q_a  +  {r\over 2} Q_c - {1\over 2} Q_d   
\,  , \quad  r = {\tilde\eps\beta^2\over  
2n_c^1\beta^1}(n_a^2+3\rho n_d^2) 
\label{hyper} 
\eeq 
and the hidden brane charge $Q_h$. Note that for $r=-1$ the above 
generator is  the standard hypercharge generator, as discussed  
in \cite{imr}. In this case only this standard hypercharge generator 
remains massless 
\footnote{As noted in \cite{imr}, if in addition one has  
$n_c^1 = 0$, then an extra generator $(1/3)Q_a-Q_d$  
(which corresponds to B-L) remains also massless.}. 
 
{\it ii) \ $m_h^3\ \neq \ 0$} 
 
In this case the hidden brane affects the form of the visible 
linear combination. The massless linear combinations are: 
\beqa 
& & Q_1\ =\ {1\over 6} Q_a -  {1\over 2} Q_c - {1\over 2} Q_d  + 
(1+r){{n_c^1\beta^1}\over {2N_hm_h^3} }Q_h \nonumber \\ 
&  & Q_2\ = \ {1\over 3} Q_a  -  Q_d   
+ {{rn_c^1\beta^1}\over {N_hm_h^3}}Q_h 
\label{uunose} 
\eeqa 
Note that the first of these charges couples like standard hypercharge  
to the SM particles, whereas the second one couples like $B-L$. Thus, in 
this more general case two massles $U(1)$'s coupling to quarks  
and leptons remain in the massless spectrum.  
 
The general solutions yielding the SM spectrum 
have the following geometrical properties. The Baryonic ($a$) 
and Leptonic ($d$) stacks are parallel in the first 
complex dimension and that is why they do not intersect. 
Thus, no lepto-quarks fields appear in our massless spectrum. 
On the other hand, Left ($b$) and Right ($c$) stacks are parallel in 
the second complex dimension, again not intersecting.   
Something similar happens for each 
brane $i$ and its mirror $i^*$, $i = a,b,c,d$. 
Strings exchanged between stacks $b$ and $c$ have the quantum numbers of 
SM Higgs (and Higgsino) fields and eventually we would be interested 
in some of these scalar states to be relatively light, so as to 
play the role of Higgs fields. So we will asume that the distance between 
$a$ and $d$ branes is bigger than that between branes $b$ and $c$, so 
that the latter can provide us with a Higgs system,  
as we will describe in more detail in section 7. 
Note that we  will consider the different distances between branes  
to be fixed quantities.  
Hence, they will play the role of external parameters of 
our models, very much like the geometrical moduli. 
 
Notice that the whole of the previous SM construction has been achieved  
by using factorizable D6-branes, that is, branes whose geometrical locus 
can be described as product of 1-cycles, each wrapping a different $T^2$.  
This kind of  
construction is preferred from a model-building point of view, since 
it has a particularly simple associated geometry. This allows us to  
easily compute some phenomenologically important quantities as, for 
instance, scalar masses at intersections. However, we could have  
also constructed a model in a more general set-up, where the 
building blocks of our configuration were D6-branes wrapping {\it general} 
3-cycles of our six-torus. In this way, we could have equally well 
realised a D6-brane tadpole-free configuration giving rise to a  
SM spectrum, the factorizable models just presented being a  
particular subfamily of the latter. 
 
For simplicity and better visualization of our constructions, we will  
give as examples models where the relevant physics arises at 
factorizable D6-branes configurations, so as to extract the  
phenomenologically interesting quantities more straightfordwardly. 
However, it is important to notice that reached this point it is not 
possible to ignore non-factorizable D6-branes anymore. In particular,  
when studying the recombination process of two branes into a third one 
(corresponding to some Higgs mechanism)  
such non-factorizable branes will naturally arise. As we will see, this  
is due to the fact that conservation of RR charge imposes that this third  
brane should wrap a 3-cycle which is the sum of the two former 3-cycles. 
In general, this will yield a non-factorizable brane. 
In order to appropiately describe these less intuitive objects, we 
have made use of a ``$q$-basis'' formalism, which is described in  
Appendix I. For some results regarding the geometry of branes wrapping 
general cycles see also \cite{bbh}. 
 
Let us also emphasize that  
in the present paper we will be dealing mostly with the open  
string sector of the theory, which is the one which may give  
rise to the SM physics. We will not discuss here the closed string  
NS potential and its associated NS tadpoles. We will rather take   
the value of NS geometric moduli as well as brane positions as frozen   
external parameters defining the geometry of the configuration. 
Some work on the NS-tadpole structure of this class 
of theories may be found in, e.g., ref.\cite{bklo,cim1,blumy}.  
 
\section{Intersecting D6-branes and  
Supersymmetry: SUSY \& Q-SUSY models} 
 
An interesting question is whether one can construct  
intersecting brane models analogous to those discussed in the 
previous section  and with $\cn = 1$ supersymmetry.  
As discussed in refs.\cite{bgkl,cim1}, 
the answer to this question  
is no, if we restrict ourself to purely toroidal models  
(no orbifolds) with factorized  
D6-branes and impose RR tadpole cancellation conditions.  
Models with $\cn = 1$ SUSY may be constructed if an additional 
$\IZ_2\times \IZ_2$ orbifolding is perfomed \cite{csu}, but 
at the cost of losing the simplicity of the spectrum, 
as exotic chiral particles beyond the SM appear. 
 
On the other hand it was shown in \cite{cim1} the  
possibility of having all brane intersections preserving 
{\it some} unbroken $\cn = 1$ SUSY, although not necessarily   
the same one. These configurations were called 
quasi-supersymmetric, (Q-SUSY).  
Roughly speaking, a Q-SUSY field 
theory consists of different subsectors each preserving at least $\cn = 1$ 
supersymmetry but, not being the same supersymmetry for every sector,  
the system as a whole has $\cn = 0$. Thus, we expect corrections  
that spoil the boson-fermion mass degeneracy appearing in a truly 
supersymmetric theory, this effects appearing only at two-loops in 
perturbation theory. There might also be some {\em massive} sectors  
not respecting any supersymmetry at all. In any case, the radiative 
corrections will be fairly supressed with respect to an ordinary $\cn = 0$ 
field theory. In particular, scalars will have a two-loop protection  
against quadratic divergences. This may be of  interest when 
dealing with scenarios where the fundamental scale of physics is two or 
three orders of magnitude above the electroweak scale \cite{cim1}. 
A two-loop protection would be enough to understand a 2-3 orders 
of magnitude hierarchy between the weak scale and a fundamental scale 
of order 10-100 TeV. 
 
Let us describe the D-brane setup that allows for an explicit realization 
of this Q-SUSY structure. Just as in the construction of the Standard Model 
presented in last section, we will be dealing with Type IIA D6-branes  
wrapping factorized 3-cycles in $T^2 \t T^2 \t T^2$. Any factorized  
D6-brane constitutes a $\oh$BPS state that preserves half  
of the supersymmetries coming from a plain toroidal compactification. 
This will be reflected in the $D6_aD6_a$ sector of the theory (that is, 
strings beginning and ending on the same $D6_a$-brane), yielding $D = 4$  
$\cn = 4$ super Yang-Mills as its low energy spectrum.  
When considering a pair of such factorized branes, 
say $a$ and $b$, each of them will preserve some $\cn = 4$ subalgebra  
of the whole $\cn = 8$ bulk superalgebra. Whether these two different 
subalgebras do have some overlap or not will determine if the $D6_aD6_b$  
sector is also supersymmetric (see figure \ref{venn}).
\EPSFIGURE{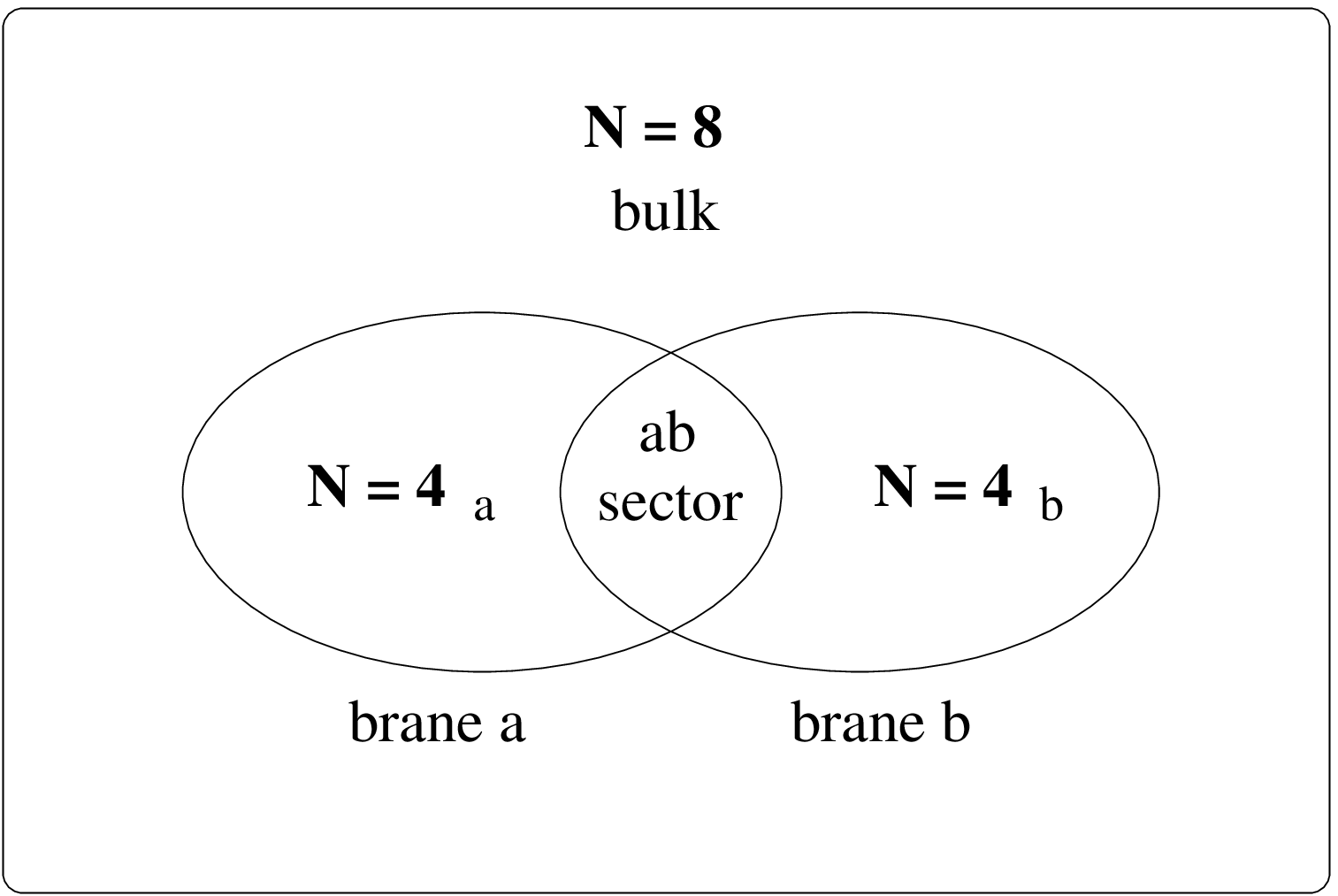, width=4in, height=2.5in}
{\label{venn}
Schematic representation of the supersymmetries preserved by 
a pair of branes in a toroidal compactification. The closed string 
sector, living on the bulk, preserves a full $D = 4$ $\cn = 8$  
superalgebra, whereas each of the branes $D6_a$ and $D6_b$ preserve  
$\cn = 4$ subalgebras, generically different ones.  
Whether the $D6_aD6_b$ sector is supersymmetric or not depends on 
the overlap that exists between these two algebras. 
In the orientifold case the bulk has only $\cn = 4$ SUSY.}
In general, the answer will be positive whenever these two branes  
are related by a $SU(3)$ rotation  
in compact dimensions \cite{bdl}. When dealing  
with D6-branes at angles, this can be easily computed in terms of the  
spectrum living at the $D6_aD6_b$ sector. Whereas the Ramond sector  
always yields a massless chiral fermion, the lightest states coming  
from the Neveu-Schwarz sector will be four scalars with masses  
\cite{afiru} 
{\small \beqa 
\begin{array}{c} 
 \quad {\bf \alpha' Mass^2} \\ 
\frac 12(-|\vartheta_{ab}^1|+|\vartheta_{ab}^2|+|\vartheta_{ab}^3|) \\ 
\frac 12(|\vartheta_{ab}^1|-|\vartheta_{ab}^2|+|\vartheta_{ab}^3|) \\ 
\frac 12(|\vartheta_{ab}^1|+|\vartheta_{ab}^2|-|\vartheta_{ab}^3|) \\ 
1-\frac 12(|\vartheta_{ab}^1|+|\vartheta_{ab}^2|+|\vartheta_{ab}^3|), 
\label{scalars} 
\end{array} 
\eeqa} 
where $\vt_{ab}^i$ represent the angles between branes $a$ and $b$  
in the $i^{th}$ torus (see fig. \ref{compact2}). The number of  
SUSY's shared by sectors $D6_aD6_a$ and $D6_bD6_b$ 
will correspond to the number of massless scalars in (\ref{scalars}). 
Indeed, if for instance $\vt_{ab}^1 = \vt_{ab}^2 + \vt_{ab}^3$, then 
the first of such scalars will be massless, and the $D6_aD6_b$ sector 
will consist of $|I_{ab}|$ $\cn = 1$ chiral multiplets. 
 
In an orientifold compactification, however, a generic D6-brane  
$a$ cannot exists on its own, but will be accompanied by its  
mirror image $a^*$. In order to achieve our Q-SUSY structure, we will require 
any brane and its mirror to preserve at least one common SUSY. In fact, 
we will require them to share a $\cn = 2$ subalgebra \footnote{From the 
model-building point of view, this is required in order to avoid 
massless chiral multiplets in the $D6_aD6_{a^*}$ sector, which does 
only provide us with matter transforming as  
symmetric and antisymmetric representations. We will forbid 
such kind of exotic spectrum right from the start.}. As discussed in  
\cite{cim1}, there exist six types of such branes. In order to describe  
them, let us define a {\em twist} vector  
$v_\a = (\th_\a^1,\th_\a^2,\th_\a^3)$  
whose three entries contain the relative angles $\th_\a^i$ between  
a D6-brane $\a$ and the horizontal axis of the $i^{th}$ torus  
(see figure \ref{twist}).
\EPSFIGURE{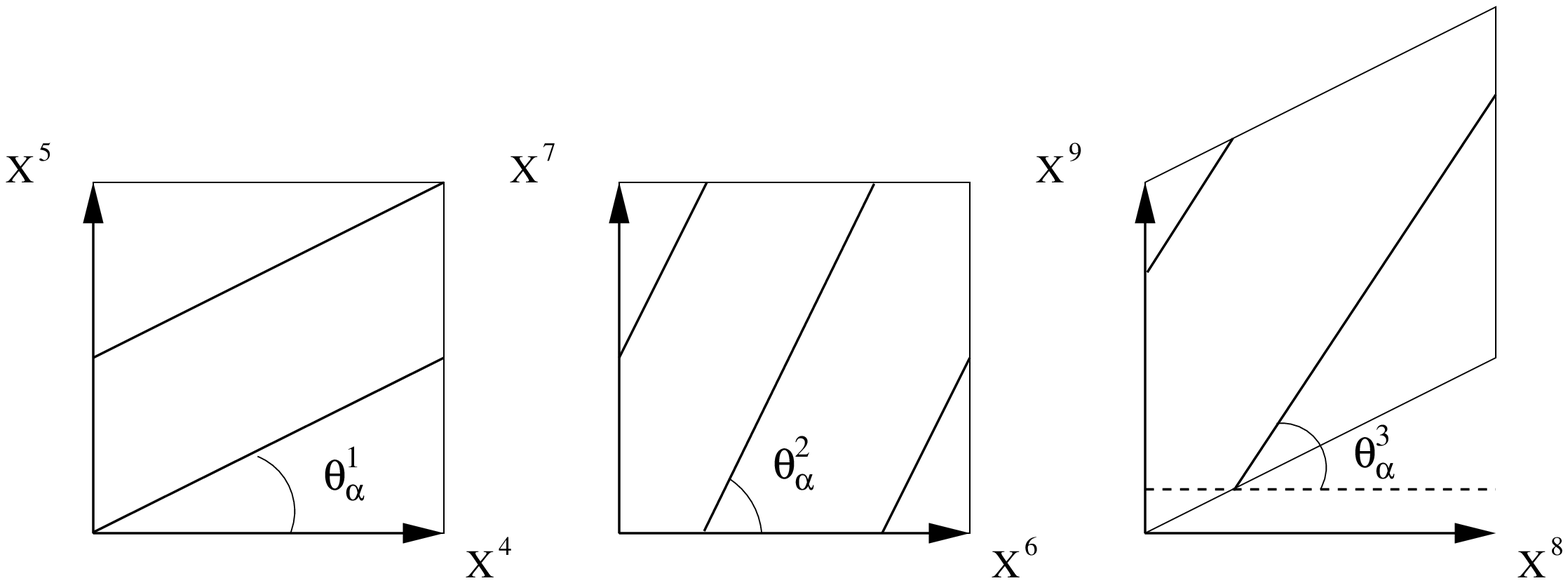, width=5in}
{\label{twist}
Definition of the twist vector $v_\a := (\th_\a^1,\th_\a^2,\th_\a^3)$.}

In terms of this twist vector, the six different types of branes can  
be listed as 
{\small 
\beq 
\begin{array}{rcc} 
{\rm type\ of\ brane} & {\rm twist\ vector} & {\rm SUSY} \\ 
a_1\ {\rm branes}\ : & v_{a1} = (0,\th_{a1},\th_{a1}) & r_2, r_3 \\ 
a_2\ {\rm branes}\ : & v_{a2} = (0,\th_{a2},-\th_{a2}) & r_1, r_4 \\ 
b_1\ {\rm branes}\ : & v_{b1} = (\th_{b1},0,\th_{b1}) & r_1, r_3\\ 
b_2\ {\rm branes}\ : & v_{b2} = (\th_{b2},0,-\th_{b2}) & r_2, r_4 \\ 
c_1\ {\rm branes}\ : & v_{c1} = (\th_{c1},\th_{c1},0) & r_1, r_2 \\  
c_2\ {\rm branes}\ : & v_{c2} = (\th_{c2},-\th_{c2},0) & r_3, r_4 
\end{array} 
\label{vectors} 
\eeq}
where we have suppressed the upper index $i$. 
We have named the generators of each $\cn = 1$ algebra by  
$r_j$, $j = 1,2,3,4$, just in order to keep track 
of the supersymmetries present on each sector (for a more detailed 
description see \cite{cim1}). Each type of brane  
in (\ref{vectors}) shares a different $\cn = 2$ subalgebra  
with their mirror brane, which can be represented by a pair of  
generators \footnote{In a D-brane language, the twist 
vector $v_a$ contains the relative angles between the D6-brane $a$ 
and the O6-plane, who lies on the cycle $(1/\b^1,0)(1/\b^2,0)(1/\b^3,0)$. 
Hence, it encodes the supersymmetries shared by both. If we  
name the generators of the $\cn =4$ algebra of the O6-plane by 
$r_1, r_2, r_3, r_4$, then the six types of branes described above  
will correspond to different choices of a $\cn = 2$ subalgebra.}.  
Notice that the twist vector of a mirror brane $\a^*$  
will be given by $v_{\a^*} = - v_\a$, whereas the angles 
between two branes $\a$ and $\b$ are the components of   
$v_{\a\b} := v_\b - v_\a$. 
 
When considering a configuration where several of such $\cn = 2$ branes  
appear, we must also consider sectors corresponding to D6-brane  
intersections. The full Q-SUSY structure can be encoded in a 
hexagonal quiver-like diagram, shown in figure \ref{hexagon}.  
Each node of this diagram represents one of the branes in  
(\ref{vectors}), and at the same time its mirror image.  
Notice that gauge groups at intersecting D6-branes models always arise 
from $D6_\a D6_\a$ sectors, hence they will be localized at these nodes. 
Just as in the toroidal case, each gauge group will correspond to a  
$\cn = 4$ subsector of the theory, while the matter content living 
on the $D6_\a D6_{\a^*}$ sector presents a non-chiral,  
generically  massive  
$\cn = 2$ spectrum. 

\EPSFIGURE{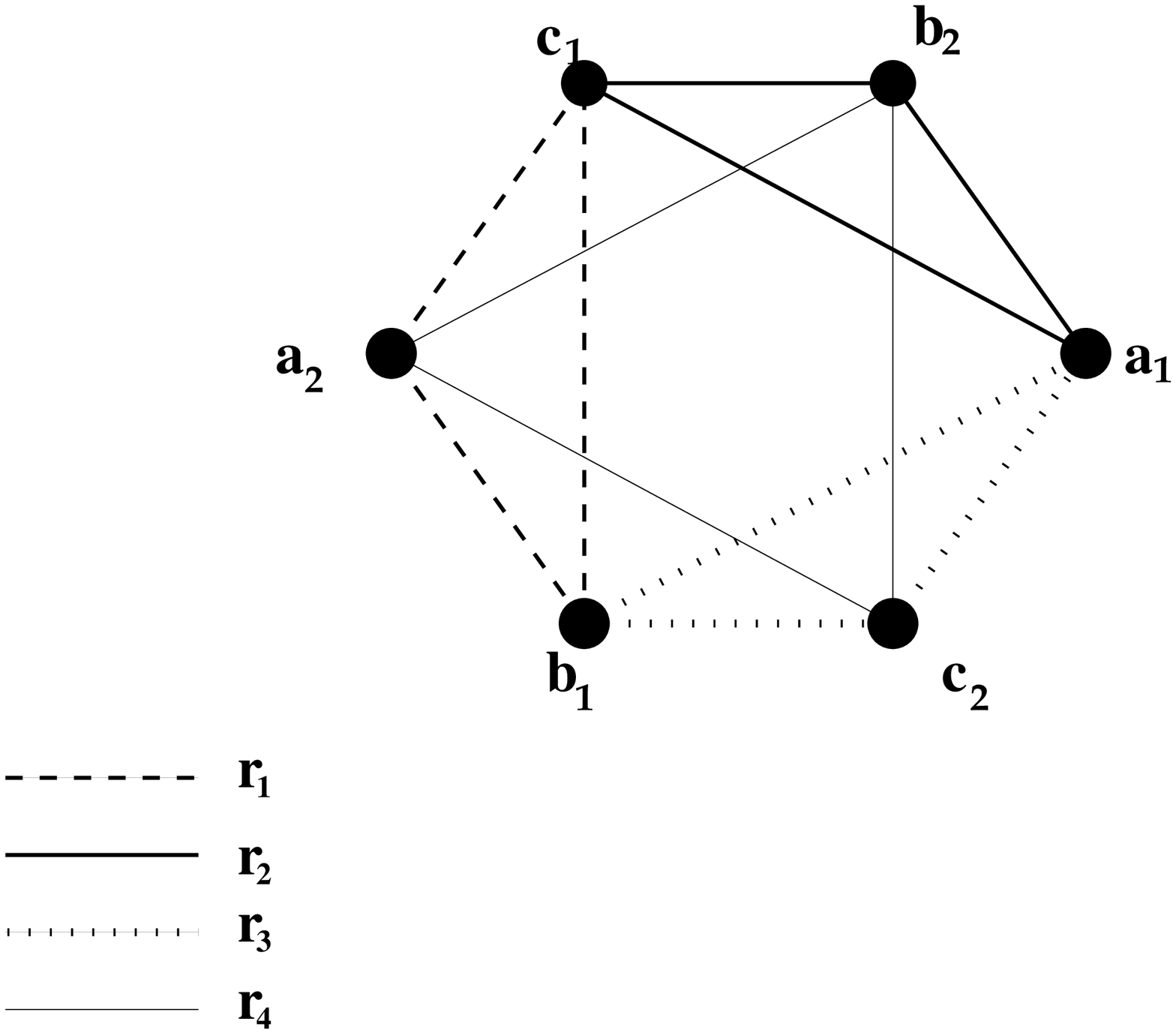, width=3in, height=2.4in}
{\label{hexagon}
General SUSY quiver diagram. Nodes represent the six different 
types of branes (plus mirrors) whereas links represent the 
chiral intersections of those branes. There are four types of links 
corresponding to the four SUSY's in the bulk.}
 
Each of the branes in (\ref{vectors}) will generically intersect with 
four other types of brane. Such intersection is represented in our  
hexagonal quiver by a link joining two nodes. In general, a non-vanishing  
intersection number $I_{\a\b}$ does signal that some chiral spectrum lives 
at the $D6_{\a}D6_{\b}$ sector. It can easily be seen that any of these  
sectors will also preserve some $\cn = 1$ SUSY, yielding chiral multiplets 
transforming in bifundamentals. Let us take, for instance, the link joining 
$a_2$ with $b_2$. Each node, as a combined system of a brane and its mirror, 
preserves two supersymmetries, having $r_4$ in common. Thus, the link  
between these two nodes will contain $I_{a_2b_2} \ (N_{a_2},\bar{N}_{b_2})\  
+\ I_{a_2b_2^*} \ (N_{a_2},N_{b_2})$ chiral multiplets under the supersymmetry 
generator $r_4$. Finally, branes corresponding to opposite nodes (as for  
instance $a_1$ and $a_2$) do never intersect, since they are parallel in one  
of the tori. This $D_{\a_1}D_{\a_2}$ sector will contain a non-chiral,  
generically massive, $\cn = 0 $ spectrum.  
 
We thus see that any low energy effective field theory coming from  
a D-brane configuration whose building blocks belong to (\ref{vectors}) 
will yield a Q-SUSY field theory. In particular, the general hexagonal 
construction depicted in fig. \ref{hexagon} will contain as massless  
sectors $\cn = 4$ vector multiplets and $\cn = 1$ chiral multiplets, 
whereas some non-chiral $\cn = 2$ and $\cn = 0$ massive sectors may  
also arise. As we mentioned at the beginning of this section, this will 
protect scalar masses from one-loop corrections, the first  
non-vanishing contributions coming from two-loops in perturbation theory 
\cite{cim1}. 
There are contributions coming both from light particle exchange 
(fig.(\ref{loop5}-a)) and heavy non-SUSY particle  
exchange (fig.(\ref{loop5}-b)).   
All will give corrections to scalar  
masses of order $\alpha_i/(4\pi)M_s$.  
For the $\cn = 4$  gauginos  the only source of masses will be the one-loop  
exchange of   massive $\cn = 0$ sectors of the theory (see  
figure (\ref{loop5}-c)).  

\EPSFIGURE{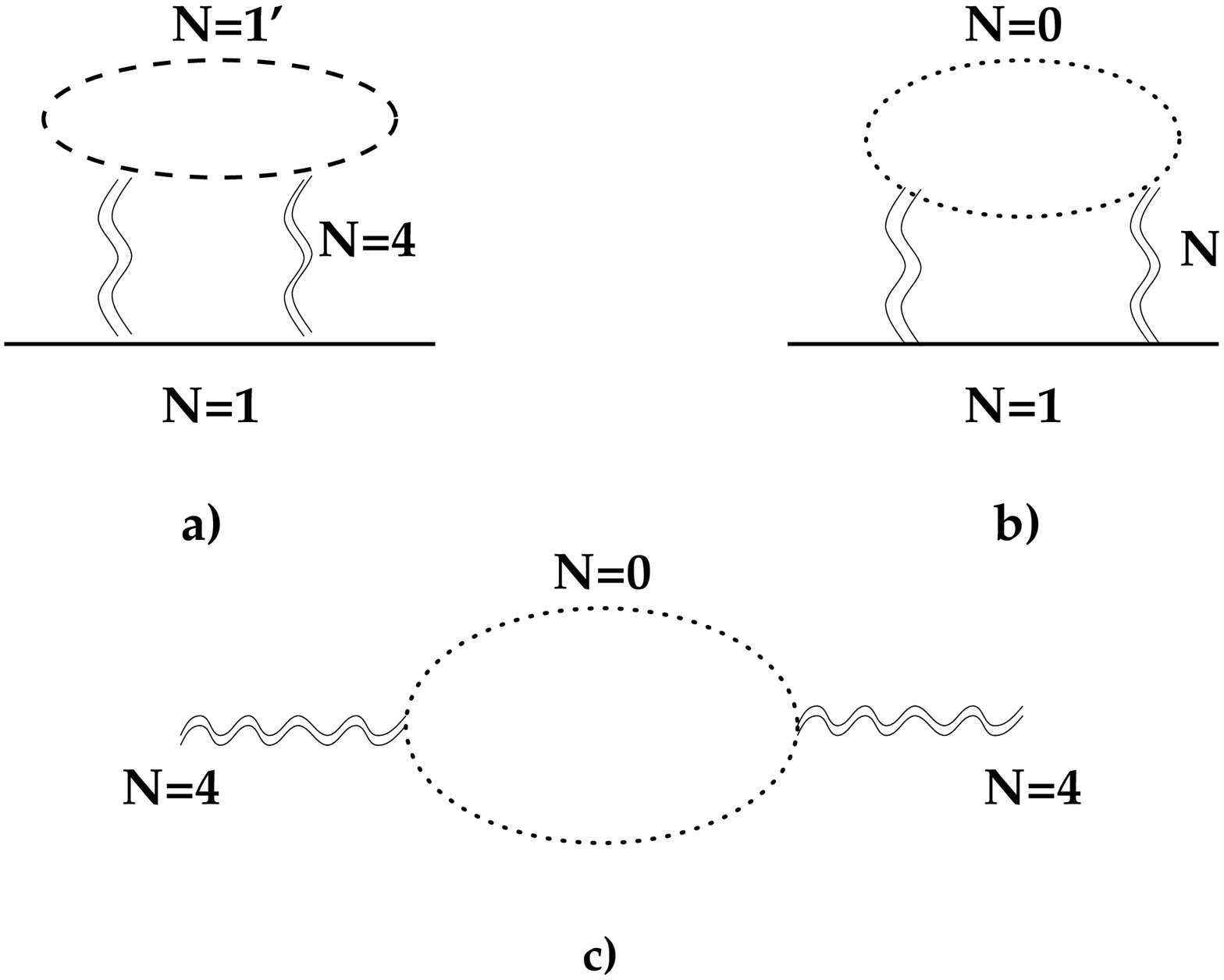, width=4in, height=2.5in}
{\label{loop5}
First non-vanishing loop contributions to the masses of 
scalar fields and gauginos   
in a Q-SUSY model: a) Quadratic divergent 
contribution present  when the upper loop contains 
{\it massless } fields 
respecting different supersymmetries than those of 
the  fields below; b) Contribution coming from 
possible  {\it massive } non-supersymmetric states 
in the upper loop, c) One-loop contribution to the masses of gauginos    
from heavy non-supersymmetric states.} 
 
These general considerations will equally well apply to any subquiver 
of the hexagonal construction just presented. In particular, we will be 
interested in models containing four stacks of branes. This is essentially 
because, just as in the explicit example presented in last section, we  
need at least four branes in order to arrange the SM chiral spectrum  
as coming from bifundamental representations.  
Let us then consider some four-stack subquivers arising from the hexagonal  
construction. Having four stacks implies that at least there is a pair 
of them with vanishing intersection, thus yielding either a $\cn = 2$  
or a $\cn = 0 $ sector. Without loss of generality, we will take two  
$a$-type branes to be such pair, and one of them to be of one  
definite kind, say $a_2$. In order to have a non-trivial chiral spectrum, 
we need the other two branes to be of type $\a_i$, with $\a \neq a$.  
Our general four-stack configuration will be of the form 
$\left(a_2, a_i^\prime, \b_j, \g_k \right)$.  
There are some inequivalent choices: 
\begin{itemize} 
 
\item Either $i = 2$, having a massive $\cn = 2$ $D6_aD6_{a^\prime}$ sector,  
\\ or $i = 1$, having a massive $\cn = 0$ $D6_aD6_{a^\prime}$ sector. 
 
\item Either $\b = \g$, having a non-chiral spectrum on the $D6_\b D6_\g$ 
sector,  
\\ or $\b \neq \g$, having a $\cn = 1$ chiral spectrum. 
 
\item Either $j = k$, so that at least one of the $a$ branes preserves  
the same supersymmetry with both $\b$ and $\g$ branes,  
\\ or $j \neq k$, so that both supersymmetries are different. 
 
\end{itemize} 
 
We thus see that we have eight inequivalent possibilities when considering 
four-stack subquivers, as the rest of them just amount of some relabeling 
of the nodes. However, performing a simple anomaly cancellation analysis, 
we can reject half of them as phenomenologically uninteresting. This comes  
from the fact that, when performing our model-building, we will usually 
choose $(a,a^\prime)$ to be the Baryonic and Leptonic branes, respectively. 
Being a non-intersecting pair of branes, this will avoid massless 
lepto-quarks in our spectrum. Now, in order to cancel $SU(3)$ anomalies, 
we need fermions transforming both in {\bf 3} and in ${\bf \bar 3}$ 
representations, same number of them on each. It can be easily shown 
that, in order for this to be true in our models, we must have either 
$\b = \g$ and $j \neq k$, or $\b \neq \g$ and $j = k$. So we are 
finally led to consider only four subquivers, which are depicted  
in figure \ref{quivers}. 
 
\EPSFIGURE{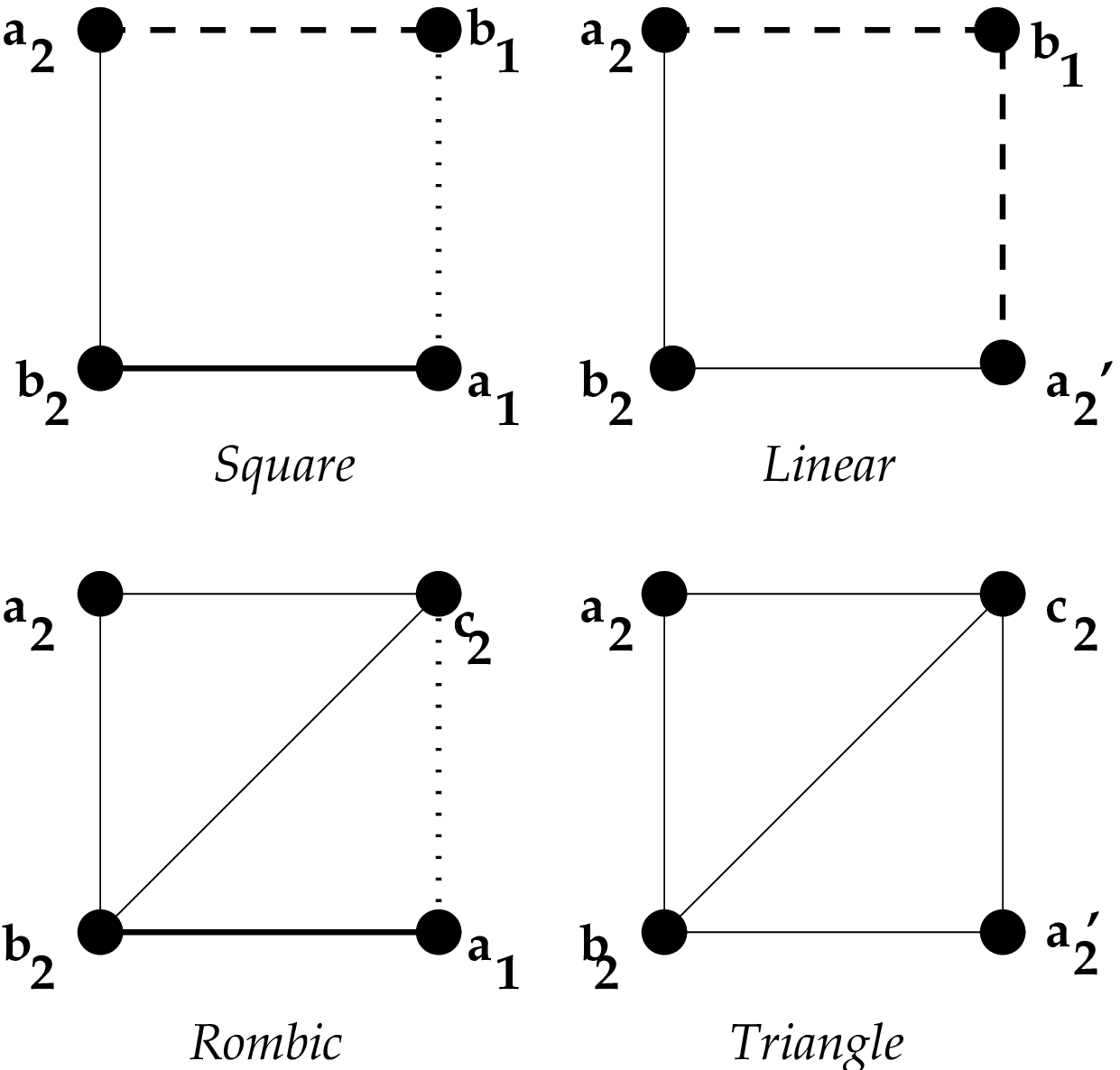, width=4in, height=3.8in}
{\label{quivers}
Simplest SUSY-quivers with four nodes. 
They are obtained from the general hexagonal quiver  
by deleting two nodes. Other equivalent quivers are obtained  
by cyclic relabeling $a\rightarrow b\rightarrow c$. 
The rightmost examples may be depicted as linear (triangle) quivers 
respectively if one 
locates the $a_2$' branes on top of their 
parallel relatives $a_2$.} 
 
In the next section we will try to combine these four Q-SUSY structures 
with phenomenologically appealing brane configurations, following the  
philosophy described in Section 2 for  getting the SM from general  
intersecting branes.

\section{Model-building: three generation models with SUSY intersections }

In this section we are going to consider the construction of  
explicit intersecting D6-brane models of phenomenological interest. 
In particular we will focus in the construction of models  
with quasi-SUSY in the sense described in previous section  
and ref.\cite{cim1}  with three quark-lepton generations and  
SM group (or some simple left-right extension). 
We will concentrate in the class of branes discussed above in which  
each individual brane preserves an unbroken $\cn = 2$ supersymmetry 
with its mirror brane, having null intersection with it. 
 
As was discussed above, in this class of brane configurations 
only up to three $U(1)$'s  may become massive by combining  
with three of the four antisymmetric fields $B_{\mu \nu }^i$ 
present in the massless spectrum \cite{imr}. We will be considering  
four stacks of  branes $a,b,c$ and $d$ each of them containing  
$N_a=3$, $N_b=2$, $N_c= 1,2$ and $N_d=1$ parallel branes. 
Thus to start with our gauge group will be either 
$U(3)_a\times U(2)_b\times U(1)_c\times U(1)_d$  
or $U(3)_a\times U(2)_b\times U(2)_c\times U(1)_d$. 
 
The most general phenomenologically interesting Q-SUSY models one  
can consider involving four stacks of branes may be depicted  
by the four SUSY-quivers shown in fig. \ref{quivers}.  
Locating the four stacks 
of branes of the SM at the four corners of those quivers 
we will get Q-SUSY models with the SM gauge group (and possibly 
some extra $U(1)$ symmetry as discussed below). 
Without loss of generality, we will take the Baryonic stack 
to to be of type $a_2$, whereas the Left stack can also be chosen to 
be of some definite type different from $a$, say $b_2$. Those two  
stacks will yield a gauge group $U(3)\times  U(2)$. 
Clearly, the leptonic brane must be non-intersecting with 
the baryonic brane, since we want to avoid lepto-quark particles. 
It must then be of the same type $a$ as the latter.  
Depending on what kind of brane one chooses for the right and  
leptonic stacks, different model building  possibilities 
will be obtained: 
 
{\it i) Square quiver (fig.(\ref{qsimr})).} 
 In this case   
there are four types of intersections respecting four different  
types of SUSY. They will correspond to Q-SUSY versions of the 
three generation configurations of ref.\cite{imr}. 
 
{\it ii) Linear quiver  (fig.(\ref{linear})).} 
This  quiver may be put into a line by 
putting the leptonic $a_2$' brane on top 
of the baryonic stack $a_2$. 
It may be considered as a variant of the square quiver 
in which the left-handed fermions have   one 
supersymmetry and right-handed leptons a different one. 
  
{\it iii) Rombic quiver  (fig.(\ref{rombic})).} 
In this third  quiver three different SUSY's are present.  
One main difference with the previous cases  
is the presence of  Higgs chiral multiplets at the intersections of 
right and left stacks. Note that the baryonic and Higgs sub-sector  
will respect the same $\cn = 1$ supersymmetry, whereas the leptonic 
sector respect other ones.  
 
{\it iv) SUSY-triangle  (fig.(\ref{MSSM})). } 
In this case baryon and lepton branes are of the same ($a_2$) type 
so that the quiver may be deformed into a triangle by locating one 
on top of each other. Note that now all intersections 
preserve the same $\cn = 1$ SUSY and, as in the rombic quiver, there  
are massless Higgs multiplets at the intersections.

It turns out that is is easy to find families of examples 
corresponding to the {\it square } and {\it  linear} quivers. 
On the contrary, the {\it triangle} and { \it rombic} 
quivers are more restricted configurations and examples  
are more scarce. 
Let us emphasize that all of these brane configurations  
lead in general to RR-tadpoles. Those may be canceled in a 
model by model basis by adding extra D-branes (with no intersection 
with the ones in the quivers, so that the massless chiral spectrum is 
not modified) and/or the addition of explicit   $H_{NS}$-flux, as described 
in Section 2.   
 
There are some general aspects of these Q-SUSY configurations  
which can be given without the need of going into the  
details of the models. One can see that the following  
 properties hold: 
 
{\it 1)} 
`` In any square, rombic and linear quiver  
there is, at least, one $U(1)$ generator which remains massless 
(does not combine with an antisymmetric B-field). 
In the triangle quiver there are two such massless 
$U(1)$'s.'' 
 
{\it 2) }  
``In any SUSY or Q-SUSY (factorizable) D6-brane configuration yielding 
SM fermion spectrum the  $U(1)_c$ (right) gauge generator is massless.'' 
 
{\it 3)} 
`If we insist that the left- and right-handed counterparts of the  
same field respect the {\it same} $\cn = 1$ SUSY, then necessarily  
the brane configuration must have intersections at which  
multiplets with the quantum numbers of SM Higgs arise'. 
 
The first  two general properties are proven in Appendix II. 
The second one follows because  
it turns out that supersymmetry at all intersections 
forces  the right-stack to have wrapping numbers $m_c^2=0,$  
$n_c^1 = n_c^3 = 0$. Notice that this massless generator $U(1)_c$ 
may or may not coincide with the one(s) mentioned in point {\it 1)}. 
Concerning the third one, it is a direct consequence of the 
quiver classification above: in order for the 
{\it Baryonic} stack (of type $a_i$) to have intersections giving rise 
to e.g., left- and right-handed quarks respecting the same 
$\cn = 1$ SUSY, the {\it Left} and {\it Right} branes must be  
of type $b_i$ and $c_i$ respectively (or viceversa, see fig.(\ref{quivers})). 
But branes of type $b_i$ and $c_i$ necessarily intersect, 
giving rise to massless states with the quantum numbers of SM Higgs 
multiplets. This is the link drawn between $b_2$ and $c_2$ in  
the lowest quivers in fig.(\ref{quivers}). 
 
Due to these properties, there is a number of limitations on the  
possible model building, which we will discuss in a model by 
model basis below. As a general consequence we will see  
that  in general Q-SUSY forces the hypercharge to come 
along with an extra $U(1)_{B-L}$ gauge boson. 
This general property can only be evaded if Q-SUSY is only 
an approximate symmetry, as will be discussed in the case 
of the square SUSY-quiver. 
Otherwise a gauged $B-L$ must be present, 
although of course it may be eventually Higgssed. 
 This is an interesting 
result by itself since it shows an intriguing connection  
between the presence of supersymmetry and the gauging of 
$(B-L)$. It is also intriguing the third general property 
above, since it gives an interesting connection between 
the presence of SUSY and the existence of Higgs multiplets. 
We will present  specific examples corresponding to the  
four types of quivers discussed above. 
 
Before presenting the models let us emphasize that these are  
not the unique models that one can construct starting 
with the four quiver configurations. Other variations may be constructed 
depending on how we cancel RR-tadpoles (by explicit  
factorizable branes and/or non-factorizable branes 
and/or explicit RR H-flux). One can also easily build models  
with extended gauge symmetries like models with  
  left-right and/or Pati-Salam symmetries.  
But we think the examples provide the reader with enough  
information to look for other interesting variations.  
A general summary of the models we construct here is 
given at the end of the paper in table  
\ref{propmodels}.

\subsection{The square quiver} 

\EPSFIGURE{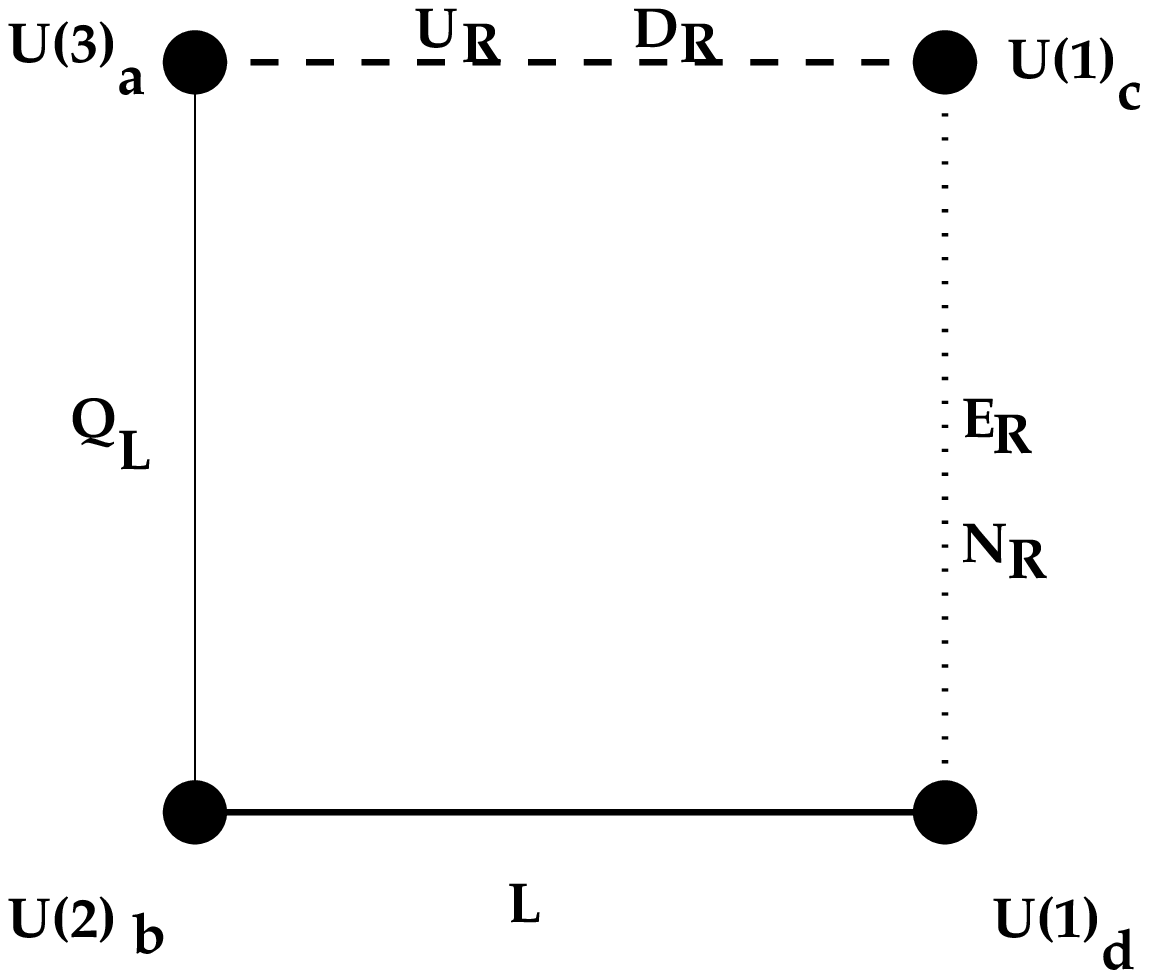, width=3in, height=2.5in}
{\label{qsimr}
A square SUSY-quiver with SM spectrum.  
Nodes represent the D6-branes giving rise to the SM gauge 
group. Links represent the brane intersections, at which 
quarks and leptons live. The different styles of the links  
denote that each of the four intersections respects  
a different $\cn = 1$ SUSY.} 

In order to describe explicit examples of square quivers
let us start from the generic family of models already 
presented in table \ref{solution}. As we previously said, this family 
is parametrized by  
two phases $\epsilon, \tilde \eps = \pm1$, the NS background 
on the first two tori $\beta^i = 1-b^i = 1, 1/2$, four integers 
$n_a^2,n_b^1,n_c^1,n_d^2$ and a parameter $\rho = 1,1/3$. 
As we also remarked, without loss of generality one can choose the 
baryonic(left)-stacks to be of type $a_2$ ($b_2$). This implies setting 
$\epsilon ={\tilde \epsilon} = 1$. 
In order to find out whether some SUSY is unbroken at the intersections, 
let us display the angles that compose the twist vector $v_\a$  
for each brane $\a = a, b, c, d$. Those angles are shown in 
table \ref{gral_angles} in terms of the ratios  
$U^i=R_2^i/R_1^i$, where $R_1^i$,$R_2^i$, $i=1,2,3$ are the 
radii of the tori.

\TABLE{\renewcommand{\arraystretch}{1.4}
\begin{tabular}{|c||c|c|c|} 
\hline 
 $Brane$    &  $\theta_\a^1$  &  $\theta_\a^2$  & $\theta_\a^3$ \\ 
\hline\hline $a$ &  0   &  $tg^{-1}({{ \beta^2U^2}\over 
{n_a^2}})$  & 
 $tg^{-1}({{-{ {}}\rho U^3}\over {2}})$  \\ 
\hline $b$ &   $tg^{-1}({{  \beta^1U^1}\over 
{n_b^1}})$   &  0   & 
 $tg^{-1}({{-3 \rho U^3}\over {2}})$ \\ 
\hline $c$ &    $tg^{-1}({{3\rho  \beta^1U^1}\over 
{n_c^1}})$   &  0  &     ${{\pi}\over 2}$  \\ 
\hline $d$ &  0     &  $tg^{-1}({{ \beta^2U^2}\over 
{\rho n_d^2}})$  &     $tg^{-1}({{3\rho U^3}\over 
{2}})$ \\ 
\hline 
\end{tabular} 
\label{gral_angles} 
\caption{\small Components of the twist vector 
$v_\a$ for each D6-brane stack of table 3.}}
In order for some SUSY to be preserved at each intersection, 
one needs to have angles  
$\vt_{\a\b}^1 \pm \vt_{\a\b}^2 \pm \vt_{\a\b}^3 \in \IZ$ for some   
choice of signs, where $\vt_{\a\b}^i$, $i=1,2,3$  are the angles  
formed by any pair of branes $\a$ and $\b$ in the three 2-tori  
\cite{cim1}. 
 Looking at table \ref{gral_angles} one can 
easily see that for certain choices 
of the $U^i$ and particular choices of some of the integers    
involved one can manage so that all intersection preserve one SUSY. 
Specifically, if the conditions  
\beqa 
n_c^1\ =\ 0 & \Rightarrow &  \rho=\frac{1}{3}, ~~ \beta^1=1 , \\ 
n_b^1 > 0 ,  &  & \ n_a^2\ =\ 3\rho n_d^2= n_d^2 \ >\ 0 , \\ 
U^1\ = \frac{n_b^1}{2} ~ U^3 ,  &   
& U^2\ =\ {{n_a^2 }\over  {6\beta^2}} ~ U^3 , 
\label{tune1} 
\eeqa 
are met,  the angles of the branes have the general form of table 
\ref{QSUSY_angles}, 
\TABLE{\renewcommand{\arraystretch}{1.25}
\begin{tabular}{|c||c|c|c|} 
\hline 
 $Brane$    &   $\theta_\a^1$  &  $\theta_\a^2$  & $\theta_\a^3$ \\ 
\hline\hline $a$ &  0   &  $\alpha _1$  & $-\alpha _1$ \\ 
\hline $b$ &   $\alpha _2$   &  0   &   $-\alpha _2$ \\ 
\hline $c$ &  ${{\pi}\over 2}$   &  0  &     ${{\pi}\over 2}$  \\ 
\hline $d$ &  0     & $\alpha _2$  &     $\alpha _2$   \\ 
\hline 
\end{tabular} 
\label{QSUSY_angles}
\caption{\small D6-brane angles when the  
conditions in (\ref{tune1}) are met.}}
where $\alpha_1=tg^{-1}(U^3/6)$ and 
$\alpha_2=tg^{-1}(U^3/2)$. 
Note that with these  choices the branes $a,b,c$ and $d$  
become  $\cn = 2$ branes of types $a_2$, $b_2$, $b_1$ and $a_1$ 
respectively. Thus we recover the square Q-SUSY structure  
of fig.\ref{qsimr}. 
Altogether the wrapping numbers of the resulting models with 
Q-SUSY properties are displayed in  
table \ref{QSUSYsquare}. One can check that there are simple choices  
of the integer parameters and extra branes with no intersection 
with the SM branes such that all RR tadpoles are canceled.  
The massless fermionic spectrum is the same as in the 
models in \cite{imr}, but now all of them come with 
a scalar superpartner. 

\TABLE{\renewcommand{\arraystretch}{2}
\begin{tabular}{|c||c|c|c|} 
\hline 
 $N_i$    &  $(n_i^1,m_i^1)$  &  $(n_i^2,m_i^2)$   & $(n_i^3,m_i^3)$ \\ 
\hline\hline $N_a=3$ & $(1,0)$  &  $(n_a^2, \beta^2)$ & 
 $(3 ,  -1/2)$  \\ 
\hline $N_b=2$ &   $(n_b^1, 1)$    &  $ (1/\beta^2,0)$  & 
$(1,-1/2)$   \\ 
\hline $N_c=1$ & $(0,1)$  & 
 $(1/\beta^2,0)$  & $(0,1)$  \\ 
\hline $N_d=1$ &   $(1,0)$    &  $(n_a^2,3\beta^2 )$  & 
$(1, 1 /2)$   \\ 
\hline $N_h$ &   $(1,0)$    &  $(1/\beta^2,0 )$  & 
$(n_h^3,m_h^3)$   \\ 
\hline 
\end{tabular} 
\label{QSUSYsquare}
\caption{\small D6-brane wrapping numbers giving rise to a Q-SUSY 
SM spectrum for a square quiver. For the sake of generality we have 
also considered the possible presence of an extra brane 
with no intersection with the SM branes.}}

We thus have obtained  
a family of square Q-SUSY models parametrized by the two integers 
$n_a^2,n_b^1$  and the NS parameter $\beta^2$, 
yielding just the spectrum of the standard model. In addition 
now each chiral fermion will have one massless scalar partner, so 
there will be the coresponding squarks and sleptons. Thus the initial spectrum 
(before taking into account the SUSY-breaking effects) will be 
rather similar to that of the MSSM. 
It turns out  
however that the choice of  parameters in order to get Q-SUSY  
(in particular setting $n_c^1=0$) changes  the structure 
of the $U(1)$'s in the model.  
Indeed, as we already mentioned above, this choice implies that  
the $Q_c$ generator is massless. In the absence of any extra brane  
(like that given by the last line in table \ref{QSUSYsquare})  
we would have just 
$Q_c$ as our only massless generator, rather than hypercharge. 
 
Let us analyze this point in more detail. 
If we substitute in equation (\ref{caplillos}) 
the wrapping numbers of the Q-SUSY SM model above, we obtain 
for the couplings of the antisymmetric fields to the  
$U(1)$'s the result 
\beqa  
B_2^1 &\wedge & \ {{2}\over { \beta^2 } }F^{b} \nonumber \\    
B_2^2 &\wedge  & \ 3  \beta^2 
(3F^a\ +\  F^{d}) \\ 
 B_2^3 &\wedge  &  \ 
(\frac{-3 n_a^2}{2} F^a\ -\ \frac{n_b^1}{\beta^2}F^{b} \ +\ 
 \frac{n_a^2}{2} F^d \ +\ N_h\frac{m_h^3}{\beta^2}F^{h}) 
\nonumber
\label{bfs} 
\eeqa 
Note that the coupling to  $B_2^0$ is zero, because $\prod_i m_a^i=0 \ 
\forall a$. Thus we see that, in the absence of additional branes 
($N_h=0$)   all $U(1)$'s but $U(1)_c$ 
gain masses by combining with antisymmetric tensors. 
On the other hand, if there are additional $h$ branes present 
the situation changes. Indeed,  
recalling eq.(\ref{uunose}), we can see that we  
would have two massless U(1)'s: $U(1)_c$ and 
the linear combination $U(1)_a-3 
U(1)_d +\frac{3n_a^2\beta^2}{N_hm_h^3}U(1)_h$. 
In fact, a linear combination of these two gives us 
a $U(1)$ charge whose coupling to SM fermions is 
precisely that of  the standard  hypercharge: 
\beq 
Q_{Y}=\frac{1}{6}Q_a-\frac{1}{2}Q_c-\frac{1}{2}Q_d- 
\frac{1}{2} \ \frac{n_a^2\beta^2}{N_hm_h^3}Q_{h}     \  . 
 \label{y} 
\eeq 
In addition to hypercharge, an extra massless $U(1)$ 
with analogous couplings to that of a $(B-L)$ generator 
 will remain in the massless spectrum. 
This is a first realization of our comment above that 
Q-SUSY requires the gauging not only of hypercharge but 
of the $B-L$ generator. This latter symmetry may  
eventually be spontaneously broken, in a way analogous to that  
discussed in subsection 7.4. 
An alternative way to guarantee the presence of just standard 
hypercharge at the massless level will be explained in section 
(5.1). 
 
An important point is in order concerning the Higgs sector  
in square quivers of the SM. In these models no Higgs  
multiplets appear at any of the intersections of the square. 
As discussed in Section 2 and in \cite{imr}, scalar fields 
with the quantum numbers of SM Higgs appear if the  
branes $b$ and $c$ (which are parallel along the 
second torus) approach to each other. However, 
since these two branes preserve opposite  
types of supersymmetries, the combined $bc$ system preserves 
no SUSY at all. This is reflected in the fact that  
the scalars in that combined system may become tachyonic  
when the branes are sufficiently close. Those tachyons  
signal the recombination of the branes and 
spontaneous symmetry breaking, as described in Section 7.  
However, since this Higgs system is not supersymmetric, 
in this case the Q-SUSY property of the rest of the spectrum will not 
be sufficient to stabilize a hierarchy of masses between the  
weak scale and the string scale, a (modest) fine-tuning  
being required to keep apart those two scales. 
This is one of the main motivations to consider the  
 rombic and triangular quivers which we  discuss 
below. We will describe more aspects of electroweak 
symmetry breaking for the square quiver in section 7.

\subsection{The linear quiver} 
 
\EPSFIGURE{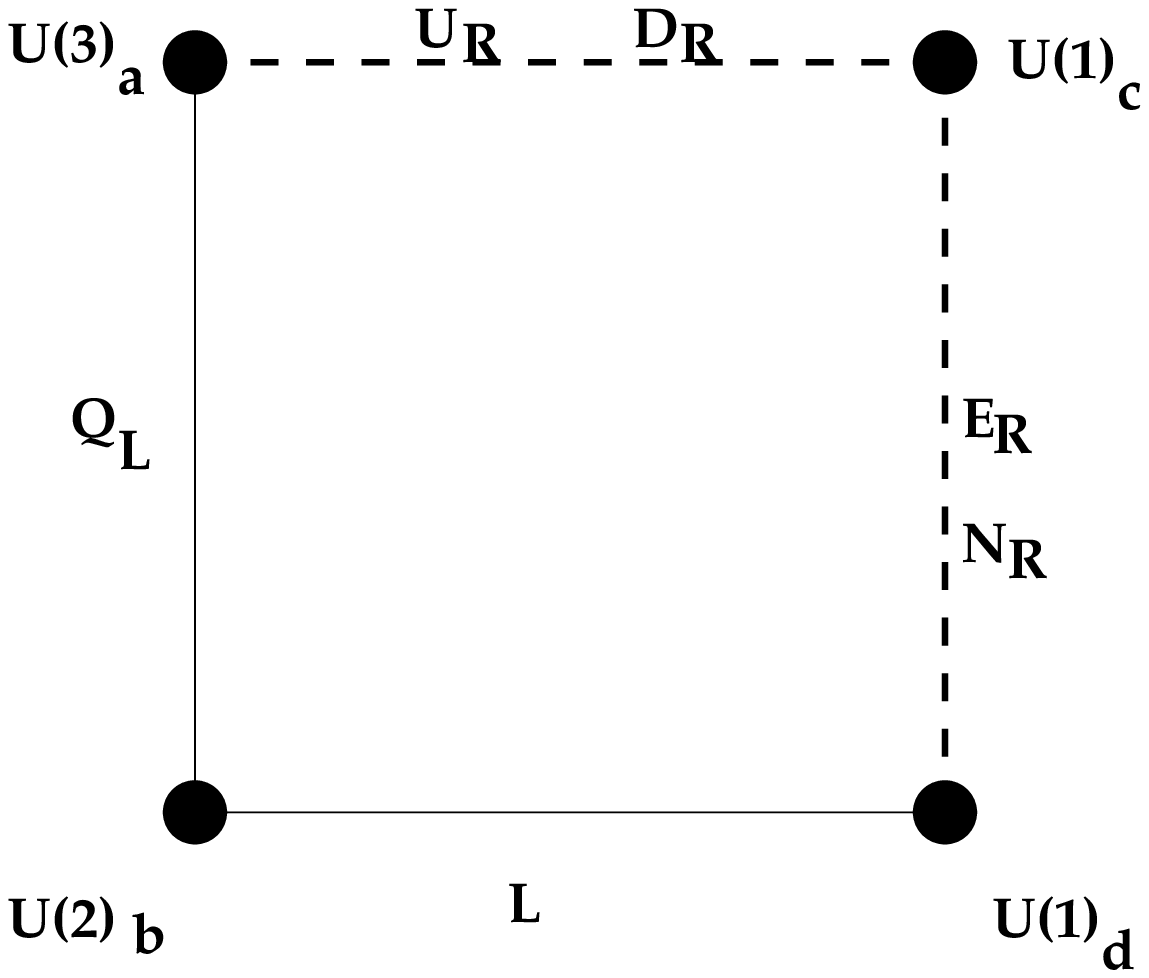, width=3in, height=2.5in}
{\label{linear}
A linear SUSY-quiver with SM spectrum. 
Note that all left-handed fermions share the same SUSY whereas 
the right-handed ones respect a different one.} 
 
In some way this may be understood as a variation of the square  
quiver in which we flip the type of leptonic brane from type $a_1$ 
to $a_2$. Due to this change the left-handed quarks and leptons share now 
the same SUSY, whereas the right-handed ones respect a different one  
(see fig.(\ref{linear})).   
 One can check that the conditions 
to get Q-SUSY at the intersections are analogous to those  
in the square quiver eq.(\ref{tune1}) and the wrapping numbers 
giving rise to the SM fermion content are shown in table  
\ref{QSUSYlinear}.  

\TABLE{\renewcommand{\arraystretch}{2}
\begin{tabular}{|c|c||c|c|c|} 
\hline 
brane\ type  & 
 $N_i$    &  $(n_i^1,m_i^1)$  &  $(n_i^2,m_i^2)$   & $(n_i^3,m_i^3)$ \\ 
\hline\hline $a_2$ & $N_a=3$ & $(1,0)$  &  $(n_a^2, \beta^2)$ & 
 $(3 ,  -1/2)$  \\ 
\hline $b_2$ & 
$N_b=2$ &   $(n_b^1, 1)$    &  $ (1/\beta^2,0)$  & 
$(1,-1/2)$   \\ 
\hline $b_1$ & 
$N_c=1$ & $(0,1)$  & 
 $(1/\beta^2,0)$  & $(0,1)$  \\  
\hline $a_2'$ & $N_d=1$ &   $(1,0)$    &  $(n_a^2,\beta^2/\rho  )$  & 
$(3\rho, -1 /2)$   \\ 
\hline  
\end{tabular} 
\label{QSUSYlinear}
\caption{\small D6-brane wrapping numbers giving rise to a Q-SUSY 
SM spectrum for a linear  quiver. Here $\rho =1,1/3$, $\beta^2=1,1/2$,  
$n_a^2, n_b^1 \in \IZ$.}} 
One can see that the wrapping numbers are identical to those of the square  
quiver except for the leptonic brane. This slight change has 
however an impact in the massless $U(1)$ spectrum. 
The couplings of the RR fields $B_2^i$ to the $U(1)$'s is 
identical to that in the square quiver eq.(\ref{bfs}) except 
for a flip in sign of the coefficient of the  
$B_2^3\wedge F^d$ term. As a consequence one observes that  
two $U(1)$'s will remain necessarily massless: $U(1)_c$  
and $U(1)_a-3 U(1)_d$. Thus in this model again we will have 
not only a massless hypercharge $Y=Q_a/6+Q_c/2-Q_d/2$ but also  
an extra $B-L$ generator. 
One can check  that this conclusion does not change even if we add  
an extra brane as we did in the square quiver case. 
Note also that, for $\rho =1$, the baryonic and leptonic 
stacks would have the same wrapping numbers and hence one can 
unify them  into a Pati-Salam stack with gauge group 
$U(4)_{PS}$ if one locates both stacks on top of each other.   
 
The simple set of branes discussed above are not enough to cancel 
RR tadpoles in this model. In fact one can check that 
for  the massive anomalous  $U(1)_b$ generator 
the Green-Schwarz cancellation mechanism will only cancel 
partially its anomalies and  one would in general 
need to add explicit  $H_{NS}$-flux to cancel tadpoles, as 
discussed in Section 2. 
This $U(1)$ being anyway massive, decouples from the 
observable massless spectrum.

Before getting into the rombic class of models let us  
make a comment concerning the possibility of building 
left-right symmetric models with gauge  
group $SU(3)\times SU(2)_L\times SU(2)_R \times U(1)_{B-L}$, 
i.e., the simplest non-Abelian extension of the SM.  
It is easy  to construct such a models starting from the 
square and linear quivers by just putting a couple of  
$(c)$-branes instead of one. We will not present here 
an analysis of these models, which is straightforward. 
Let us just make a comment concerning the connection 
between the property of Q-SUSY and the existence of a 
gauged $B-L$. 
In these models, in order to get the 
desired quantum numbers,  we have to ensure that there is a  
massless $U(1)_{B-L}$ remaining at the massless level (i.e., no couplings 
to $B_2^i$ fields) and no other linear combination  
of $U(1)$'s. 
It turns out that in the linear quiver models imposing the 
Q-SUSY conditions analogous to eq.(\ref{tune1})  
automatically guarantees that there is a   
$U(1)_{B-L}$  gauge boson in the massless spectrum. 
%
\EPSFIGURE{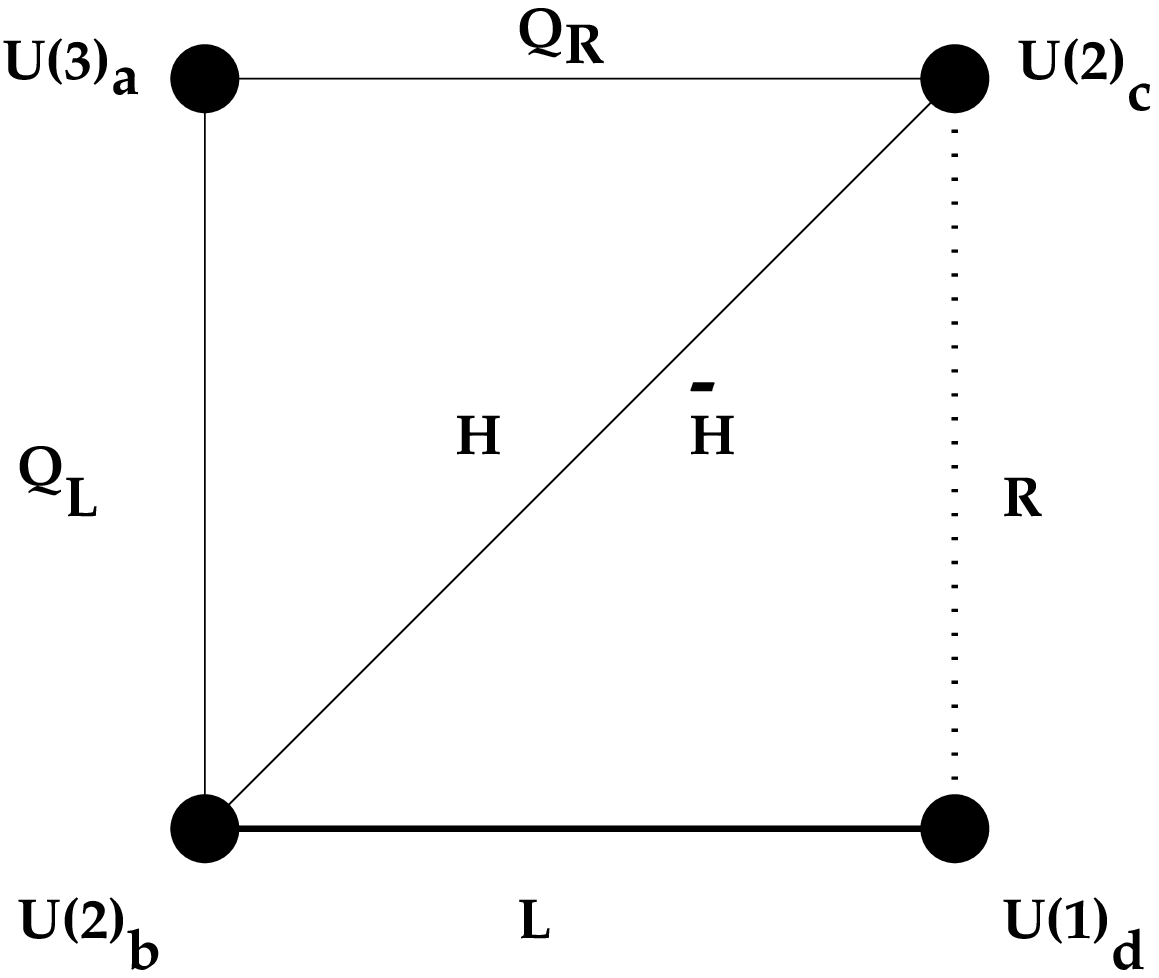, width=2.5in, height=2in}
{\label{rombic}
Rombic SUSY-quiver. The quark sector and Higgs fields respect 
the same $\cn = 1$ SUSY, while the leptonic sector respects different 
ones. In the figure a left-right symmetric model has been depicted. 
Unlike the square case, there are Higgs chiral multiplets at some 
intersections.}
%
This connection between  Q-SUSY and the presence of 
$B-L$ is another example of the intriguing connection 
that we find upon model building in between the presence 
of Q-SUSY (or SUSY) and the existence of a massless  
$B-L$ generator.

\subsection{The rombic quiver} 
 
The general structure of the rombic quiver is presented in 
figure \ref{rombic}. 
In order to give an example of such Q-SUSY configuration
we are going to present a model with a left-right 
symmetric $SU(2)_L\times SU(2)_R\times U(1)_{B-L}$ gauge group.
Getting such gauge group will again imply considering
four stacks of branes, following the same general philosophy 
described in section 2 in order to get the Standard Model.
In particular, we will be considering four sets of branes 
with the wrapping numbers of table \ref{trombic}. 

\TABLE{\renewcommand{\arraystretch}{2} 
\begin{tabular}{|c|c||c|c|c|} 
\hline brane\ type & 
 $N_i$  &  $(n_i^1,m_i^1)$  &  $(n_i^2,m_i^2)$   & $(n_i^3,m_i^3)$ \\ 
\hline\hline $a_2$ & $N_a=3$ & $(2,0)$  &  $(3,1/2)$ & 
 $(3, -1/2)$  \\ 
\hline $b_2$ & $N_b=2$ &   $(3, -1/2)$    &  $ (2,0)$  & 
$(3,1/2)$   \\ 
\hline $c_2$ & $N_c=2$ & $(3,1/2)$  & 
 $(3,-1/2)$  & $(2,0)$  \\ 
\hline $a_1$ & $N_d=1$ &   $(2,0)$    &  $(3,1/2 )$  & 
$(3, 1 /2)$ \\ 
\hline  
\end{tabular} 
\label{trombic}
\caption{\small D6-brane wrapping numbers giving rise to a Q-SUSY 
Left-Right symmetric  spectrum from rombic quiver.} } 

Note that in this brane configuration all the three tori 
have NS background (i.e., $\beta^i=1/2$, for all $i$). 
From the above configuration one finds for the intersection numbers: 
\beq 
\begin{array}{lcl} 
I_{ab}\ =   \ 3, & & I_{ab*}\ =\ 0, \\ 
I_{ac}\ =   \ -3, & & I_{ac*}\ =\ 0, \\ 
I_{bc}\ =   \ 3,  & & I_{bc*}\ =\ 0, \\ 
I_{cd}\ =   \ -3, & & I_{cd*}\ =\ 0, \\ 
I_{bd}\ =   \ 0, & & I_{bd*}\ =\ 3, 
\end{array} 
\eeq 
whose associated massless chiral spectrum is presented in table 
\ref{trombic2}. 
\TABLE{\renewcommand{\arraystretch}{1.25}
\begin{tabular}{|c|c|c|c|c|c|c|c|} 
\hline Intersection & 
 Matter fields  &   &  $Q_a$  & $Q_b $ & $Q_c $ & $Q_d$  & B-L \\ 
\hline\hline (ab) & $Q_L$ &  $3(3,2,2)$ & 1  & -1 & 0 & 0 & 1/3 \\ 
\hline (ac) & $Q_R$   &  $3( {\bar 3},1,2)$ &  -1  & 0  & 1  & 0 & -1/3 \\ 
\hline (bd*) & $L$    &  $3(1,2,1)$ &  0   & 1   & 0  & 1 &  -1  \\ 
\hline (cd) & $R$   &  $3(1,1,2)$ &  0  & 0  & -1  & 1  &  1 \\ 
\hline (bc) & $H$   &  $3(1,2,2)$ &  0 & 1 & -1 & 0  & 0 \\ 
\hline \end{tabular} 
\label{trombic2}
\caption{\small Spectrum of the Left-Right symmetric 
model from a rombic quiver.} } 
Unlike the square quiver case, this model has electroweak  
Higgs fields in the massless chiral spectrum.  
The property of Q-SUSY is obtained for 
\beq 
U^1\ =\ U^2\ =\ U^3 
\label{tune2} 
\eeq 
in which case each brane respects some $\cn = 2$ supersymmetry  
and the branes $a,b,c$ and $d$ are of type $a_2,b_2,c_2$ and $a_1$  
respectively. Let us analyze the structure of the $U(1)$'s. 
In this case one has that the RR fields $B_2^{i}$, $i=1,2,3$ 
couple to the $U(1)$'s as follows: 
\beq
\begin{array}{rcl}
B_2^1 &\wedge & \   6(-\ F^b\ +\ F^c) \\ 
B_2^2 &\wedge  & \ 3 (\ 3F^a\ -\ 2F^c\ +\ F^d\ ) \\ 
B_2^3 &\wedge  &  \ 3(\ -3F^a\ +\ 2F^b\ +\ F^d\ )
\end{array}
\label{brombic} 
\eeq
From here one concludes that there is a unique massless 
$U(1)$ given by: 
\beq 
Q_{B-L}\ =\  -{{2}\over 3}Q_a\ -\  Q_b\ -\ Q_c  
\eeq 
which couples to the fermionic spectrum as the standard $(B-L)$ generator of  
left-right symmetric models. 
 
An advantage of this model with respect to the square  
and linear quivers  is that 
the subsector formed by $Q_R,Q_L$ and the electroweak 
Higgs fields $H$ all respect the same $\cn = 1$ supersymmetry. 
Thus the quark subsector of the theory is supersymmetric. 
Since we know experimentally that the Yukawa couplings of 
the leptons are in general very small, it is enough to have 
an approximate SUSY in the quarks sector in order to  
stabilize the hierarchy between the weak scale and a 
possible cut-off scale of order 10-100 TeV. 
 
As in the linear model, 
the set of branes discussed above are not enough to cancel 
RR tadpoles in this model. In fact one can check that  
for  the massive anomalous  $U(1)_d$ generator  
the Green-Schwarz cancellation mechanism will only cancel 
partially its anomalies. 
As in the linear models,  
  one would in general  
need to add explicit  $H_{NS}$-flux to cancel tadpoles. 
Again $U(1)_d$ is massive and decouples anyway 
at low energies.  
  
Let us finally mention that the left-right symmetry  
may eventually break down to the Standard Model  
by brane recombination, in a way analogous to that  
explained for the other models in Section 7. 
The reader can verify, following the methods developed 
in that section, that after the recombination of one of 
the $(c)$ branes with the $(d)$ brane into a single one,  
the resulting configuration yields the fermion content and  
the gauge group of the SM. This would correspond, from the  
effective action point of view, to giving a vev to the right-handed 
sneutrinos, what eventually would break $B-L$ into the hypercharge.

\subsection{The $\cn = 1$ triangle quiver} 
 
\EPSFIGURE{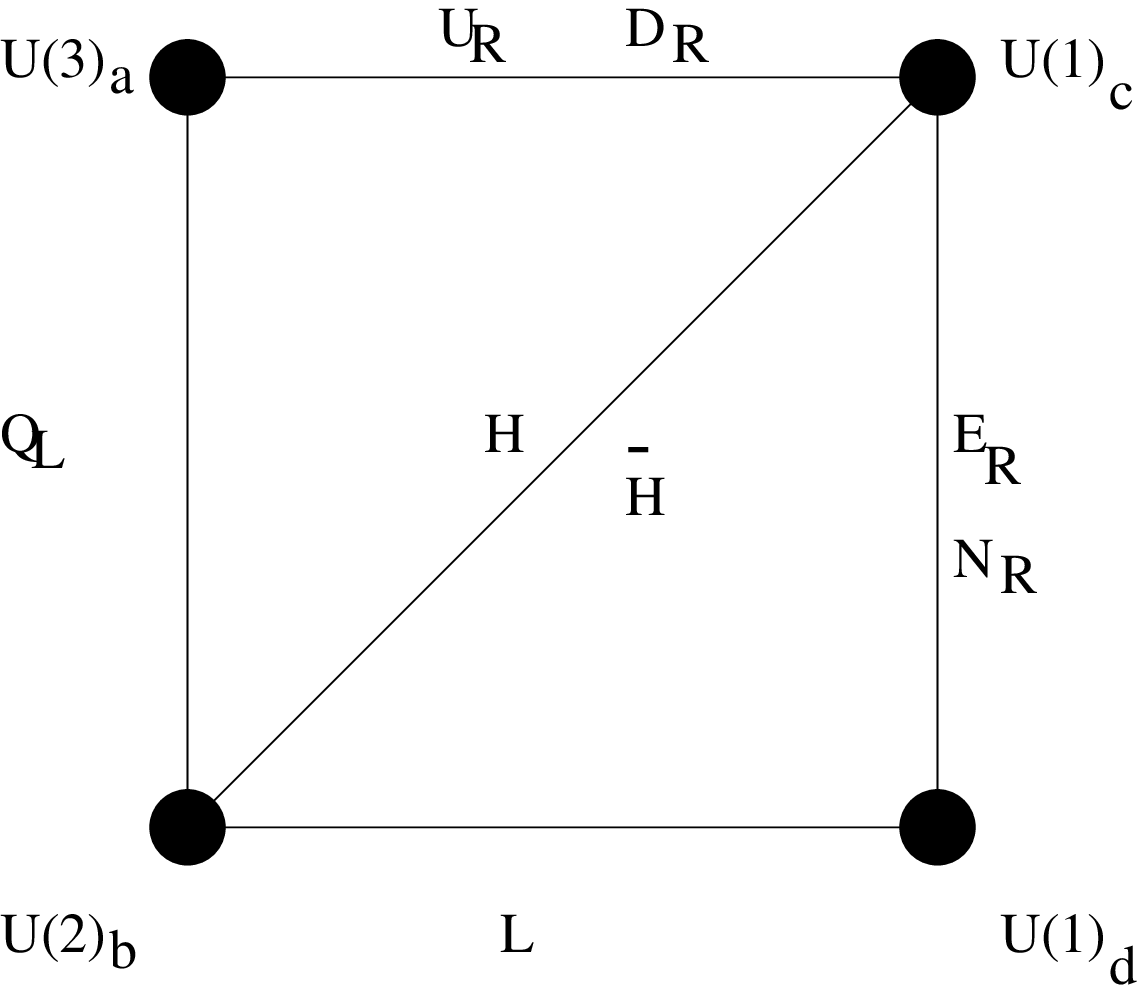, width=2.5in, height=2in}
{\label{MSSM}
Triangular SUSY-quiver with SM spectrum. 
The $a$ and $d$ branes are parallel so that actually this  
may be depicted as a triangle quiver.} 
 
Let us now show an example based in the triangle 
quiver in fig. \ref{MSSM}. Note that, as shown in fig. \ref{quivers}, 
the baryonic ($a$)  and leptonic ($d$) branes are of the same ($a_2$) 
type and could be put on top of each other in the quiver, leading to a 
triangular shape. In this example all of the intersections will 
preserve the same $\cn = 1$ SUSY so that, at least locally, the 
models will look fully supersymmetric.  
As we remarked at the beginning of this section, this quiver 
predicts the necessary presence of Higgs multiplets  
appearing at the intersections of {\it Left } and {\it Right } 
brane stacks. Imposing the that left- and right-handed 
fermion multiplets respect the same $\cn = 1$ SUSY forces the  
existence of these Higgs multiplets. 
 
Consider the  
wrapping numbers in table \ref{SUSYmodel}. It is easy to check that if 
the conditions 
\beq 
2U^1\ =\ 2\beta^2\ U^2\ =\ U^3 
\label{tune3} 
\eeq 
are met, the brane configuration indeed respects the same supersymmetry 
at all intersections. Here again  
$U^i=R_2^i/R_1^i$, where $R_1^i$,$R_2^i$, $i=1,2,3$ are the 
radii of the tori. From the brane wrapping numbers above one can obtain the 
intersection numbers 
\beq 
\begin{array}{lcl} 
I_{ab}\ =   \ 1, & & I_{ab*}\ =\ 2, \\ 
I_{ac}\ =   \ -3, & & I_{ac*}\ =\ -3, \\ 
I_{bd}\ =   \ -1,  & & I_{bd*}\ =\ 2, \\ 
I_{cd}\ =   \ 3, & & I_{cd*}\ =\ -3, \\ 
I_{bc}\ =   \ -1/\beta^2 , & & I_{bc*}\ =\ -1/\beta^2 , 
\end{array} 
\label{intersicsm}  
\eeq 
which gives rise to the massless spectrum in 
 table \ref{espectrosm}.  

\TABLE{\renewcommand{\arraystretch}{2}
\begin{tabular}{|c|c||c|c|c|} 
\hline 
brane\ type  & 
 $N_i$    &  $(n_i^1,m_i^1)$  &  $(n_i^2,m_i^2)$   & $(n_i^3,m_i^3)$ \\ 
\hline\hline   $a_2$ 
 &  $N_a=3$ & $(1,0)$  &  $(3,\beta^2)$ &  $(3 ,  -1/2)$  \\ 
\hline $b_2$ &  $N_b=2$ &   $(1,1)$   &  $ (1/\beta^2 ,0)$ & $(1,-1/2)$   \\ 
\hline $c_2$ &  $N_c=1$ & $(0,1)$  & $(0,-1)$  & $(2,0)$  \\ 
\hline $a_2$' &  $N_d=1$ &   $(1,0)$    &  $(3,\beta^2 )$ & $(3, -1/2)$   \\ 
\hline 
\end{tabular} 
\label{SUSYmodel}
\caption{\small Wrapping numbers of a three generation 
 SUSY-SM with $\cn = 1$ SUSY locally.} }

\TABLE{\renewcommand{\arraystretch}{1.3}
\begin{tabular}{|c|c|c|c|c|c|c|c|} 
\hline Intersection & 
 Matter fields  &   &  $Q_a$  & $Q_b $ & $Q_c $ & $Q_d$  & $Q_Y$ \\ 
\hline 
\hline $ab$ & $Q_L$ &  $(3, 2)$ & 1  & -1 & 0 & 0 & 1/6 \\ 
\hline $ab^*$  & $q_L$ & $2(3,2)$ &  1  & 1  & 0  & 0  & 1/6 \\ 
\hline $ac$ & $U_R$ &  $3( {\bar 3},1)$ &  -1  & 0  & 1  & 0 & -2/3 \\ 
\hline $ac^*$  & $D_R$   &  $3({\bar 3},1)$ &  -1  & 0  & -1  & 0 & 1/3 \\ 
\hline $bd$ & $ L $ &  $(1,2)$ &  0   & -1 & 0  & 1 & -1/2 \\ 
\hline $bd^*$ & $ l $ &  $2(1,2)$ &  0  & 1   & 0  & 1 & -1/2 \\ 
\hline $cd$ & $N_R$ & $3(1,1)$ &  0  & 0  & 1  & -1  & 0   \\ 
\hline $cd^*$ & $E_R$  & $3(1,1)$ &  0  & 0  & -1  & -1  & 1 \\ 
\hline $bc$ &  $H$ & ${1\over {\beta^2}}(1,2)$ &  0 & -1 & 1  &  0 & -1/2 \\ 
\hline $bc^*$ & $ {\bar H}$ &  ${1\over {\beta^2}}(1,2)$ & 0 & -1 & -1 & 0  
& 1/2 \\ 
\hline 
\end{tabular} 
\label{espectrosm}
\caption{\small Chiral spectrum of the SUSY's SM obtained from the  
triangular quiver. The hypercharge generator is defined as  
$Q_Y = \frac 16 Q_a - \frac 12 Q_c - \frac 12 Q_d$.}}
 
Let us now study the structure of the couplings of the $U(1)$  
fields to the antisymmetric B-fields, which determine  
which of them become massive. One finds the  
couplings: 
\beq
\begin{array}{rcl}
B_2^1 &\wedge & \   {2\over {\beta^2}}  F^b \\ 
B_2^2 &\wedge  & \ 3\beta^2  ( 3F^a\  +\ F^d ) \\ 
 B_2^3 &\wedge  &  \ 
- {1\over 2}( 9F^a\ +\ {2\over \beta^2} F^b\ +\ 3F^d ) \ . 
\end{array}
\label{bmssm}    
\eeq
One concludes from here that both $U(1)_b$ and $3U(1)_a+U(1)_d$ 
generators become massive. 
The massless anomaly-free generators are thus $U(1)_c$ and  
$U(1)_a-3U(1)_d$, which correspond with the third component 
of right-handed isospin and B-L, respectively. The hypercharge  
is given by the linear combination of generators: 
\beq 
U(1)_Y\ =\ { {U(1)_a}\over 6 } \  
-\ { {U(1)_c}\over 2} \ -\ { { U(1)_d} \over 2} \ .  
\eeq 
Thus, as expected, the massless hypercharge comes along with a  
massless B-L generator, if supersymmetric intersections are imposed. 
Note also that, 
as in the linear quiver case, if one locates the leptonic stack  
of branes $d$ on top of the baryonic stack one would  
get an enlarged $U(4)_{PS}$ Pati-Salam gauge symmetry. 
 
Note that there are two versions of the model with $\beta^2 = 1,1/2$. 
Interestingly enough, for the $\beta^2=1$ case  
{\it the massless chiral spectrum 
is exactly the same of the MSSM}. For $\beta^2=1/2$ the Higgs sector is 
doubled, although the rest of the spectrum remains the same.  
As in the rombic example, the simple brane configuration by itself 
would give rise to RR tadpoles. Those may be canceled by the addition 
 of  either some (non-factorizable) 
brane system with no intersection with the SM branes 
(for the $\beta^2=1/2$ case)  or some 
explicit $H_{NS}$-flux (for the $\beta^2=1$ case). 
 
Before  ending this model-building section let us make a final comment. 
The square quiver models have an attractive point which the others do not 
have. Square quiver models are just a subclass of those presented 
in \cite{imr},  and the latter have the attractive feature that 
the number of generations is related to the number of colours  
by cancellation of  $U(1)_b$ anomalies. In order for this to work there 
cannot be extra fermion  doublets, like the Higgsinos appearing in the 
triangle and rombic models, which contribute to $U(1)_b$ 
anomalies. In the case of the linear quiver models one can check that 
the $U(1)_b$ anomalies do not cancel and Wess-Zumino  
terms (induced by the presence of $H_{NS}$-flux)  
 must appear (in addition to the Green-Schwarz mechanism terms) 
to complete $U(1)_b$ anomaly cancellation. Thus also in this case 
the number of generations/colours argument is not present. 
On the other hand, as we mentioned above, 
the triangle and rombic quivers predict the generic presence of 
massless Higgs multiplets to give rise to electroweak symmetry breaking.

\section{SUSY-breaking and Fayet-Iliopoulos terms}

The Q-SUSY examples discussed in the previous section
may present three different sources of supersymmetry breaking:

{\it 1)} Soft SUSY-breaking masses from loop graphs, as 
discussed in Section 3. These will in general be present in the 
four types of models considered, since all of them will have 
non-SUSY massive sectors with masses generically of order $M_s$.  
They will induce generically  
one-loop gaugino masses as well as two-loop scalar 
masses$^2$. These effects will be of order $\alpha/(4\pi)M_s$, 
$M_s$ being the string scale. 
 
{\it 2)} Fayet-Iliopoulos terms. These may be present in any 
of the models and have a nice geometrical interpretation, as 
discussed in \cite{csu, cim1}. Indeed, in order to obtain SUSY 
at the brane intersections, we had to appropriately tune 
the $U^i$ complex structure parameters (see eqs. 
(\ref{tune1}),(\ref{tune2}), (\ref{tune3})). A slight departure from  these 
adjustments is seen in the effective Lagrangian as the  
turning on of FI-terms for the $U(1)$ fields of the models. 
This source of SUSY-breaking may be large or  small, 
depending on the values given to the $U^i$.

{\it 3)} Some models have explicit SUSY-breaking in  
some sector.  
In particular we already mentioned that in the square and linear  
quivers  the Higgs sector is non-SUSY and so will be the 
Yukawa couplings. In the case of rombic models the Higgs, and 
quark sectors respect SUSY at tree level but there will be  
explicit SUSY-breaking in the leptonic sector  
(non-SUSY Yukawa couplings). This is expected because, although all 
intersections preserve some SUSY, Yukawa couplings involve 
in general different intersections respecting different supersymmetries. 
As we mentioned, since the leptonic 
Yukawas are known to be small, the Higgs mass will be sufficiently protected  
from loop corrections so as to avoid fine-tuning. 
Finally, in the triangle quiver case all intersections respect the same  
SUSY and there is a supersymmetric superpotential. 
 
The first source of SUSY-breaking will depend on the particular  
massive non-SUSY sector of the given model. On the other hand  
following ref.\cite{cim1}   
we can give a relatively model-independent discussion of the  
FI-terms in this class of theories.  
Consider  a D6-brane which is forming angles with the  
orientifold plane given by  $(\vt^1, \vt^2, \vt^1+\vt^2 + \d, 0)$, 
$\vt^i > 0$. Then, for $\d =0$ one would recover one unbroken supersymmetry. 
For small non-vanishing $\d $    
one can approximate the effect of this by turning  
on a Fayet-Iliopoulos term for  the $U(1)$ field associated to the  
brane and given by 
\beq 
  \xi_a \approx  - {\d \over 2}M_s^2 . 
\label{Fayet4}  
\eeq 
In order to compute the mass of a scalar living at the intersection of two 
branes, let us take two D6-branes $D6_a$ and $D6_b$ whose separation from 
the same SUSY limit  is given by $\d_a$ and $\d_b$, respectively. 
Then  the mass of this scalar is  given by 
\beq 
  m_{ab}^2 =  (- q_a \xi_a - q_b \xi_b) \approx 
{1 \over 2} (\d_a - \d_b) M_s^2\ . 
\label{FImass} 
\eeq 
and in fact one can check that this result  
is in agreement with the masses for the 
scalars lying at a intersection obtained from the string mass 
formulae (see ref.\cite{cim1} for details). 
Thus for small deviations from a SUSY configuration, 
the masses of the scalars at an intersection may be understood as 
coming from  Fayet-Iliopoulos terms.

Let us emphasize some points to try to avoid confusion. 
Normally, when one talks about a FI-term in a SUSY  
field theory,  one is taking about a massless  
$U(1)$. On the other hand in our models with  
exact Q-SUSY it is the massive $U(1)$'s 
(masses of order $M_s$) that get FI-terms. 
This may be understood because a FI-term  
is the SUSY partner of a $B\wedge F$ coupling \cite{Dine}, and those  
$U(1)$'s with a non-vanishing coupling to some $B$-field  
are the ones which become massive. 
On the other hand 
these massive $U(1)$'s  decouple well below 
the string scale. 
 This means that, e.g., there are no quartic scalar 
couplings proportional to the gauge coupling constants of massive 
$U(1)$'s. 
Thus from the  
effective low-energy point of view the masses of scalars coming 
from FI-terms of  these massive $U(1)$'s   
will look rather like standard soft SUSY-breaking 
masses. It is about these scalar masses that we will be talking 
in this section.

\subsection{FI-terms and sfermion masses in the square and linear quivers}

Let us now consider the turning of FI-terms in the square quiver. 
In fact we are going to consider a slight generalization  
of the square quiver which has the relative advantage that 
the only unbroken $U(1)$ is hypercharge. 
 
Indeed, in order to get exact SUSY at all intersections 
 we imposed $n_c^1=0$.  
This was required so that  the brane $c$ makes 
a ${\pi\over2}$ angle with the O-plane in tori 1 and 3, preserving in this 
way two supersymmetries with the orientifold and one supersymmetry in each 
intersection with branes $a$ and $d$. 
This had as a consequence that the single $U(1)$ massless  
(in the absence of extra branes) was $Q_c$, 
rather than hypercharge. 
 On the other hand, if we are going to break SUSY anyway by adding  
FI-terms,  we can also relax  the Q-SUSY  condition 
and allow for approximate  SUSY at  the intersections 
involving the $c$-brane.

Let us start from the model of table \ref{solution} and let us impose  
the conditions 
\beqa
n_b^1 > 0 , & & n_a^2\ =\ 3\rho n_d^2  >\ 0 ,  \nonumber \\ 
U^1\ =\frac{3 \rho n_b^1}{2\beta^1}U^3 ,  & & 
 U^2\ =\ {{n_a^2\rho}\over  {2\beta^2}} U^3. 
\label{ello2} 
\eeqa 
Note that we have not imposed the condition $n_c^1=0$, and thus we do not need 
to impose any condition on $\beta^1$, $\rho$. 
The conditions (\ref{ello2}) will ensure SUSY  
for all intersections not involving brane $c$. 
Now, the condition to get a massless hypercharge generator is: 
\beq 
n_c^1=- {\beta^2\over2 \beta^1}(n_a^2+3\rho n_d^2)  
= - {\beta^2\over \beta^1}3\rho n_d^2\ , 
\eeq 
\TABLE{\renewcommand{\arraystretch}{2}
\begin{tabular}{|c||c|c|c|} 
\hline 
 $N_i$    &  $(n_i^1,m_i^1)$  &  $(n_i^2,m_i^2)$   & $(n_i^3,m_i^3)$ \\ 
\hline\hline $N_a=3$ & $(1/\beta ^1,0)$  &  $(n_a^2, \beta^2)$ &    
 $(3 ,  -1/2)$  \\ 
\hline $N_b=2$ &   $(n_b^1, \beta^1)$    &  $ (1/\beta^2,0)$  & 
$(1,-1 /2)$   \\ 
\hline $N_c=1$ & $(-{\beta^2\over \beta^1}n_a^2,  \beta^1)$  & 
 $(1/\beta^2,0)$  & $(0,1)$  \\ 
\hline $N_d=1$ &   $(1/\beta^1,0)$    &  $(n_a^2, 3\beta^2 )$  & 
$(1, 1 /2)$   \\ 
\hline 
\end{tabular} 
\label{solution2}
\caption{\small D6-brane wrapping numbers giving rise to an 
approximate square Q-SUSY Standard Model.} } 
since we have chosen $\eps = \tilde \eps = 1$.  
This implies that in order to avoid multiple wrappings \cite{imr} 
we also need to set  $\rho=1/3$ (and hence $n_a^2 = n_d^2$). 
We present in table \ref{solution2} the wrapping numbers of these models. 

Consider now a slight departure of the moduli $U^{1,2}$ 
from their assigned values in (\ref{ello2}):  
\beq
\begin{array}{c}
U^1\ =\frac{n_b^1}{2\beta^1} U^3 +\delta^1 \\  
U^2\ =\ {{n_a^2}\over  {6\beta^2}}U^3+\delta^2 
\end{array}
\label{tune4} 
\eeq 
and let us also define 
\beq
\Delta \ \equiv \ 
{\pi\over2}-tan^{-1}\left(\frac{(\beta^1)^2}{\beta^2n_d^2}U^1\right) 
\ = \
{\pi\over2}-tan^{-1}\left[\frac{(\beta^1)^2}{\beta^2n_d^2}\left(\frac{n_b^1} 
{2\beta^1} U^3 +\delta^1\right)\right] \ \ . 
\label{deltona} 
\eeq

This parameter $\Delta $ will measure the departure from supersymmetry 
of the brane $c$. Under these conditions the angles formed by the 
branes with the orientifold plane are the ones shown in table  
\ref{apLSUSY_angles}, where 
$\alpha_1=tan^{-1}(\frac{ U^3}{6})$ and 
$\alpha_2=tan^{-1}(\frac{ U^3}{2})$. 
\TABLE{\renewcommand{\arraystretch}{1.25}
 \begin{tabular}{|c||c|c|c|} 
\hline 
 Brane  &  $\theta_\a^1$  &  $\theta_\a^2$   & $\theta_\a^3$ \\ 
\hline\hline $a$ &  0   &  $\alpha_1+\d_a$ 
& $-\alpha_1$ \\ 
\hline $b$ &   $\alpha_2+\d_b$   &  0   & 
 $-\alpha_2$ \\ 
\hline $c$ & 
 ${{\pi}\over 2}+\Delta $   &  0  & 
   ${{\pi}\over 2}$  \\ 
\hline $d$ &  0     & 
 $\alpha_2+\d_d $  & 
   $\alpha_2$   \\ 
\hline 
\end{tabular} 
\label{apLSUSY_angles} 
\caption{\small Angles that the D6-brane stacks form with the 
orientifold axis in the approximate square configuration in the text.} }

There one has (for small $\d^i$) 
\beqa
\delta_a & = & 
\frac{{\beta^2 \over {3n_a^2}}\delta^2}{1+({U^3\over 6})^2} \ ;
\nonumber \\ 
\delta_b & = & \frac{{1 \over n_b^1}\delta^1}{1+({U^3\over 2})^2} \ ;\\
\delta_d & = & \frac{{\beta^2 \over n_a^2}\delta^2}{1+({U^3\over 2})^2}  
\nonumber
\label{deltas} 
\eeqa

Note also that (again for $\delta^1\ll1$): 
\beq 
tan\Delta\simeq\frac{2n_a^2\beta^2}{ n_b^1\beta^1}\frac{1}{U^3} 
\left(1-\frac{2\beta^1}{ n_b^1}\frac{\delta^1}{U^3}\right) 
\eeq  
so that one can check that $\Delta $ can be made quite small for not too 
large $U_3$. From table \ref{apLSUSY_angles} one sees that for  
all practical purposes the effect of not setting $n_c^1=0$ 
induces SUSY-breaking which (for small $\Delta$) can be parametrized 
as a FI-term for the $U(1)_c$ generator.

We can now compute the masses of the sparticles in terms of  
the FI-terms  following the discussion given in 
ref.\cite{cim1}. The results are  
shown in table \ref{sparticles2}. Notice that we could formally put the 
masses depending on $\Delta$ as coming from a Fayet-Iliopoulos term. 
However, and unlike the other cases, supersymmetry in these 
intersections would only be recovered in the $U^3\to\infty$ limit. 
\TABLE{\renewcommand{\arraystretch}{1.25}
\begin{tabular}{|c|c|c|c|} 
\hline Sector & 
  $(\theta_{\alpha \beta}^1, 
\theta_{\alpha \beta}^2,\theta_{\alpha \beta}^3)$  
  & sparticle  & mass$^2$ \\ 
\hline\hline (ab) & $(\alpha_2+\delta_b,-\alpha_1-\delta_a, 
-\alpha_2+\alpha_1) $   & 
 $1\times\tilde Q_L$ & ${1\over2}(\delta_a-\delta_b)$  \\ 
\hline (ab*) &$(-\alpha_2-\delta_b,-\alpha_1-\delta_a, 
\alpha_2+\alpha_1) $   & 
$ 2\times\tilde q_L$ & ${1\over2}(\delta_a+\delta_b)$  \\ 
\hline (ac) &$({\pi\over2}+\Delta,-\alpha_1-\delta_a, 
{\pi\over2}+\alpha_1) $    & 
 $3\times\tilde U_R$ & ${1\over2}(\delta_a+\Delta)$  \\ 
\hline (ac*) &$(-{\pi\over2}-\Delta,-\alpha_1-\delta_a, 
-{\pi\over2}+\alpha_1) $   & 
$3\times\tilde D_R$ & ${1\over2}(\delta_a-\Delta)$  \\ 
\hline (bd) & $(-\alpha_2-\delta_b,\alpha_2+\delta_d, 
2\alpha_2) $   & 
  $3\times\tilde L$ & ${1\over2}(\delta_b+\delta_d)$  \\ 
\hline (cd) & $(-{\pi\over2}-\Delta,\alpha_2+\delta_d,    
\alpha_2-{\pi\over2})$  & 
  $3\times\tilde N_R$ & ${1\over2}(\delta_d-\Delta)$  \\ 
\hline (cd*) & $(-{\pi\over2}-\Delta,-\alpha_2-\delta_d, 
-\alpha_2-{\pi\over2})$  & 
  $3\times\tilde E_R$& ${1\over2}(\delta_d+\Delta)$  \\ 
\hline 
\end{tabular} 
\label{sparticles2}
\caption{\small Squark and slepton masses from FI-terms  
in the square quiver model.} } 

An important point to note is that there is a wide range of parameters 
for which all squark and slepton (mass)$^2$ are positive and hence 
there are no unwanted charge and colour-breaking minima. 
It is enough to satisfy $\delta_a,\delta_d> \Delta$ and $\d_a > |\d_b|$  
in order to have all (mass)$^2$  positive. Note also that  
(for a given choice of discrete parameters) the squark and slepton 
masses come determined from just three independent parameters 
$\d_1, \d_2$  and $U^3$. In some limits the expressions for these 
masses becomes particularly simple. For example, if $\d_1=0$ one gets 
\beqa 
m^2_{{\tilde Q}_L}\ =\ { {\d_a}\over 2}M_s^2 \ &;&\  
m^2_{{\tilde L}}\ =\ { {\d_d}\over 2}M_s^2 \nonumber  \\  
m^2_{{\tilde U}_R,{\tilde D}_R}\ =\ \left( { {\d_a}\over 2}\ \pm \ 
{ {\Delta}\over 2} \right)  M_s^2\ & ;&\  
m^2_{{\tilde E}_R,{\tilde N}_R}\ =\ \left( { {\d_d}\over 2}\ \pm \ 
{ {\Delta}\over 2}\right) M_s^2  
\label{msql} 
\eeqa 
Note that in this case the ratio of {\it average } squark masses versus  
slepton masses is controlled only by the value of $U^3$: 
\beq 
{ {m^2_{squark}}\over { m^2_{slepton}} } \ =\  
{ {\d_a} \over {\d_d} }\ =\ {1\over 3} \ { {1+(U^3/2)^2} \over  
{1+(U^3/6)^2 } } \ =\ {{g_a^2} \over {g_d^2} } 
\label{massessql} 
\eeq 
The last  equality follows from eqs.(\ref{coup})and (\ref{longis}) 
in the next subsection. In this connection note that  
using those equations we can rewrite eqs.(\ref{deltas}) as 
\beq 
\delta_a\ =\  g_a^2(\beta^2\delta^2) \ 
;\ 
\delta_b\ =\ g_b^2({{\delta^1}\over {\beta^2}}) \ ;\ 
\delta_d\ =\ g_d^2(\beta^2\delta^2) 
\label{deltas2} 
\eeq 
Thus the (masses)$^2$ of sfermions from FI-terms are proportional 
to the square  of the $U(1)$ coupling constants, 
very much like if they were coming from a one-loop effect 
\footnote{In fact in some sense they do, since tree-level closed string  
couplings are one-loop from the open string channel point of view.}. 
In any event,  
it is interesting how one can obtain the masses of squarks  
and sleptons in this scheme in terms of a few 
geometrical parameters. Notice however that one should add to these masses 
the contributions coming from the loops involving massive non-SUSY  
particles, as described above.

One  interesting point concerns the right-handed sneutrino.  
From the above mass structure one can check that there is  
a region of parameter space for which one might have 
a tachyonic mass for the ${\tilde {\nu }}_R$ but a positive 
one for the rest of the scalars. It is enough to have  
$\d_a> \Delta $ but $\d_d < \Delta$. One can check that this is 
possible provided $U^3 > \sqrt{12}$. If this happens lepton number 
is broken and the $B-L$ gauge boson becomes massive swallowing a 
linear combination of ${\tilde {\nu }}_R$'s and 
antisymmetric $B$-fields. From the brane point of view the 
$c$ and $d^*$ branes will recombine into a single one and 
the obtained structure is similar to the models discussed  
in subsection 7.4 .

Another point is in order concerning FI-terms. We are used to  
the fact that in {\it supersymmetric} theories   
the masses that scalars get from a FI-term are proportional to 
the $U(1)$ charges of their fermionic partners. Note that in  
the case of the Q-SUSY theories that is in general not the case, 
as observed in ref.\cite{cim1}.  
Indeed, consider for example the FI-term associated to the  
baryonic symmetry $U(1)_a$. One can see from table  
\ref{sparticles2} that both left- and right-handed squarks 
get positive masses, although their fermionic partners have opposite 
baryon number. This is a reflection of the fact that  
different $\cn = 1$ SUSY's are preserved at the $ab$ and at the $ac$ 
intersections.

One can repeat an analogous discussion for the linear  
quiver case. It is easy to check that one obtains the  
same results for the masses coming from FI-terms  
than the ones in the square quiver, table \ref{sparticles2}. 
The only difference is that one necessarily has $\Delta =0$ 
in the linear case. Indeed, if $\Delta \not=0$ is taken, then 
the $Q_c$ generator is massive and the unbroken  
generator would be $B-L$, rather than hypercharge. 
This is different to what happens in the  square quiver case in which 
one can chose parameters so that it is hypercharge which  
remains massless.

\subsection{FI-terms 
and sfermion masses in the  triangle and rombic quivers }

One can also easily compute the masses of the scalars  
coming from turning on FI-terms in the other triangle 
and rombic  type of models. Note 
that in these cases there 
are several (or all) intersections respecting the {\it same}  
$\cn = 1$ supersymmetry. In these $\cn = 1$ subsectors the usual  
fact will hold and the masses of scalars from FI-terms 
will be proportional to the $U(1)$ fermionic charges. 
Consider for example the triangular case. By going  
slightly away from the supersymmetric configuration 
\beq 
U^1\ = \ {1\over 2 }U^3 \ +\ \d_1 \ ;\ U^2\ =  
\ {1\over {2\beta^2} }U^3 \ +\ \d_2 
\  
\label{tune5} 
\eeq 
it is easy to check that  
one induces FI terms for the $U(1)$'s as follows: 
\beq  
\xi_a \ =\ \xi_d \ = -  { {\d_2 \b^2/6} \over {1+(U^3/6)^2 } } M_s^2\ ;\   
\xi_b \ = - { {\d_1/2} \over {1+(U^3/2)^2 } } M_s^2 
\label{fitriangle} 
\eeq 
and $\xi_c = 0$ (recall that $U(1)_c$ is massless). 
Note that there are only two independent parameters for the 
FI-terms, which corresponds to the existence of two massive (anomalous)  
$U(1)$'s, $U(1)_b$ and $3U(1)_a+U(1)_d$ (more physically 
$(9B+L)$). Now,  
unlike the case of the square quiver,  
the masses for the scalars at the intersections 
will be just proportional to these charges. Looking at 
the particle spectrum in the triangle models  
(table \ref{espectrosm}) we see that left- and 
right-handed squarks will have {\it opposite} masses from the 
$U(1)_a$ baryonic FI-term. Thus, FI-terms are unable  
to give positive masses for 
all squarks and sleptons at the same time.  
There is nothing wrong with that, only that the FI-terms 
contribution cannot be the dominant source of scalar masses if 
we want to avoid unwanted charge/colour breaking minima, 
the one-loop contributions should also be important.

A similar general conclusion holds for the rombic model, 
since the quiver contains a triangle with three  
intersections respecting the same SUSY.

\section{Gauge coupling constants}

Unlike the heterotic case, the gauge coupling constants of 
brane models in which each gauge factor lives in a different  
stack of branes have no unification. Rather, the size of each 
coupling constant square is inversely proportional to the 
volume wrapped by the corresponding brane. 
The physical gauge couplings will depend on 
 
{\it i)} The size of the gauge couplings at the string scale. 
 
{\it ii)} The running between the string scale $M_s$ and the  
weak scale. 
 
For a string scale of order 10-100 TeV there is running for 
two to three orders of magnitude. The effect of this running may be 
enhanced  due to the fact that in between the string scale,  
and the weak scale there is not only the spectrum of 
the SUSY SM but also of the adjoint scalars and fermions  
corresponding to the $\cn = 4$ structure of the gauge sectors.  
Those fields are expected to get masses of order $\alpha/(4\pi) M_s$, 
which may be  
well below the string scale. In addition close to the string scale 
there can be important  threshold corrections 
due to the excitation of KK , winding and ``gonion''  
states \cite{afiru2}  
some of which may be below the string scale. 
All these running effects are difficult to evaluate in a model  
independent way. However one can obtain easy and closed  
formulae for the size of the gauge couplings at the string scale. 
 
The formulae for the size of gauge couplings get simplified 
in the case of Q-SUSY and SUSY models compared to 
the general non-SUSY intersection models. 
As noted in e.g. \cite{afiru2}, the tree-level value of the 
different  
gauge coupling constants at the string scale 
are controlled by the 
length of the wrapping cycles, i.e., 
\beq 
{{4\pi ^2}\over {g_i^2} } \ =\ 
{{M_s^3}\over (2\pi)^2 {\lambda_{II}}} \ ||l_i|| 
\label{coup} 
\eeq 
where $M_s$ is the string scale, $\lambda_{II}$ is the Type II string 
coupling, and $||l_i||$ is the length of the cycle of the i-th set of 
branes 
\beq 
{||l_i||^2 \over (2\pi)^6}\ =\ 
((n_i^1R_1^{(1)})^2+(m_i^1R_2^{(1)})^2) 
((n_i^2R_1^{(2)})^2+(m_i^2R_2^{(2)})^2) 
((n_i^3R_1^{(3)})^2+(m_i^3R_2^{(3)})^2) \ . 
\label{length} 
\eeq 
The ratio  of the $SU(3)$ and $SU(2)_L$ couplings 
is thus given by : 
\beq 
\frac{\alpha_{QCD}}{\alpha_L}=\frac{||l_b||}{||l_a||} 
\label{relacion1} 
\eeq 
For models without a non-Abelian extended gauge  
structure (like the square and triangle quiver examples above) 
 the hypercharge is a linear combination  
of $U(1)_a$, $U(1)_c$ and $U(1)_d$. 
One has for the hypercharge coupling: 
\beq 
\alpha_Y^{-1}=\frac{1}{6^2}\alpha_{a}^{-1}+\frac{1}{2^2}(\alpha_{c}^{-1}+ 
\alpha_{d}^{-1}) 
\eeq 
where $\alpha_a$ is the coupling of the $U(1)_a$, which in our 
normalization 
verifies at the string scale $\alpha_a=\alpha_{QCD}/6$. We thus have: 
\beqa
\frac{\alpha_{QCD}}{\alpha_Y} & = & \frac{1}{6}+ 
\frac{1}{4}\left(\frac{\alpha_{c}^{-1}}{\alpha_{QCD}^{-1}}+ 
\frac{\alpha_{d}^{-1}}{\alpha_{QCD}^{-1}} \right ) \nonumber \\ 
& = & \frac{1}{6}+\frac{1}{4}\left(\frac{||l_c||}{||l_a||}+ 
\frac{||l_d||}{||l_a||}  \right) 
\label{hiperra} 
\eeqa
We can now evaluate these ratios for the different models. 
It turns out that for Q-SUSY models the complicated non-linear 
expression for the volume of the cycles (\ref{length}) 
is substantially simplified and the dependence on  
moduli becomes linear \cite{cim1}. 
 
{\it i) Square and linear quiver} 
 
In this case we have for the wrapped volume of each 
brane 
\beq
\begin{array}{rcl} 
||l_a|| & = & {3n_a^2}R_1^{(1)}R_1^{(2)}R_1^{(3)} (1+(U^3/6)^2) \\ 
||l_b||& = & \frac{n_b^1}{\beta^2} 
 R_1^{(1)}R_1^{(2)}R_1^{(3)} (1+(U^3/2)^2) \\    
||l_c|| & = & \frac{n_b^1}{2\beta^2}R_1^{(1)}R_1^{(2)}R_1^{(3)}(U^3)^{2} \\ 
||l_d|| & = & {3\rho n_a^2}R_1^{(1)}R_1^{(2)}R_1^{(3)} (1+(U^3/2)^2) 
\label{longis} 
\end{array}
\eeq
Here we have $\rho =1/3$ in the square quiver and both 
choices $\rho=1,1/3$ in the linear quiver case. 
With these values for the volumes of the cycles one has 
 for the ratio of $SU(3)$ and $SU(2)_L$ couplings: 
\beq 
\frac{\alpha_{QCD}}{\alpha_L}=\frac{||l_b||}{||l_a||}= 
{{n_b^1}\over {3n_a^2\beta^2}}  { {1+(U^3/2)^{2}}\over 
{1+(U^3/6)^{2}} } 
\label{relacion2} 
\eeq 
and for the hypercharge coupling 
\beq 
\frac{\alpha_{QCD}}{\alpha_Y}=  \frac{1}{6}+ 
\frac{1}{12}\left(  { { 3\rho  +(3\rho /4+{{n_b^1} 
\over {2n_a^2\beta^2}}) (U^3)^2 } 
\over { 1+(U^3/6)^2 } } \right) 
\eeq 

{\it ii) Triangle quiver } 
 
In this case we have for the wrapped volume of each brane 
\beq
\begin{array}{rcl} 
||l_a|| & = & 9 R_1^{(1)}R_1^{(2)}R_1^{(3)} (1+(U^3/6)^2) \\ 
||l_b|| & = & \frac{1}{\beta^2} 
 R_1^{(1)}R_1^{(2)}R_1^{(3)} (1+(U^3/2)^2) \\ 
||l_c|| &  = &  \frac{1}{2\beta^2}R_1^{(1)}R_1^{(2)}R_1^{(3)}(U^3)^{2} \\ 
||l_d|| & = & ||l_a||
\end{array}  
\label{longissm} 
\eeq
so that we find for the ratio of strong and weak couplings 
\beq 
\frac{\alpha_{QCD}}{\alpha_L}=\frac{||l_b||}{||l_a||}\ =\ 
{ {1}\over {9\beta^2} } 
 { {1+(U^3/2)^2}  \over 
{1 + (U^3/6)^2 } } 
\label{relacion3} 
\eeq 
and for the hypercharge coupling 
\beq 
\frac{\alpha_{QCD}}{\alpha_Y}=  \frac{1}{6}\ +\  \frac{1}{4}\  
\left( 1+\ {1\over {18\beta^2}} { {(U^3)^2}\over {1 + (U^3/6)^2} } 
\right) 
\label{relacion4} 
\eeq 
As we said, this gives us the ratios of the coupling constants 
at the string scale, and one should compute the running of the 
couplings in the region in between the weak scale and the 
string scale.  
There can be important effects if there are extra particles 
in that energy region from e.g., KK or winding states, gonions 
etc. One can in fact check that if one neglects the effect 
of the running, it is not possible to find values for $U^3$  
such that one reproduces   
at the string  scale the  
experimental  weak scale ratios 
$\alpha_{QCD}^{-1} \ : \ \alpha_L^{-1} \ : \ \alpha_Y^{-1} \ = 
\ 8.3 \ : \ 29.6 \ : \ 98.4$. 
In the case of the square quiver it is easy to obtain  
the weak scale result for $ \frac{\alpha_{QCD}}{\alpha_L}$ 
but the ratio $\frac{\alpha_{QCD}}{\alpha_Y}$ turns out to 
be too small. In the case of the triangle model presented above it is 
also difficult to obtain $ \frac{\alpha_{QCD}}{\alpha_L}$ 
large enough. This is due to the fact that this particular 
example has very large winding numbers $n_a^i$ for the  
baryonic brane, which then tends to give too small  
$\alpha _{QCD}$. It should be interesting to do a 
systematic search for other triangle models  
giving rise to larger values for $\alpha _{QCD}$.

Note that these statements concern these  
particular models with just 4 stacks of branes 
discussed above.  
If for example   
 there is an extra brane 
$h$ contributing to the hypercharge group 
as in eq.(\ref{y}) 
there is an extra piece  
${1\over 4}\frac{||l_h||}{||l_a||}$ to be added to eq.(\ref{hiperra}) 
and in the square quiver case one can then easily    
adjust $\frac{\alpha_{QCD}}{\alpha_Y}$ by varying 
the extra brane parameters.

\section{The SM Higgs mechanism as a brane recombination process}

All the above constructions involve the unbroken electroweak 
gauge symmetry $SU(2)_L\times U(1)_Y$ which we know  
is spontaneously broken by the Higgs mechanism. The different models  
have scalars in their spectrum with the quantum numbers of  
SM Higgs fields to do the job. Whereas the mechanism  
is understood at the field theory level, there should exist  
a string version of the SM Higgs mechanism in terms of the  
underlying branes. Specifically, from the string theory  
point of view we know  that the gauge group  
of the SM originates from open strings starting and ending  
on D-branes. The gauge group $SU(2)_L$ originates on two 
parallel branes whereas hypercharge comes from a linear combination 
of $U(1)$'s attached to different branes. In the SM Higgs 
mechanism the rank of the gauge group is reduced. As we will 
now discuss, the stringy counterpart of this rank-reduction 
is brane recombination. 
Brane recombination is a process in which two intersecting 
branes fuse into a single one (see fig.\ref{fussion}).     
 In particular, it is known that tachyons appearing 
in string theory signal an instability with respect to the decay of 
the system into another one with lower energy. 
Tachyons appearing at a pair of interesting branes 
show the instability with respect to the recombination of 
both branes into a single one with less energy, i.e., less 
wrapped volume and lying in the same homology class than the initial pair.

 In a brane recombination process the number of massless chiral 
fermions decreases. Consider a couple of intersecting 
branes $\alpha $ and $\beta $ which recombine into a single  
final brane $\gamma$. Consider an spectator brane $\rho $ 
which intersects both branes $\alpha $ and $\beta $ ( 
i.e., $I_{\rho \alpha }\not= 0$, $I_{\rho \beta }\not= 0$) but 
not participating in the recombination. 
The net number of chiral fermions before recombination $n_i$  
\beq 
n_i\ =  \ |I_{\rho   \alpha}|+|I_{\rho \beta}| \ \geq 
|I_{\rho   \alpha} + I_{\rho \beta}| \  =\ |I_{\rho \gamma }| \ =\ n_f 
\label{ctheorem} 
\eeq 
is bigger than the net number of chiral fermions after 
recombination $n_f$. 
 The effective field theory interpretation 
is that after a Higgs mechanism some of the chiral fermions  
may acquire masses from (not-necessarily renormalizable) 
Yukawa couplings.  
 
\EPSFIGURE{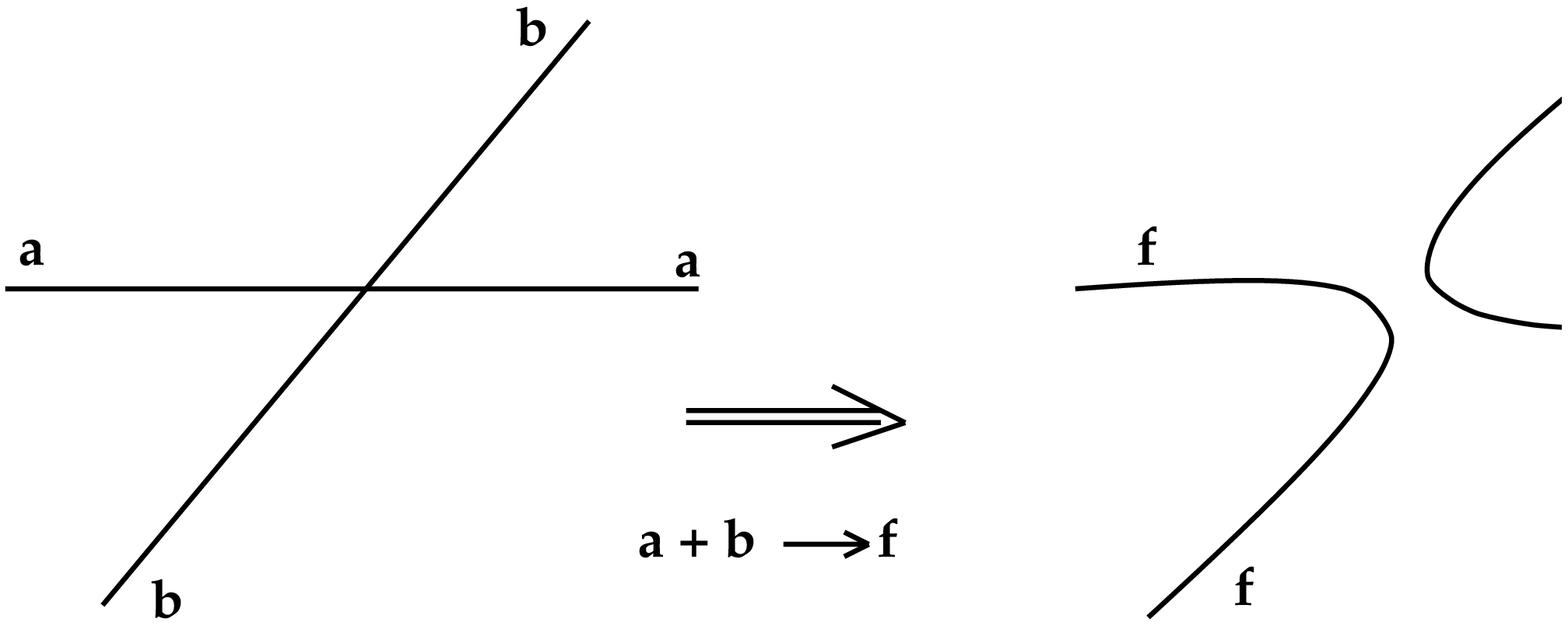, width=6in, height = 3in}
{\label{fussion}
Recombination of two 
intersecting branes $a,b$ in a compact space. 
Initially the gauge group is $U(1)_a\times U(1)_b$. 
At the intersection a tachyon scalar triggers the recombination 
into a single brane $f$. The final gauge symmetry 
is $U(1)_f$, corresponding to the Higgsing 
$U(1)_a\times U(1)_b\rightarrow U(1)_f$ which is 
induced by the tachyon.} 
 
In what follows we will show how  
general arguments about  brane recombination   
as above yield results which are  
in agreement with the low-energy field theory expectations. 
However we will also see that brane recombination has some extra 
stringy ingredients: it cannot just be described by the lightest 
tachyon getting a vev, higher excitations of the tachyon must 
be involved in the full process of brane recombination.   
This seems to indicate that a full description of the  
process would require string field theory, rather 
than an effective field theory Lagrangian. 
Our argumentation on recombination will only make  
use of the computation of intersection numbers before  
and after brane recombination, imposing conservation of 
RR charge. Thus it will be mostly topological 
rather than geometrical in nature, i.e., we will 
be able to say what fermions and gauge bosons  
become massive in a  
brane recombination process, but not the size of the 
masses acquired.

The understanding of the Higgs mechanism as a brane 
recombination process has nothing to do with the presence  
or not of some SUSY at the brane intersections. 
So our discussion below also applies to general  
non-SUSY intersecting models. 
 However  
for concreteness we will first discuss how the  
Higgs/recombination process occurs in the square quiver models 
and will later describe the case  of  the triangle quiver  
in which an interesting departure between brane  
recombination and effective field theory Higgsing occurs.

\subsection{The electroweak Higgs system in the square quiver }

 \TABLE{\renewcommand{\arraystretch}{1.25}
\begin{tabular}{|c|c|c|c|} 
\hline 
 Higgs   &  $Q_b$  &  $Q_c$   & Y \\ 
\hline\hline $h_1$ & 1  &  -1 & 1/2  \\ 
\hline $h_2$ &   -1    &  1  &  -1/2   \\ 
\hline\hline $H_1$ & -1  &  -1 & 1/2  \\ 
\hline $H_2$ &   1    &  1  &  -1/2   \\   
\hline 
\end{tabular} 
\label{higgsses}
\caption{\small Electroweak Higgs fields.} } 

We will consider for simplicity the version of the square 
quiver discussed in section (5.1) in which only the hypercharge 
remains massless, although the discussion in this subsection applies 
in fact to the larger class of models discussed in \cite{imr}. 
In the square quiver there are not generic intersections giving rise 
to massless scalars with the quantum numbers of the SM Higgs 
fields. On the other hand  open strings stretched between the `$b$' and `$c$' 
stacks of branes do have the quantum numbers of Higgs fields appropriate 
to yield electroweak symmetry-breaking. These two stacks of branes are  
parallel on the second torus, and that is why they generically do not  
intersect. However if one approaches those stacks to each other 
scalar Higgsses with the quantum numbers of  
table \ref{higgsses} appear in the light spectrum 
\cite{imr}.  
These states corresponding to open strings stretching 
between branes $b$ and $c$ (denoted by $h^\pm$) and between branes 
$b$ and $c^*$ (denoted by $H^\pm$) have masses (see \cite{imr}) 
\beqa 
{m_{H^{\pm}}}^2 \ &=&\ {X^2_{bc^*} \over 4\pi} M_s^2 
\ \pm  \ \frac{M_s^2}{2}\left||\theta^1_{bc^*}|-|\theta^3_{bc^*}|\right| 
\ =\  M_s^2 \ \left({X^2_{bc^*}\over 4\pi}\ \pm   
\ \frac{1}{2} (2\alpha_2-\Delta +\d_b) \right),   
\nonumber \\ 
 {m_{h^{\pm}}}^2 \ &=&\ {X^2_{bc} \over 4\pi} M_s^2 
 \ \pm \ \frac{M_s^2}{2}\left||\theta^1_{bc}|-|\theta^3_{bc}|\right| 
\ =\  M_s^2 \ \left({X^2_{bc}\over 4\pi}\ \pm   
\ \frac{1}{2} (2\alpha_2+\Delta +\d_b) \right), 
\label{mashigs} 
\eeqa 
where $X_{bc^*}$ ($X_{bc}$) is the distance  
(in $\alpha^{\prime \frac 12}$ units) 
 in transverse space along the second torus. One also has  
$\alpha_2=tg^{-1}(U^3/2)$ and $\Delta$, $\d_b$ were defined in 
(\ref{deltona},\ref{deltas}).  
There are also fermionic partners (``Higgsinos'') of these two types of 
complex scalar fields. 
The above scalar mass spectrum can be interpreted as arising from 
a field theory mass matrix 
\beq 
(H_1^* \ H_2)  
\left( 
\bf {M^2} 
\right) 
\left( 
\begin{array}{c} 
H_1 \\ H_2^* 
\end{array} 
\right) 
+(h_1^* \ h_2) 
\left( 
\bf {m^2} 
\right) 
\left( 
\begin{array}{c} 
h_1 \\ h_2^* 
\end{array} 
\right) + h.c. 
\eeq 
where 
\beq 
{\bf M^2} =\ M_s^2 
\left( 
\begin{array}{cc} 
X_2^{(bc^*)}& 
\frac{1}{2}(2\alpha_2-\Delta +\d_b)  \\ 
\frac{1}{2}(2\alpha_2-\Delta +\d_b) & 
X_2^{(bc^*)}\\ 
\end{array} 
\right), \ {\bf m^2}=(bc^*,-\Delta  \leftrightarrow bc,\Delta ) 
\eeq 
\vspace{1cm} 
The fields $H_i$ and $h_i$ are thus defined as 
\beq 
H^{\pm}={1\over2}(H_1^*\pm H_2); \ h^{\pm}={1\over2}(h_1^*\pm h_2)  \ . 
\eeq 
It is clear from the above formulae that 
 when the distance between each pair of stacks is 
small enough, some of the scalars $H^{\pm}$ and $h^{\pm}$ 
become tachyonic, which will be the signal of 
spontaneous symmetry breaking in the SM. 
 In this process the weak vector bosons 
$Z_0$ and $W^{\pm}$ get a mass but also the fermions get masses 
proportional to their Yukawa couplings to the different Higgs fields. 
The form of the Yukawa couplings among the SM fields in table \ref{tabpssm}  
and the different 
Higgs fields are essentially fixed by conservation of  
gauge symmetries and  
 have  the general form \cite{imr}: 
\beqa 
y^U_jQ_LU_R^j h_1 \ +\ y^D_jQ_LD_R^jH_2 \ +  \nonumber \\ 
y^u_{ij}q_L^iU_R^j H_1 \ +\ y^d_{ij}q_L^iD_R^jh_2 \ + \\ 
y^L_{ij}L^iE_R^jH_2  \ +\  y^N_{ij}L^iN_R^jh_1 \ +\  h.c. \nonumber 
\label{yuki} 
\eeqa 
where $i=1,2$ and $j=1,2,3$.  
With this Yukawa structure one observes that, 
 for example, if only the Higgs fields 
of type $H_i$ get vevs, two down-like and  one up-like quarks still 
remain massless, as well as all  neutrinos. If in addition the 
Higgs of type $h_i$ get a vev, all fermions get a mass. 
 
As discussed in previous sections, it is clear that this Higgs sector of the  
square quiver is not supersymmetric. However one can obtain a  
electroweak scale well below the string scale $M_s$ by  
``modest tuning'' the distances $X^2_{bc^*}/2\pi = 2\alpha_2 - \Delta + \d_b$, 
$X^2_{bc}/2\pi = 2\alpha_2 + \Delta + \d_b$.  
Let us turn now to a description of the electroweak Higgs 
mechanism in the brane recombination language.

\subsection{The mechanism of brane recombination} 
 
We are now going to show how the process of brane recombination 
yield results consistent with the above field theory 
description of the Higgs mechanism. For this to happen we have to 
show that 
 
{\it i)} The final gauge group after brane recombination 
is indeed just $SU(3)\times U(1)_{em}$.  
 
{\it ii)} The quarks and leptons become massive in the 
expected way. 
 
Before proceeding let us first make a few comments about the 
recombination process. 
 In general, even if the 
two initial intersecting branes $\alpha$ and $\beta$ wrap 
factorizable cycles, the resulting recombined brane $\gamma$ will not 
be factorizable. In this case the form of the recombined cycle   
will not have an easy geometrical description. Still, many 
properties of the recombination process will only depend 
on the general homology class of the cycles. In the case of 
non-factorizable cycles one has to work  with  a 
$2\times 2\times 2=8$-dimensional basis for the RR-charges 
of the cycles. Instead of wrapping numbers $n^i,m^i$, one has 
to work with 8-dimensional vectors $ \vec{q}_k$, $k=1-8$ 
(see  Appendix I). 
For a given system of branes to be recombinable into another 
 the initial and final 
configurations must lie in the same homology class,  
the recombination must preserve RR charge: 
\beq 
\sum_{s=initial~branes} N_s (1+\Omega) \vec{q}_s 
\ =\ \sum_{f=final~branes} N_f 
(1+\Omega) \vec{q}_f 
\eeq 
where  $\Omega \vec{q}_s$ gives the charge vector of the orientifold mirror 
of each brane. 
Consider for example the case of 
the Higgs fields $H_i$ in our class of models. These states come from 
the open string exchange between branes of type $b$ and $c^*$. 
Thus this Higgs field taking a vev should be related to the 
recombination of one of the two branes of type $b$ (let us call it $b_1$) with 
the brane $c^*$ into 
a single  non-factorizable `$e$' brane with charge:   
\beq 
2 \vec{q}_b+\Omega \vec{q}_c=(\vec{q}_{b_1}+\Omega 
\vec{q}_c)+\vec{q}_{b_2}\equiv 
\vec{q}_e 
+\vec{q}_{b_2} 
\eeq 
Instead of two $b$-branes (which contained  the $SU(2)_L$ 
charged current electroweak interactions) we are now left with 
only one. Thus the electroweak symmetry has been broken. 
We can now check which chiral fermions, if any, remain in the 
massless spectrum by computing the new intersection numbers. 
The latter are easy to compute 
knowing that $I_{\alpha \beta}$ is a 
bilinear quantity in terms of 
$\vec{q}_\alpha $ and $\vec{q}_\beta$ 
(see Appendix I). In our case we simply have 
 $I_{ae}=I_{ab_1}+I_{ac^*}$ and 
$I_{ae^*}=I_{ab_1^*}+I_{ac}$. Thus the new intersection 
numbers $after$ recombination are: 
\beq 
\begin{array}{lcl} 
I_{ab_2}\ =   \ 1, & & I_{ab_2^*}\ =\ 2, \\ 
I_{ae}\ =   \ -2, & & I_{ae^*}\ =\ -1, \\ 
I_{b_2d}\ =   \ -3,  & & I_{b_2d^*}\ =\ 0, \\ 
I_{de}\ =   \ 0, & & I_{de^*}\ =\ -3, 
\end{array} 
\label{intersec3}
\eeq 
Note that the number of chiral fermions has been reduced, 
only three quark flavours and some three leptons remain massless. 
This result is as expected from field theory arguments.
Indeed, looking at (\ref{yuki}) we see that if 
Higgs fields of type $H_i$ get vevs, two up-like and one down-like   
quarks become massive and also charged leptons do, due to 
Yukawa couplings. 
 
\EPSFIGURE{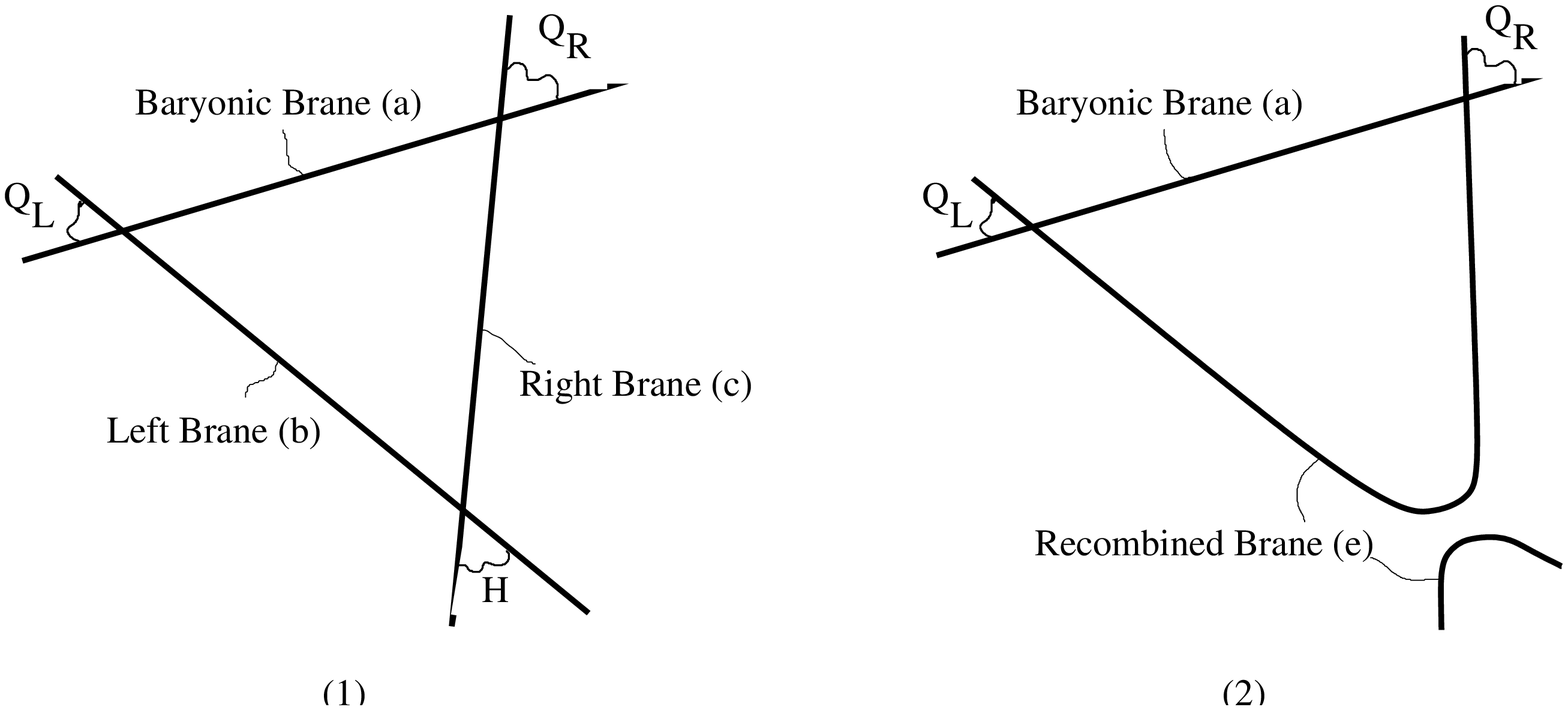, width=6in}
{\label{recombination}
Picture of the recombination/symmetry breaking. (1) Before the 
recombination, the worldsheets connecting the $Q_L$, $Q_R$ and 
Higgs multiplets have a trianglular shape with sides embedded  
in the branes. (2) After the recombination, branes '$b$' and 
'$c$' have recombined into brane '$e$', giving a non-factorizable cycle 
with lower volume. Note that the intersection number between '$a$' and  
'$e$' is zero, but these branes are not parallel.} 
 
In figure~\ref{recombination} we can obtain an intuitive image of what 
may be happening. The Yukawa coupling amplitudes are obtained 
from correlators involving vertices of quarks and Higgs fields 
on the vertices  of a  triangle in the extra compact dimensions. 
The different Yukawa couplings are expected to 
be exponentially suppressed \cite{afiru2,imr}  
by the area of the corresponding 
triangle  i.e., $Y_{abc^*}\ =\  \exp{(- A_{abc^*})}$. 
 After the 
brane $c^*$ has recombined with one of the branes $b$, the triangle is 
smoothed out at the $bc^*$ vertex. This corresponds to the Higgs fields 
$H_i$ getting a vev. 
Note 
that the net contribution to $I_{ae}$ due to the quarks at the other two 
vertices is zero, these quarks are no-longer chiral, have acquired masses.

At the end of the above recombination process we have four stack 
of branes $a$, $b_2$, $e$ and $d$, suggesting a gauge group 
$SU(3)\times U(1)_a\times U(1)_b\times U(1)_e \times U(1)_d$. 
Now, if our interpretation of the Higgs mechanism as a recombination 
process is correct, we should just obtain electromagnetic charge 
as our only surviving $U(1)$ boson in the recombined system. 
Thus as a test we should check that indeed it is the 
electromagnetic charge with survives the recombination process. 
The $U(1)$ combination with remains light $U(1)_{massless}$ 
is the one with no coupling to the RR closed string 
fields $B_2^I$, $I=0,1,2,3$. This is the combination 
\beq 
U(1)_{massless}\ =\ \sum_{k=a,b_2,e,d} c^kU(1)_k, 
\eeq 
with coefficients $c^k$ such that 
\beq 
\sum_kc^k{\tilde d}_k^I\ =\ 0\ ,\  \,  I=0,1,2,3 
\label{cons} 
\eeq 
where ${\tilde d}_k^I$ are the coefficients of the couplings of RR fields 
to the 
four recombined $U(1)$'s, i.e., the coefficients 
in the couplings ${\tilde d}_k^I\ B_2^I\wedge F_k$. The 
corresponding coefficients for the starting cycles    
$d_k^I$  are  given in 
eq.(\ref{caplillos}). Using linearity of the RR charge in the 
recombination one can compute the new 4 
 ${\tilde d}_k^I$ 
coefficients in terms of the ones  
before recombination $d_k^I$ as follows:   
\begin{itemize} 
\item{$\tilde d_k^I= d_k^I$} for those   brane stacks  $k$ not taking  part in 
the recombination. 
\item{$\tilde 
d_k^I=\frac{1}{N_\alpha}d_\alpha^I+\frac{1}{N_\beta}d_\beta^I$, for 
the new brane resulting from  the recombination.} 
\end{itemize} 
Using this and  
imposing eq.(\ref{cons}) before and after recombination 
one can compute explicitly the new  ${\tilde d}_k^I$'s and 
obtain which $U(1)$'s remain massless. 
In the present case, if  
before recombination we had as the only surviving $U(1)$ the 
hypercharge of eq.(\ref{hyper}) (with $r=-1$), after 
recombination there is only one remaining massless 
generator given by 
\beq 
Q_{em}=\frac{1}{6}Q_a+\frac{1}{2}Q_e-\frac{1}{2}Q_d- 
\frac{1}{2}Q_{b_2}   \,  \  . 
 \label{em} 
\eeq 
One can compute the charges of the chiral fermions that remain massless 
and one finds the residual spectrum 
(obtained  from the intersection numbers (\ref{intersec3}) ) 
after recombination to be: 
\vspace{.5cm} 
 
{\bf Gauge Group:}~~SU(3)$\times$ $U(1)_{em}$ 
 
{\bf Matter Content:}~~$2(3)_{-\frac{1}{3}}+2(\bar3)_{\frac{1}{3}}+ 
1(3)_{\frac{2}{3}}+ 
1(\bar3)_{-\frac{2}{3}}+3(1)_{0}+3(1)_{0} $ 
 
\vspace{.5cm} 
Thus we  see we have one massless u-quark, two massless 
d-like  quarks (i.e. down and strange) 
and massless (left and right)   
neutrinos, as expected from the field 
theory arguments  from the standard Yukawa coupling of the Higgs fields 
$H_i$ to fermions. This is interesting and in fact not  far 
from  the experimental situation in which three quark flavours 
are relatively light (leading to Gellman's flavour $SU(3)$ ) 
and neutrinos are  almost massless. 
 
From the effective Lagrangian point of view, we could have in addition 
a vev for the other Higgs fields of type $h_i$. If this is the case 
the rest of the fermions would now become massive, but still a charge 
generator should remain unbroken. From the brane recombination point of view 
this  corresponds to the recombination of the new brane `$e$' 
with the `$b_2^*$' brane which was an spectator in the 
first recombination. The new recombined brane `$f$' will have 
a RR charge vector 
$\vec{q}_f \equiv(\vec{q}_e+\Omega \vec{q}_{b_2})$. We are thus 
only left with three 
stacks of branes, $a$,$d$ and $f$. One can easily check that the 
intersection numbers between these branes is zero. 
For example, $I_{af}=I_{ae }+I_{ab_2^*}=-2+2=0$. 
 Thus there are 
no massless fermions left after this second recombination, 
again as expected from field theory arguments. 
 One might worry that, since we have now only three stacks of branes 
but we still have four RR fields $B_2^I$, all $U(1)$'s could become 
massive and we would be left with no photon in the low 
energy spectrum. This turns out not to be the case. 
One can easily check that a $U(1)$ generator 
\beq 
Q_{massless}\ =\ 
\frac{1}{6}Q_a+\frac{1}{2}Q_f-\frac{1}{2}Q_d   \,  \ 
 \label{em2} 
\eeq 
remains unbroken. This massless generator can be identified with 
electromagnetism by noting that e.g., the massive fields    
stretched between branes $a$ and $d$ have charges which correspond 
to the electric charge they had before this last recombination.

Note that from the effective field theory point of view 
there are other (less interesting) field directions 
of the scalar potential, particularly if we include in the complete  potential 
additional massive scalars which appear at the intersections of branes $b$ and 
$c$ with their mirrors $b^*$ and $c^*$.  
In particular, open strings stretched between 
the branes $b$ and $b^*$  contain (massive) scalars transforming like symmetric 
(triplet) and antisymmetric (singlet) $SU(2)_L$ representations  
which couple to the SM Higgs doublets $H_i$ and $h_i$ and modify the 
scalar potential. 
These other field directions could in principle lead to unwanted vacua 
with breaking of electromagnetic $U(1)$. From the field theory point of view 
this can be controlled by making the unwanted fields heavy, by  
appropriately 
separating each brane from its mirror. 
 
These different Higgsing possibilities also exist in the brane recombination 
language as coming from different choices for brane recombination. 
The brane recombination discussed above describing the SM Higgs mechanism 
correspond to the recombination of branes: 
\beq 
b_1\ +\ b_2^* \ +\ c^* \, \ \rightarrow \, \  e \ +\ b_2^*  \, \ \rightarrow 
\, f 
\label{cadena} 
\eeq 
Recombining e.g., $b_1 + b_2  + c^*$  or $b_1 + b_2  + c$ would have lead to 
other less interesting final configurations with e.g., broken  
electromagnetic charge. 
Thus from the phenomenological point of view these other possibilities 
should be somehow energetically disfavoured. 
  

We have seen how the Higgs mechanism in the SM may be understood 
as a brane recombination process in which the three branes 
giving rise to the $U(2)_b\times U(1)_c$ recombine into a single 
brane giving rise to a final (in general non-factorizable) brane $f$. 
There is only one massless Abelian generator which can be identified 
with the standard electromagnetic charge. Thus the actual string vacuum 
after recombination involves only three stacks of branes $a$,$d$ and $f$ 
and the gauge group is just $SU(3)_{QCD}\times U(1)_{em}$. 
 
There is a couple of questions which appear in such an interpretation: 
 
{\it i)} 
The recombination language shows  us how the quarks and leptons 
become massive at each step, but how can one explain the 
observed existence of large hierarchies of fermion masses? 
We already mentioned that in the initial picture 
in which the $a,b,c,d$ branes wrap factorizable cycles, 
there are Yukawa couplings  between Higgs fields 
and  chiral fermions (see fig.(\ref{recombination})). The worldsheet 
with one right-handed fermion, one left-handed fermion  and 
one Higgs field have a triangular shape with those 
particles at the corners \cite{afiru2}. As we mentioned above, the 
corresponding coupling may be exponentially 
suppressed.  After the branes recombine (i.e., after the Higgs get 
vevs) the vertex where the Higgs field lies is smoothed out 
(see fig.(\ref{recombination})). Still, if the vev of the Higgs is small 
compared to the 
string scale, this will amount to a small perturbation, fermion 
masses would still be exponentially suppressed by the area left between 
the recombined brane $e$  and the baryonic $a$-stack ($d$-stack in the 
case of leptons), which is the smoothed out triangle. 
Thus the fermion masses may still have hierarchical ratios. 
 
{\it ii)} 
The actual final vacuum contains only the stacks $a$, $d$ and $f$ 
and only $SU(3)_{QCD}\times U(1)_{em}$ as gauge group. 
If we put energy in the system, we should be able to see the 
electroweak gauge bosons $Z_0$ and $W^{\pm }$ produced, with masses 
of order the vev of the Higgs fields. Which are those states 
in the final recombined system? The $Z_0$ is neutral 
and should correspond to open strings beginning and ending 
on the same recombined brane `$f$'. In particular, it should 
correspond to an open string stretching between the opposite 
 smoothed  portions of that brane in fig.(\ref{recombination}). 
On the other hand open strings stretching between the recombined brane $f$ and 
its mirror $f^*$ have charge $= \pm 1$ (see eq.(\ref{em2})) and should  
give rise  to the charged $W^{\pm}$ bosons.

\subsection{Higgs mechanism and  
brane recombination in the triangle  quiver }

Let us now discuss for comparison the Higgs mechanism and 
brane recombination in the triangle quiver. 
 Those models 
have an extra massless $U(1)_{B-L}$ but we will concentrate 
first on  electroweak symmetry breaking. 
One important difference is that in these models there are 
massless chiral multiplets corresponding to Higgs fields  
at the intersections of the stacks $(b)$ and $(c)$ 
(see table \ref{espectrosm}).   
In the case $\beta^2=1$ there is in fact a single Higgs set 
and the charged chiral massless spectrum is just that of the 
MSSM (plus right-handed neutrinos). 
Looking at the charges in table \ref{espectrosm} one sees that the 
only allowed Yukawa superpotential couplings are: 
\beqa 
y^u_{ij}q_L^iU_R^j {\bar H} \ +\ y^d_{ij}q_L^iD_R^j H \ + \nonumber \\ 
y^L_{ij}l^iE_R^jH   \ +\  y^N_{ij}l^iN_R^j{\bar H} \ +\  h.c.  
\label{yuki2} 
\eeqa 
with $i=1,2$  and $j=1,2,3$. Thus  
one would say that one of the generations of quarks  
and leptons does not get masses at this level. 
 
Let us now see how would be the Higgs 
electroweak  symmetry breaking in the brane recombination language.  
As in the square quiver case, let us assume that 
one of the two $(b)$ branes, e.g., $b_1$ recombines with $c^*$. 
Looking at table \ref{espectrosm} we see that this should correspond to 
${\bar H}$ getting a vev. After the recombination $b_1+c^*\rightarrow e$ 
into a single brane $e$ the electroweak symmetry is broken.  
We are left with four stacks of branes $a$, $b_2$, $e$ and $d$ and 
the intersection numbers are given by 
\beq 
\begin{array}{lcl} 
I_{ab_2}\ =  \ 1, & & I_{ab_2^*}\ =\ 2,  \\ 
I_{ae}\ = \ -2, & & I_{ae^*}\ =\ -1, \\ 
I_{db_2}\ = \ 1, & & I_{db_2^*}\ =\ 2, \\ 
I_{de}\ = \ -2, & & I_{de^*}\ =\ -1, \\ 
I_{b_2e}\ = \ -{1\over\beta^2}, & & I_{b_2e^*}\ =\ -{1\over\beta^2},\\ 
I_{ee^*}\ =\ {2\over\beta^2} 
\label{intersec5} 
\end{array} 
\eeq 
with $\beta^2=1,1/2$. It is easy to check now that  
(apart from the $B-L$ symmetry which remains unaffected by this  
process and will be discussed later) there is an unbroken massless 
gauge generator which can be identified with electromagnetism 
\beq 
U(1)_{em}\ =\ { {U(1)_a}\over 6 } \ -\ { {U(1)_d}\over 2} \ -\ { { U(1)_{b_2}} 
\over 2} \ +\ { { U(1)_{e}} \over 2} \ . 
\eeq 
and is not rendered massive by couplings to RR B-fields. 
The massless chiral spectrum after recombination obtained from the  
above intersection numbers  
\footnote{In order to compute the chiral fermion spectrum from  
an intersection of a brane $j$ with its mirror $j^*$ one 
should use the general formula (\ref{specori2}) of Appendix I. 
Since in the present case  $I_{j,ori}=0$, the number of chiral 
fermions is just half the intersection number. This is also true 
for the spectrum in table (\ref{espectrosm2}).}  
is (for $\beta^2=1$) 
\vspace{.5cm} 
 
$\begin{array}{ll}
{\bf Gauge Group:} & SU(3) \times U(1)_{em} \times [U(1)_{B-L}] \\
{\bf Matter Content:} & 
2(3)_{-\frac{1}{3}}\ +\ 2(\bar3)_{\frac{1}{3}}  
+\ 1(3)_{\frac{2}{3}}\ +\  1(\bar3)_{-\frac{2}{3}}\ + 2(1)_{-1}\ +\ 2(1)_{1}
\\ &  +\ (1)_{0}+(1)_{0}\ +\ [(1)_{1}+(1)_{0}+(1)_{-1}] 
\end{array}$

\vspace{.5cm} 
Thus at this level, with only $b_1+c^*$ recombination,  
we are left with two  D-quarks, one  
U-quark,  two charged leptons and one (Dirac) neutrino. 
In addition there are charged and neutral Higgsinos (last three 
states in brackets).  
 
Comparing this massless spectrum with the one expected from the 
field theory Yukawa couplings with a vev for the Higgs ${\bar H}$,  
we see that  one extra D-quark and one charged lepton 
became massive after recombination, which was  not expected 
from the effective field theory Lagrangian.  
It seems that this puzzling result can be understood 
as follows. Consider the brane intersection 
$bc^*$ giving rise the the  Higgs field ${\bar H}$ (see table  
\ref{espectrosm}). In addition to the ${\bar H}$ massless 
chiral multiplet, there are at this intersection other {\it massive}  
vector-like pairs with the opposite $Q_b$ and $Q_c$ 
charges to ${\bar H}$. These are in some way 
$\cn = 4$ partners of the massless Higgs multiplet (see section  
(4.1) of ref.\cite{cim1}). Thus at the intersection nothere are  
massive $Y=-1/2$ fields $H'$ with $Q_{b,c}$ charges qual to $(1,1)$. 
 Note that such $H'$ fields have precisely the  
gauge quantum numbers required to couple to $Q_L$ and $D_R$ 
(see table \ref{espectrosm}), so that if a vev is induced for $H'$ 
a D-quark (and a charged lepton) will become massive. 
 Thus what it seems is happening here  
is that when the branes $b_1$ and $c^*$ recombine at this intersection 
it is not only the lightest (tachyonic)  scalar which is  
involved but  also massive excitations. Recall however  
that our recombination arguments are purely topological in character,  
they tell us who becomes massive and who remains massless, 
but it does not tells us how big are the masses. It is reasonable to expect 
that since massive modes are involved, the masses of these 
D-quark and charged lepton are small, so that they can perhaps be 
identified with the d-quark and the electron. 
 
As in the square quiver example, 
further recombination of the spectator brane $b_2^*$ with the brane 
$e$, into a final brane $f$, $b_2^* + e\rightarrow f$ renders 
massive all of the fermions. Indeed, using bilinearity of the  
intersection numbers it is easy to check that all intersection 
numbers  vanish. Still an electromagnetic generator 
\beq 
U(1)_{em}\ =\ { {U(1)_a}\over 6 } \ -\ { {U(1)_d}\over 2} \ +\ { { U(1)_{f}} 
\over 2} 
\ . 
\eeq 
can be checked to remain in the massless spectrum, i.e., it does not 
receive any mass from couplings to B-fields.  
Thus the electroweak symmetry breaking process is completed. 
 
We have not addressed here the question of what triggers  
electroweak symmetry breaking in the triangle quiver.  
The Higgs multiplets are massless to start with but  
the scalars get in general masses from loops and FI-terms, 
as explained in the previous section. In addition there 
is the one-loop contribution from top-quark loops which  
will tend to induce a negative mass$^2$ to the  
${\bar H}$ higgs in the usual way 
\cite{ir}. We are assuming here that 
those effects combined yield a negative $mass^2$ to the 
Higgs fields.

\subsection{Breaking of B-L and neutrino masses from  
brane recombination}

We have seen that in all of the four classes of models constructed 
(except for the square quiver with $\Delta \not=0$) there is  
an extra gauge boson corresponding to $B-L$ in the massless 
spectrum. This extra $U(1)$ may be Higgssed in a variety of ways 
but perhaps the simplest  would be to give a vev to some 
right-handed sneutrino, which are always present in these 
models. We would like to discuss in this section  
how this process would be interpreted in  
the brane recombination language and how it implies the 
appearance of Majorana neutrino masses 
\footnote{Note that in the models of ref.\cite{imr} and in 
the subset of them formed by the models of section (5.1) 
in the present paper, there is no extra $U(1)_{B-L}$  
which should be spontaneously broken. Thus it is not required to 
spontaneously break lepton number 
by  giving a vev to a ${\tilde {\nu }}_R$. In such a case there  
can only be Dirac neutrino masses, as argued in ref.\cite{imr}. 
In models with an extra $U(1)_{B-L}$ we will be forced  
to break lepton number conservation anyhow  
and at some level 
Majorana masses will appear, as discussed below.}. 
 
Let us consider for definiteness the triangle quiver model. 
Its massless chiral spectrum is the one of the MSSM  
with right-handed neutrinos. In addition to hypercharge  
we have the $B-L$ generator at the massless level.  
The breaking of this extra gauge symmetry could be obtained  
if some right-handed sneutrino ${\tilde {\nu }}_R$ gets a vev. 
The stringy counterpart of this process would be the recombination of 
branes $c$ and $d$ in the model. So let us assume  
there is a recombination $c + d \rightarrow j$ into a final brane 
$j$. We will be left with three stacks of branes now $a$,$b$ and $j$. 
Using the results in (\ref{intersicsm}) and the linearity of intersection  
numbers one gets  
\beq 
\begin{array}{lcl} 
I_{ab}\ = \ 1, & & I_{ab_2^*}\ =\ 2, \\ 
I_{aj}\ = \ -3, & & I_{aj^*}\ =\ -3, \\ 
I_{bj}\ = \ -1 - \frac{1}{\b^2}, & & I_{bj^*}\ =\ 2 - \frac{1}{\b^2},\\ 
I_{jj^*}\ = \ -6.     
\label{intersec8} 
\end{array} 
\eeq 
One can then easily check that there is only one $U(1)$ with no 
couplings to RR-fields and hence remains massless. It is given 
by 
\beq 
U(1)_Y\ =\ { {U(1)_a}\over 6 } \ - \ { {U(1)_j}\over 2} \ . 
\eeq 
and corresponds to standard hypercharge. 
The resulting fermionic spectrum is shown in table \ref{espectrosm2}, 
for the case  $\b^2 = 1$. 
\TABLE{\renewcommand{\arraystretch}{1.25}
\begin{tabular}{|c|c|c|c|c|c|c|} 
\hline Intersection & 
 Matter fields  &   &  $Q_a$  & $Q_b $  & $Q_j$  & Y \\ 
\hline\hline $ab$ & $Q_L$ &  $2(3, 2)$ & 1  & -1 & 0  & 1/6 \\ 
\hline $ab^*$  & $q_L$   &  $(3,2)$ &  1  & 1  & 0   & 1/6 \\ 
\hline $aj$ & $U_R$   &  $3({\bar 3},1)$ &  -1  & 0  & 1 & -2/3  \\ 
\hline $aj^*$  & $D_R$  &  $3({\bar 3},1)$ &  -1  & 0  & -1 & 1/3 \\ 
\hline $bj$ & $ L $    &  $2(1,2)$ &  0   &  -1 & 1 & -1/2 \\ 
\hline $bj^*$ & $ l $    &  $(1,2)$ &  0   &  1 & 1 & -1/2 \\ 
\hline $jj*$ & $ E_R $    &  $ 3(1,1)$ &  0   & 0 & -2 &  1  \\ 
\hline 
\end{tabular} 
\label{espectrosm2}
\caption{\small Chiral spectrum in the triangle quiver after $cd^*$  
brane recombination.} } 

Note that this fermion spectrum is the one of the SM  
{\it without right-handed neutrinos}.  
The breaking of electro-weak symmetry may then proceed as in the 
previous subsection. Doublet  Higgs scalars  may be provided both by 
the scalar partners of the fermionic doublets in the table but 
also they  could be provided by the $H,{\bar H}$ fields of the original 
model in table \ref{espectrosm} if they remained relatively light 
upon recombination, which is something which will depend  
on the detailed geometry of the recombined branes.  
 
Note also that  in the recombination 
process not only $U(1)_{B-L}$ has acquired a mass but also 
the right-handed neutrinos have dissapeared from the  
massless spectrum. In addition, compared  to the 
spectrum in table \ref{espectrosm},  a pair of $SU(2)_L$ doublets  
have gained masses. The latter was expected from the  
point of view of the effective field-theory Higgs mechanism. 
Looking at the superpotential in (\ref{yuki2}) one observes  
that a vev for ${\tilde {\nu }}_R$'s mixes leptons  
with Higgsinos. On the other hand the fact that the  
${ {\nu }}_R$'s get massive is less obvious from 
the effective field theory point of view, since there  
are apparently no renormalizable  Yukawa couplings giving  
(Majorana) masses to them. As in the case of the unexpected 
massive fermions upon EW symmetry breaking discussed in the 
 previous section, an understanding of this fact seems to imply that 
brane recombination involves  in the process also the effects of  
 massive fields. There are massive chiral fields with opposite charges  
to those of the ${\tilde {\nu }}_R$'s at the intersections. 
In general dim=5 couplings of type $(\nu_R\nu_R{\tilde {\nu }}_R^*{\tilde {\nu 
}}_R^*)$ will give masses to $\nu_R$'s once the sneutrinos get vevs 
\footnote{In addition there will be mixing of the $\nu_R$'s  
with the gauginos of the $U(1)_{B-L}$ generator.}. 
Whatever the low-energy field theory interpretation, it is a fact that 
the right-handed neutrinos get massive upon $c+d$ brane recombination. 
This is interesting in itself since it has always been difficult  
to find mechanisms in string-theory giving rise to Majorana 
masses for  neutrinos. It seems that brane recombination  
gives one possible answer.

One interesting question is what happens now with the 
left-handed neutrinos. 
The gauge group of the above model after this recombination  
is just $SU(3)\times SU(2)\times U(1)_Y$. Once the 
${\tilde {\nu }}_R$'s get vevs, lepton number is violated and 
there is no longer distinction between sleptons and Higgs  
fields \footnote{This is rather similar  
to R-parity violating models with lepton number violation.}. 
Since there are no $\nu_R$'s left in the massless spectrum 
one may argue that there are no possible Dirac fermions  
for neutrinos and an effective field theory analysis would  
suggest that the left-handed neutrinos should  remain massless 
after electroweak symmetry breaking. 
The brane recombination analysis tells us that this will not be  
the case and after full brane recombination   
of the branes involved in lepton number violation {\it and} electroweak 
symmetry 
breaking: 
\beq 
b_2\ +\ b_1^*\ +\ c\ +\ d\rightarrow \ f 
\eeq 
there are no massless fermions left. Indeed, using  
bilinearity of intersection numbers plus eq.(\ref{intersicsm}) 
it is easy to check that $I_{af}=I_{af^*}=I_{ff^*}=0$.  
So somehow left-handed neutrinos have managed to become massive also 
after full brane recombination. 
As in previous cases a possible field theory interpretation  
of the neutrino masses is the importance of the massive states 
at the intersections. Dimension 5 operators of the form 
$(LL{\bar H}{\bar H})$ can give Majorana masses to 
left-handed neutrinos if the scalars in ${\bar H}$ get a vev.

An important point would be to know the size of  
neutrino masses after the full recombination/Higgsing  
process. Unfortunately the intersection numbers are topological 
numbers which count the net number of massless fermions but 
do not give as any information on the size  of the masses 
of the non-chiral fermions. The masses 
(Yukawa couplings)  are geometrical quantities which depend on the 
precise locations of the wrapping branes. Note in particular that 
although the initial $a,b,c,d$ stacks of branes of this model  
are factorizable branes with intersections respecting the 
same supersymmetry, after any recombination takes place  
the resulting recombined brane is in general non-factorizable and 
we do not know the precise shape of the cycle that the  
brane is wrapping. Thus we cannot compute in detail aspects 
like Yukawa couplings unless we get that geometrical information. 
On the other hand it is reasonable to expect that, if the  
 ${\tilde {\nu }}_R$-vevs are of order the string scale, since the  
left-handed neutrinos get their masses involving massive modes  
of order $M_s$,  
see-saw-like Majorana masses of order $m_{\nu}\propto |H|^2/M_s$,  
with a model dependent coefficient which will depend on the  
geometry and can perfectly be rather small. 
So neutrino masses within the experimental indications could 
be obtained. 
 
\EPSFIGURE{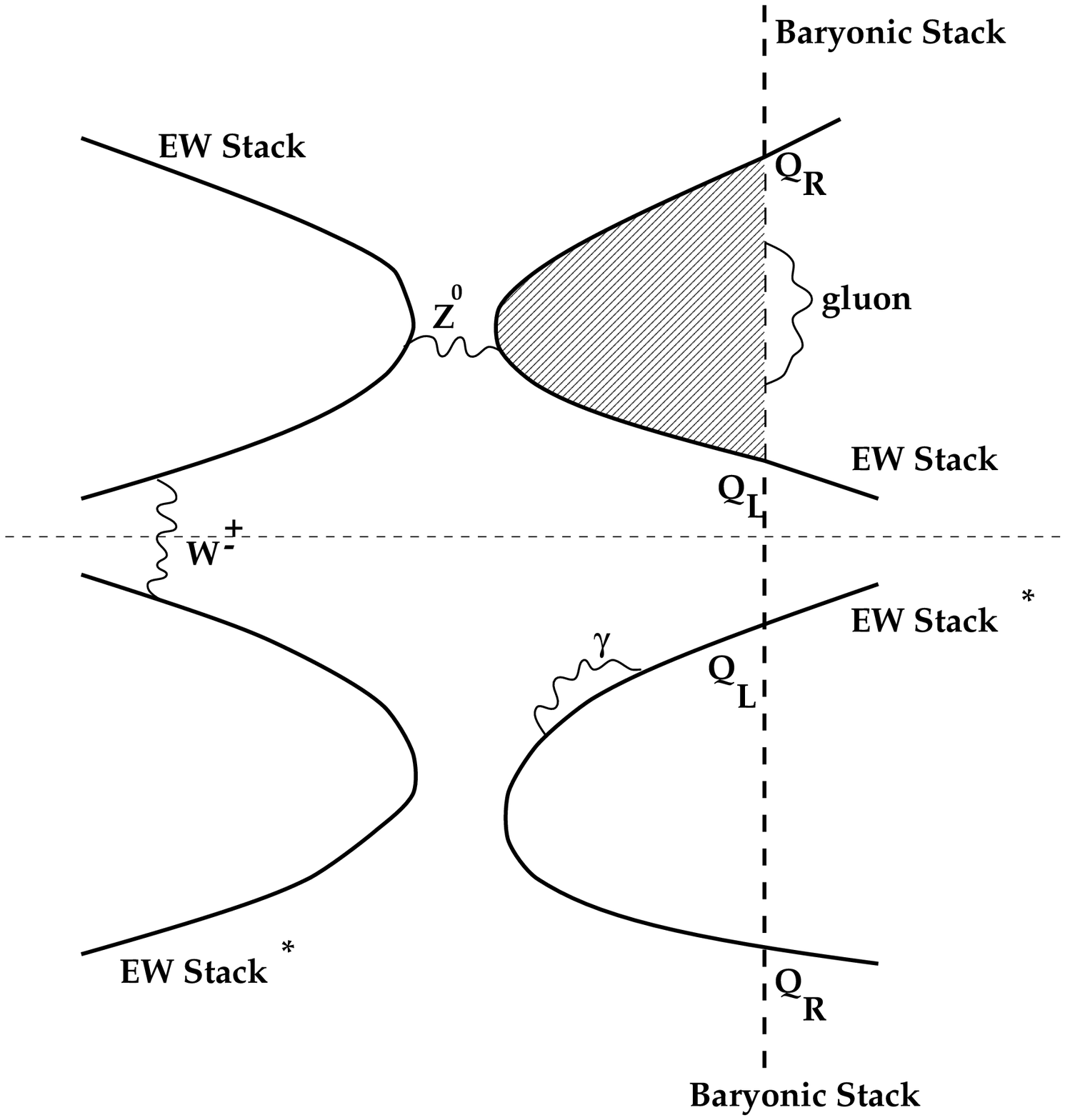, width=4.5in}
{\label{artistic}
Artist's view of a portion of the {\it baryonic } (thick dashed line) 
and {\it EW} (continuous line) brane configurations after full 
brane recombination. The mirror of the EW-brane 
is also depicted below. Massless gluons and photons come from 
open strings starting and ending on the baryonic and EW branes. 
The $Z^0$ and $W^{\pm}$ bosons come from massive open strings 
states stretching in between different regions of the same 
EW and/or its mirror EW$^*$. The mass of the  quark is 
exponentially suppressed by the area of shaded world-sheet.} 
 
In the scheme we are discussing, 
 after full recombination of the {\it left }, 
{\it right} and {\it lepton} branes into a single brane, 
the observed SM particles and interactions 
would come from open string exchange between a couple of 
brane stacks (see fig.\ref{artistic} for an artist's  view): 
 
{\it Baryonic  stack}. It contains three parallel branes 
(and mirrors)   and the gauge group $SU(3)\times U(1)_a$, 
with  $U(1)_a$ gauging 
baryon number. QCD gauge bosons originate in this stack. 
 
{\it Electro-Weak stack (f) }. It contains only one brane (plus mirror) 
with a $U(1)_{EW}$ gauge boson in its worldvolume. It results 
from the recombination of {\it left },  
{\it right} and {\it lepton} branes.   
 
A linear combination of $U(1)_a$ and $U(1)_{EW}$ may be identified 
with the photon and it  is massless. The orthogonal combination    
gets massive by combining with a RR B-field, but its symmetry      
remains as a global symmetry in perturbation theory, 
guaranteeing proton stability, as in the examples in  
previous sections.   
 
The  $W^{\pm}$,$Z^0$ bosons correspond to massive string states 
stretching between different sectors of the 
electroweak  brane ($f$)  or its mirror ($f^*$). 
The masses of  fermions of the SM 
will depend on the detailed geography of the configuration. 
 All quarks and leptons are massive 
but some of their masses may be very small due to the 
fact that they are located at different distances from  the 
location of the Higgs fields,  and 
some Yukawa couplings are naturally exponentially 
suppressed, as discussed in ref.\cite{afiru2}. 
Thus every SM parameter would have a reflection in terms  
of the detailed geometry of the underlying 
{\it baryonic} and {\it electro-weak} stacks.

\section{Final comments and conclusions} 
 
Let us make a number of general comments about the 
explicit intersecting D6-brane models constructed in the previous  
sections. They consist of Type IIA string  
compactifications on $T^6$ along with an orientifold operation. 
 We have concentrated on configurations with four stacks  
of D6-branes :  $a$ ({\it baryonic}), $b$ ({\it left}),  
$c$ ({\it right}), and $d$ ({\it leptonic}).  
This is the simplest structure 
capable of giving rise to all fermions of the SM in  
bifundamental representations of the underlying group. 
By varying the geometry of the tori, one  
can obtain models in which  all brane intersections  
respect {\it some } $\cn = 1$ supersymmetry. There are four 
classes of models with four stacks of branes with 
SUSY-quiver structures  shown in fig.(\ref{quivers}). 
They are called respectively square, linear, rombic and triangle quivers 
depending on the SUSY-quiver structure.  
We have built explicit models  
corresponding to these four classes.  
Some general characteristics of 
these models are displayed in table \ref{propmodels}. 

 \TABLE{\renewcommand{\arraystretch}{1.6}
\begin{tabular}{|c|c|c|c|c|} 
\hline 
Property  & Square & Linear  & Rombic  &  Triangle  \\ 
\hline 
\hline 
\# different  $\cn = 1$ SUSY's & $4$ &  2 & 3 & 1 \\ 
\hline 
Gauge group   & SM ($+ U(1)$)& SM$+U(1)$  &  LR   &   SM$+U(1)$ \\ 
\hline 
SUSY Higgs &  no  & no  &  yes  &  yes\\ 
\hline 
Minimal  Higgs system & no &  no  &  no   &  yes \\ 
\hline 
$m^2_{{\tilde q},{\tilde l}}>0 $ from FI & yes  &  yes  &  no   &  no \\ 
\hline 
String  Scale   & $ \sim $ 10 TeV & $ \sim $ 10 TeV  & 
$\sim $ 10 TeV     &  $\sim $ 10 TeV -$ M_{Planck}$ \\ 
\hline 
Additional RR sources  &  no  & yes   &  yes  &  yes\\   
\hline 
\end{tabular} 
\label{propmodels} 
\caption{Some general aspects of the four classes of SUSY-quiver 
models with three quark/lepton generations    
discussed in the text. The linear and triangle quivers lead to 
an extra $U(1)_{B-L}$ in the massless spectrum. In the square  
quiver case the presence of such an extra $U(1)$ is optional.  
The rombic example provided is a left-right symmetric model.}} 
The four classes of models have  
quarks, leptons and Higgs multiplets respecting in general  
different SUSY's, as recorded in the table. 
In general, in addition to the SM gauge group, the 
property of Q-SUSY (or SUSY) seems to force the presence of an additional  
$U(1)$ generator corresponding to $U(1)_{B-L}$,
an abelian symmetry well known from left-right symmetric models.  
This property is quite intriguing since there is in 
principle no obvious connection between the 
SUSY properties of a theory and the gauge groups one is gauging. 
In the case of the square quiver one can give mass to that  
additional $U(1)$ by departing from the SUSY limit (i.e. $\Delta \not=0$). 

The models from the square and linear quiver have non-SUSY Higgs  
sectors. If, as suggested in ref.\cite{cim1}, 
 one insists in getting one-loop protection of the 
Higgs particles to stabilize the weak scale, models with triangle  
or rombic quiver structure would be required. In particular, in the  
case of the triangle quiver, specific configurations with  
the minimal SUSY Higgs sector can be obtained. The latter triangle  
models have (for $\beta^2 = 1$) the chiral content of the MSSM 
(with right-handed neutrinos). From this point of view   
the triangle class of models are particularly attractive. 
They are also attractive because they predict the presence of a 
SM Higgs sector. Indeed we showed how, in order for left- and right-handed  
multiplets to respect the same $\cn = 1$ SUSY, brane configurations 
are forced to have intersections at which Higgs sectors arise. 
This is an interesting property which is not present in 
schemes which follow the unification route  
like SUSY-GUT, $CY_3$ and/or Horava-Witten heterotic compactifications  
etc. 
\footnote{Also note that in those unification schemes the stability of  
the proton relies on the asumed presence of symmetries like R-parity or 
generalizations as well as some doublet-triplet splitting 
mechanism. In the present case proton stability is a consequence 
of the gauge character of baryon symmetry.}. 
 
On the other hand, we have seen that going slightly away from the 
Q-SUSY limit, one can parametrize the corresponding SUSY-breaking in terms of 
FI-terms. As remarked in the table, one can check that in the 
square and linear quivers all squarks and sleptons can get  
positive $(mass)^2$ from FI-terms whereas that is not the case for the 
triangle and rombic quivers. In the latter cases the leading source 
of SUSY-breaking scalar masses should come from loop effects, 
as in fig.(\ref{loop5}).  

The general consistency of any of these D6-brane configurations require  
global cancellation of the total RR charge. In the case of the square quiver 
it is easy to find simple choices of D6-branes wrapping factorized   
cycles on the torus and with no  intersection with the SM branes, so that 
all RR-tadpoles cancel. The other three types of quivers are equally  
consistent from this point of view, although in general   
non-factorizable extra branes and additional RR H-flux may be required 
to  cancel tadpoles. 

In D6-brane models like this, the standard way \cite{aadd} 
to lower the string scale 
$M_s$ compared with the  Planck scale $M_{Planck}$ by making 
large some of the torus radii cannot be performed 
\footnote{For a discussion of this point see \cite{bgkl,afiru,cim1} 
and references therein.}. 
Note also that the intersecting brane  structure is not necessarily 
linked to a low string scale (i.e., $M_s\approx $ 10-100 TeV) 
hypothesis. Consider in particular a triangle quiver configuration, 
where all intersections preserve the {\it same } $\cn = 1$ SUSY.  
In principle one can consider this triangle as  
part of some bigger $\cn = 1$  brane configuration somewhat 
analogous to the class of models in ref.\cite{csu}.  
One could then translate most of our discussion  
(FI-terms, brane recombination, gauge coupling constants) on triangle 
quivers to that configuration. This is what we mean by indicating in  
the table that the string scale for the triangle quiver case  
is $\sim 10$ TeV-$M_{Planck}$. 

Independently of the particular models discussed, we have  
presented a brane interpretation of the SM Higgs  
mechanism. It is important to realize that the familiar  
brane interpretation of the Higgs mechanism in terms of the 
separation of parallel branes is not appropriate for the 
Higgs mechanism of the SM. Brane separation does not lower 
the rank of the gauge group and corresponds to adjoint  
Higgsing. We claim that the appropriate brane interpretation  
of the SM Higgs mechanism (and analogous Higgsings lowering the  
rank) is brane recombination 
\footnote{An analogous proposal in the context of D4-brane models was 
put forward in ref.\cite{afiru2}.} 
. In our approach the non-Abelian weak 
interactions $SU(2)_L$ live on the worldvolume of two parallel 
branes i.e., $b_1$,$b_2$.  
If the Higgs field comes  
from open strings exchanged between branes $b$ and $c^*$ 
(as in the models constructed), what will happen is that one 
of the two parallel $b$-branes (say, $b_1$) will fuse  
with $c^*$ giving rise to a single brane $b_1+c^*\rightarrow e$. 
This is  the brane recombination mechanism. Since each brane comes 
along with its own $U(1)$, it is obvious that the rank has  
been  reduced in the process. At the same time the number  
of chiral fermions (computed from the intersection numbers of the 
residual branes left)  
may be shown to decrease (or vanish), corresponding to the  
Higgs field giving masses to chiral fermions through  
Yukawa couplings. 

An analogous brane recombination interpretation exists for a  
process in which a right-handed sneutrino gets a vev. 
In this case lepton number is broken and at the same time  
it is shown that both right-handed and left-handed  
neutrinos get independent (i.e., Majorana masses). If 
one consider this process in the specific models constructed (like the  
square or the triangle models) one finds that the distinction 
between Higgs multiplets and sleptons dissappear, and the  
properties of the models are somewhat similar to 
R-parity violating SUSY models with lepton number 
violation. In this respect we note that it seems quite difficult  
within the brane intersection scheme to have lepton number  
violating neutrino masses without having at the same time  
L-violating dimension-four couplings, both things come  
along once brane recombination takes place. 
 
At the end of the day, in these schemes the  
observed SM would have a description in terms of two 
final recombined brane-stacks : a {\it baryonic} stack 
and an {\it electroweak stack} supporting a  
$SU(3)\times U(1)_{em}$ gauge group.  
Every SM parameter would have a reflection in terms 
of the detailed geometry of the underlying 
{\it baryonic} and {\it electro-weak } stacks. 
  
\vspace{2cm} 
 
\centerline{\bf Acknowledgements} 
We are grateful to G.~Aldaz\'abal, C.~Kokorelis, 
R.~Rabad\'an and  A.~Uranga for useful discussions. 
The research of D.C. and F.M. was  supported by 
 the Ministerio de Educaci\'on, Cultura y Deporte (Spain) through FPU grants. 
This work is partially supported by CICYT (Spain) and the 
European Commission (RTN contract HPRN-CT-2000-00148).

\newpage

\section{Appendix I: Branes wrapping general cycles.} 
 
As we have mentioned in  the text, when dealing with more general  
intersecting D-brane constructions or considering a generic  
configuration after some brane recombination has taken place, 
we are naturally led to consider branes wrapping general cycles. 
 
In our particular setup, a D$6_a$-brane wrapping a general cycle 
will be located in a 3-submanifold of  
$T^6 = T^2 \times T^2 \times T^2$, thus 
corresponding to an element $[\Pi_a]$ of $H_3(T^6, \IZ)$, 
which is the group of homology classes of 3-cycles \cite{nakahara}. 
It turns out that $H_3(T^6, \IZ)$ is a discrete vector space, 
so any of its elements can be represented by a vector  
with integer entries. This vector space has dimension 20. 
One particular subset of $H_3(T^6, \IZ)$ is given by what we have  
called {\it factorizable cycles}. These are 3-cycles that can be  
expressed as products of 1-cycles on each $T^2$  
(see figure \ref{compact2} for an example)\footnote{Notice that 
for this definition to make sense we also need our $T^6$ 
to be factorized as $T^2 \times T^2 \times T^2$.}. 
Any of those cycles can be expressed by 6 integers 
as  $[\Pi_a] = \prod_{i = 1}^3 [(n_a^i, m_a^i)]$,  
where $n_a^i, m_a^i \in \IZ$ describe the 1-cycle the D$6_a$-brane 
is wrapping on the $i^{th}$ torus. Factorizable cycles can be 
easily described geometrically, which allows us to compute  
many phenomenologically interesting quantities.
For instance, we can compute the lightest bosonic 
spectrum living at the intersection of two factorizable D6-branes 
$a$ and $b$ by simply computing the angles they form on each torus  
and using (\ref{scalars}). 
Thanks to this, we were able to check whether certain  
$\cn = 1$ supersymmetries were preserved at each intersection  
and define Q-SUSY or SUSY models in this way. 
Unfortunately, we will be unable to study which  
supersymmetries are left, if any, after a brane 
recombination process has taken place. 
 Due to this fact, we have constructed our  
particle physics models mostly using branes wrapping factorizable cycles, 
although non-factorized branes will be generically  
unavoidable after brane recombination.  
 
It turns out that factorizable 3-cycles are not a vector  
subspace of the homology group $H_3(T^6, \IZ)$. Indeed, the sum of two 
factorizable cycles is not, in general, a factorizable cycle. 
This is an important point, since the homology class $[\Pi_a]$  
where a D$6_a$-brane lives determines its RR charges. When 
two branes $a$ and $b$ fuse into a third one $c$ in a brane  
recombination process the total RR charge should be conserved,  
which implies that the final brane will lie in a 3-cycle such that 
$[\Pi_c] = [\Pi_a] + [\Pi_b]$. Even if we start with a configuration 
where every brane is factorizable, we are finally led to consider  
non-factorizable branes as well. To this regard, we will consider 
the smallest vector subspace of $H_3(T^6, \IZ)$ that contains 
factorizable 3-cycles. This is $[H_1(T^2, \IZ)]^3$, and its dimension 
is $2^3 = 8$. Following \cite{torons}, we define a basis on this 
subspace by 
\begin{center}
\begin{tabular}{ccc} 
$q$ comp. & 3-cycle & factor. comp.  \\ \hline 
$q_1$ & $[a_1] \times [a_2] \times [a_3]$ & $n^1 n^2 n^3$ \\ 
$q_2$ & $[b_1] \times [b_2] \times [b_3]$ & $m^1 m^2 m^3$ \\ 
$q_3$ & $[a_1] \times [b_2] \times [b_3]$ & $n^1 m^2 m^3$ \\ 
$q_4$ & $[b_1] \times [a_2] \times [a_3]$ & $m^1 n^2 n^3$ \\ 
$q_5$ & $[b_1] \times [a_2] \times [b_3]$ & $m^1 n^2 m^3$ \\ 
$q_6$ & $[a_1] \times [b_2] \times [a_3]$ & $n^1 m^2 n^3$ \\ 
$q_7$ & $[b_1] \times [b_2] \times [a_3]$ & $m^1 m^2 n^3$ \\ 
$q_8$ & $[a_1] \times [a_2] \times [b_3]$ & $n^1 n^2 m^3$ \\ 
\end{tabular}
\end{center}
where each element of this basis can be expressed as a product 
of 1-cycles $[a_i]$, $[b_i]$ of the $i^{th}$ $T^2$. Each general cycle 
$[\Pi_a]$ under consideration can then be expressed by a vector $\q_a$, 
whose 8 integer components are defined above. In addition, a  
factorizable 3-cycle will correspond to vector $\q$ whose  
components are given in the third column above  
\footnote{In the construction of the $q$-basis we are not  
considering fractional wrapping numbers, so both $n$ and $m$ are integer, 
see below.}. 
See \cite{torons} for more details on this construction and some 
other features involving non-factorizable cycles. 
 
The usefulness of this $\q$-basis formalism comes from the fact  
that quantities as the intersection number of two branes 
$a$ and $b$ can be easily expressed as bilinear products involving 
an intersection matrix. That is, it can be expressed as 
\beq 
I_{ab} \equiv [\Pi_a]\cdot [\Pi_b] = \q_a^{\ t\ }I \q_b, 
\label{numinter} 
\eeq 
where the intersection matrix is given by 
\beq 
I = 
\left( 
\begin{array}{cccccccc} 
0 & 1 & & & & & & \\ 
-1 & 0 & & & & & & \\ 
 & & 0 & 1 & & & & \\  
 & & -1 & 0 & & & & \\ 
 & & & & 0 & 1 & & \\ 
 & & & & -1 & 0 & & \\  
 & & & & & & 0 & 1 \\ 
 & & & & & & -1 & 0 \\ 
\end{array} 
\right)  
\label{matrizD6} 
\eeq 
 
When dealing with orientifold compactifications each generic brane $a$  
must be accompanied by its mirror image $a^*$. In this $q$-basis  
formalism, the corresponding vectors can be 
related under the action of a linear operator $\Om$, such that 
\beq 
\Om \q_a = \q_{a^*}, \quad \Om^2 = {\rm Id}. 
\label{accionom} 
\eeq 
 
As we already mentioned, the geometrical action associated to $\Om$ amounts 
to a reflection on each complex internal dimension. In terms of a 1-cycle 
$(n^i, m^i)$ wrapping on the $i^{th}$ $T^2$, this translates into 
\beq 
\Om : (n^i, m^i) \mapsto (n^i, - m^i - 2b^i\ n^i), 
\label{action} 
\eeq 
where $b^i = 0, \oh$ is the T-dual discrete NS background defined in 
Section 2, and related to the complex structure of the $i^{th}$ torus. 
From this we can deduce the action of the operator $\Om$ on a general 
vector $\q$ describing a D6-brane: 
\beq 
\Om =  
\left( 
\begin{array}{cccccccc} 
1 & 0 & 0 & 0 & 0 & 0 & 0 & 0\\ 
-8b^{1}b^{2}b^{3} & -1 & -2b^{1} & -4b^{2}b^{3} & -2b^{2}  
& -4b^{1}b^{3} & -2b^{3} & -4b^{2}b^{1} \\ 
4b^{2}b^{3} & 0 & 1 & 0 & 0 & 2b^{3} & 0 & 2b^{2} \\  
-2b^{1} &0 & 0 & -1 &0 &0 &0 &0 \\ 
4b^{1}b^{3} &0 &0 & 2b^{3} & 1 & 0 &0 &2b^{1} \\ 
-2b^{2} &0 &0 &0 & 0 & -1 &0 &0 \\  
4b^{1}b^{2} &0 &0 &2b^{2} &0 &2b^{1} & 1 & 0 \\ 
-2b^{3} &0 &0 &0 &0 &0 & 0 & -1 \\ 
\end{array} 
\right)  
\eeq 
Notice that this linear operator should leave fixed the cycle where 
the O6-plane lies, which is 
\beq 
[\Pi_{ori}] =  
\bigotimes_{i=1}^3 \left({1 \over 1 - b^{i}}[a_i] - 2 b^{i} [b_i]\right), 
\label{cicloori} 
\eeq 
and that can be easily translated into a vector $\q_{ori}$. 
The chiral massless spectrum arising at general cycles intersections 
is given by 
\beq 
\begin{array}{c} 
\sum_{a<b} \left[I_{ab} (N_a, \bar N_b)   
+ I_{ab^*} (N_a, N_b)\right]\\ 
\sum_a \left[8\b^1\b^2\b^3 I_{a,ori} \ ({\bf A}_a)\ + 
\oh \left(I_{aa^*} - 8\b^1\b^2\b^3 I_{a,ori} \right)  
({\bf A}_a + {\bf S}_a)  \right]  
\label{specori2} 
\end{array} 
\eeq 
where ${\bf S}_a$ (${\bf A}_a$) stands for the (anti)symmetric  
representation of the gauge group $U(N_a)$. Here $\b^i = 1 - b^i$, 
as defined in the text. All intersections involved in (\ref{specori2}) 
can be computed in this $q$-basis formalism, as for instance 
\beq 
I_{ab} = \q_a^{\ t\ }I \Om \q_b, \quad  
\quad I_{a,ori} = \q_a^{\ t\ }I \q_{ori}. 
\eeq 
 
Remarkably enough, RR tadpoles cancellation have an extremely  
simple expression in this formalism 
\beq 
\sum_a N_a \left(\q_a + \Om \q_a \right) = Q_{ori}\ \q_{ori} 
\label{tadpolesq} 
\eeq 
where $Q_{ori}$ represents the relative charge between the O6-plane 
and the D6-branes and is given by $Q_{ori} = 32 \b^1\b^2\b^3$. Notice 
that we can define the operators $P_\pm = \oh (1 \pm \Om)$, which 
satisfy 
\beqa 
P_\pm^2 = P_\pm, & & P_\pm \cdot P_\mp = 0, 
\label{proyectores} 
\eeqa 
thus being projector operators on the $q$-basis space. Notice that  
the projector $P_+$ is involved in condition (\ref{tadpolesq}), 
which means that only some of the components of $\q_a$ are 
relevant for tadpoles. On the other hand, the projector $P_-$ 
is involved on the coupling to branes of the RR $B_2$ fields  
that mediate the generalized GS mechanism (see Section 2 and  
\cite{afiru,imr} for a proper definition of these fields). 
Indeed, a general D6$_{a}$-brane whose vector is $\q_a$ will 
couple to these antisymmetric four-dimensional fields as 
$P_- \q_a$. Notice that since $P_-$ is a projector, it will 
only couple to four fields. In the same way that was done 
for factorizable branes in (\ref{caplillos}), we can give 
an explicit expression for these couplings in terms of  
the components of the vector $\q$ of this brane. 
\beq 
\begin{array}{l} 
B_2^0 \ : N_a \left( b^1b^2b^3 q_1 + q_2 + b^1 q_3 + b^2b^3 q_4 + 
b^2 q_5 + b^1b^3 q_6 + b^3 q_7 + b^1b^2 q_8 \right) F_a, \\ 
B_2^1 \ : N_a \left( b^1 q_1 + q_4 \right) F_a, \\  
B_2^2 \ : N_a \left( b^2 q_1 + q_6 \right) F_a, \\  
B_2^3 \ : N_a \left( b^3 q_1 + q_8 \right) F_a. 
\label{caplillos5} 
\end{array} 
\eeq 
 
In order to relate these expressions with the ones presented in the text, 
let us recall that in (\ref{caplillos}), we were expressing our D6-brane 
configurations in terms of {\it fractional} 1-cycles. These fractional 
1-cycles are defined as \cite{bkl,cim1} 
\beq 
(n^i,m^i)_{{\rm frac}} \equiv (n^i,m^i) + b^{i} (0,n^i), 
\label{fwrapping} 
\eeq 
so when computing the coupling of a fractional brane to, say,  
the field $B_2^1$, we have the coefficient 
\beq 
N_a \left( b^1 q_1 + q_4 \right)  
= N_a \left( b^1 n^1n^2n^3 + m^1n^2n^3 \right) 
= N_a m^1_{{\rm frac}} n^2_{{\rm frac}} n^3_{{\rm frac}}, 
\label{demofrac} 
\eeq 
so it reduces to the previous expression. Notice that, apart 
from this appendix, we have used the fractional notation 
on the whole text, without any subindex. 
 
\section{Appendix II: Extra $U(1)$'s and Q-SUSY structure} 
 
One of the most interesting aspects regarding intersecting brane  
world models involves the massive $U(1)$ structure, arising from  
couplings with antisymmetric $B_2$ fields, as shown in Section 2 
and more extensively in \cite{imr}. 
 In particular, we are interested in the abelian gauge symmetries  
that remain after those couplings have been taken into account.  
When dealing with Q-SUSY models of  
factorizable branes there are some general results that can be  
stated regarding such massless $U(1)$'s. Indeed, in this second 
appendix we will try to elucidate the number of massless $U(1)$'s  
in terms of the different Q-SUSY structures presented in Section 3. 
  
Let us start from a generic brane content consisting of four stacks 
of factorizable D6-branes, which contains each of the SUSY-quivers 
considered in Section 3. This brane content is presented in table 
\ref{generic}. 
\TABLE{\renewcommand{\arraystretch}{2}
\begin{tabular}{|c|c||c|c|c|} 
\hline 
brane\ type  & 
 $N_i$    &  $(n_i^1, m_i^1)$  &  $(n_i^2, m_i^2)$   & $(n_i^3, m_i^3)$ \\ 
\hline\hline $a$ & $N_\a$ & $(n_\a^1, 0)$  &  $(n_\a^2, m_\a^2)$ & 
 $(n_\a^3, \eps_\a m_\a^3)$  \\ 
\hline $b$ & $N_\b$ &   $(n_\b^1, m_\b^1)$ & $(n_\b^2, 0)$  & 
$(n_\b^3, \eps_\b m_\b^3)$   \\ 
\hline $b$ or $c$ & 
$N_\g$ & $(n_\g^1, m_\g^1)$ & $(n_\g^2, \eps_\g^2 m_\g^2)$ &  
$(n_\g^3, \eps_\g^3 m_\g^3)$ \\ 
\hline $a$ & $N_\d$ &$(n_\d^1, 0)$  &  $(n_\d^2, m_\d^2)$ & 
 $(n_\d^3, \eps_\d m_\d^3)$  \\  
\hline  
\end{tabular} 
\label{generic} 
\caption{\small D6-brane wrapping numbers giving rise to  
a generic Q-SUSY model of those presented in Section 3. Here 
every discrete parameter $n_i^j$, $m_i^j$ is taken positive, whereas 
the phases $\eps_i^j = \pm 1$ determine the type of brane ($a1$ or $a2$ 
etc$\dots$) we are dealing with. }}

Notice that we must impose $m_\g^2\cdot m_\g^3 = 0$ for the brane 
$\g$ to belong to the hexagonal structure depicted in  
figure \ref{hexagon}. 
We should also impose this condition in order to avoid chiral exotic matter 
appearing in the $cc^*$ sector. This brane content will yield the  
most general Q-SUSY quiver of four stacks of factorizable branes, 
modulo renumbering of the tori. However, as we already mentioned in 
the text, without loss of generality we can take the branes $\a$,  
$\b$ to be of type $a_2$, $b_2$, respectively. This amounts to take 
$\eps_\a = \eps_\b = -1$ in table \ref{generic}. 
 
Given a specific brane content, we can easily compute which of the 
$U(1)$'s will remain massless in our configuration by looking at 
the couplings (\ref{caplillos}). For our purposes, it will be useful 
to encode such information in matrix notation. In general,  
for a configuration of $K$ stacks of branes we can 
define the $\B$ as a $4 \times K$ matrix, containing on each 
column the coupling of the $i^{th}$ brane to the four $B_2$ fields. 
When dealing with factorizable branes, such matrix has the form 
\beq 
\B = 
\left( 
\begin{array}{c} 
\dots \\ 
\end{array} 
\begin{array}{c} 
N_i\ m_i^1 m_i^2 m_i ^3 \\ 
N_i\ m_i^1 n_i^2 n_i ^3 \\ 
N_i\ n_i^1 m_i^2 n_i ^3 \\ 
N_i\ n_i^1 n_i^2 m_i ^3 \\ 
\end{array} 
\begin{array}{c} 
\dots \\ 
\end{array} 
\right). 
\label{B} 
\eeq 
 
Given a general linear combination of $U(1)$ fields, whose generator is 
\beq 
Q_X = \sum_{j = 1}^K c_X^j Q_j, 
\label{combi} 
\eeq 
it will remain as a massless linear combination of the low energy spectrum 
whenever it does not couple to any of the $B_2$ fields, that is, if 
\beq 
\B \cdot \q_X = 0, \quad \quad \q_X^{\ t} = (\cdots \ c_X^j \ \cdots). 
\label{kernel} 
\eeq 
So we can see that each massless combination of $U(1)$'s corresponds 
to a vector $\q_X$ belonging to the kernel of $\B$, now seen as a  
linear operator. In particular the number of massive $U(1)$'s  
on each configuration equals {\it Rank}($\B$). This simple observation  
will help us to elucidate how many $U(1)$'s remain massless on each  
Q-SUSY configuration arising from table \ref{generic}. Let us 
distinguish two different cases 
 
\begin{itemize} 
 
\item {\it \ $m_\g^3\ =\ 0$} 
 
This choice contains both rombic and triangular quiver structures. 
Our $\B$ matrix will have the form 
\beq 
\B = 
\left( 
\begin{array}{cccc} 
0 & 0 & 0 & 0 \\ 
0 & N_\b\ m_\b^1 n_\b^2 n_\b^3 & N_\g\ m_\g^1 n_\g^2 n_\g^3 & 0 \\ 
N_\a\ n_\a^1 m_\a^2 n_\a^3 & 0 & \eps_\g^2 N_\g\ n_\g^1 m_\g^2 n_\g^3 & 
 N_\d\ n_\d^1 m_\d^2 n_\d^3 \\ 
- N_\a\ n_\a^1 n_\a^2 m_\a^3 & - N_\b\ n_\b^1 n_\b^2 m_\b^3 &  
0 & \eps_\d N_\d\ n_\d^1 n_\d^2 m_\d^3 \\ 
\end{array} 
\right), 
\label{B2} 
\eeq 
from which it can easily be seen that the rank of $\B$ will be 
always lower than 4, so at least one $U(1)$ remains massless. 
Now, having a Q-SUSY structure of any sort will imply some topological  
restrictions\footnote{Notice that, although supersymmetry between two  
stacks of branes always implies a geometrical condition given by  
the angles they form, the {\it ability} for a full configuration 
of branes to be Q-supersymmetric does imply some topological  
restrictions.}, such as  
\beq 
{m_\a^3/n_\a^3 \over m_\a^2/n_\a^2} =  
{m_\d^3/n_\d^3 \over m_\d^2/n_\d^2}. 
\eeq 
Indeed, if we define 
\beqa 
\lam_1 = {n_\a^2 m_\a^3 \over m_\a^2 n_\a^3}, & &  
\lam_2 = {n_\b^1 m_\b^3 \over m_\b^1 n_\b^3},  
\label{defs} 
\eeqa 
then it can easily be seen that having a Q-SUSY structure implies 
$\B$ taking the form 
\beq 
\nonumber 
\B = 
\left( 
\begin{array}{cccc} 
0 & 0 & 0 & 0 \\ 
0 & N_\b\ m_\b^1 n_\b^2 n_\b^3 & N_\g\ m_\g^1 n_\g^2 n_\g^3 & 0  \nonumber \\ 
N_\a\ n_\a^1 m_\a^2 n_\a^3 & 0 &  
\eps_\g^2 (\lam_2/\lam_1)\ N_\g\ m_\g^1 n_\g^2 n_\g^3  
& N_\d\ n_\d^1 m_\d^2 n_\d^3 \nonumber  \\ 
- \lam_1 N_\a\ n_\a^1 m_\a^2 n_\a^3 & - \lam_2  N_\b\ m_\b^1 n_\b^2 n_\b^3 & 
0  &  \eps_\d \lam_1 N_\d\ n_\d^1 m_\d^2 n_\d^3  \nonumber 
\end{array} 
\right). 
\nonumber 
\eeq 
The minor determinants of this matrix will be proportional to 
\beq 
\begin{array}{l} 
{\rm det}_{(1,2,3)} \propto 1 + \eps_\g^2 \\ 
{\rm det}_{(1,2,4)},\ {\rm det}_{(1,3,4)}  \propto 1 + \eps_\d \\ 
{\rm det}_{(2,3,4)} \propto \eps_\g^2\eps_\d -1 
\end{array} 
\label{minors} 
\eeq 
 
For a triangular quiver we should impose $\eps_\g^2 = \eps_\d = -1$, 
so every minor determinant will vanish and the rank of $\B$ will be two. 
Thus, we will have precisely 2 massless surviving $U(1)$'s, at least 
if none of the entries in (\ref{B2}) vanishes. In any case, when trying 
to build either Standard or Left-Right symmetric models, one extra  
abelian group will arise. When dealing with a rombic quiver,  
however, we should impose instead $\eps_\g^2 = - \eps_\d = -1$, 
and as a result there will be two nonvanishing minor determinants.  
Thus, we will have just one massless $U(1)$. The other choices of 
phases correspond to some other SUSY-quivers not considered in this 
paper. 
 
\item {\it \ $m_\g^2\ =\ 0$} 
 
This second choice will contain square and linear Q-SUSY structures. 
Proceeding in the same manner as done above, we find that in order 
to have a Q-SUSY structure our $\B$ matrix should take the form 
\beq 
\B = 
\left( 
\begin{array}{cccc} 
0 & 0 & 0 & 0 \\ 
0 & N_\b\ m_\b^1 n_\b^2 n_\b^3 & N_\g\ m_\g^1 n_\g^2 n_\g^3 & 0 \\ 
N_\a\ n_\a^1 m_\a^2 n_\a^3 & 0 & 0  
& N_\d\ n_\d^1 m_\d^2 n_\d^3 \\ 
- \lam_1 N_\a\ n_\a^1 m_\a^2 n_\a^3 & - \lam_2  N_\b\ m_\b^1 n_\b^2 n_\b^3 & 
\eps_\g^3 \lam_2 \ N_\g\ m_\g^1 n_\g^2 n_\g^3  &  
\eps_\d \lam_1 N_\d\ n_\d^1 m_\d^2 n_\d^3  
\end{array} 
\right), 
\nonumber 
\eeq 
the minor determinants now being proportional to 
\beq 
\begin{array}{l} 
{\rm det}_{(1,2,3)},\ {\rm det}_{(2,3,4)} \propto 1 + \eps_\g^3 \\ 
{\rm det}_{(1,2,4)},\ {\rm det}_{(1,3,4)}  \propto 1 + \eps_\d 
\end{array} 
\label{minors2} 
\eeq 
 
The square quiver case amounts to take $\eps_\g^3 = \eps_\d = 1$, 
which implies {\it Rank}($\B$) = 3 and just one massless $U(1)$. 
For the Linear quiver, in turn, we must take $\eps_\g^3 = - \eps_\d = 1$, 
again with the same result. 
 
\end{itemize} 
 
Apart from these general considerations, let us notice that, when 
trying to get the Standard Model from a general orientifold configuration 
whith the brane content of table \ref{generic}, the hypercharge 
generator will have an associated vector of the form 
\beq 
\q_Y = \left( 
\begin{array}{c} 
1/6 \\ 0 \\ \eps / 2 \\ \tilde\eps /2 
\end{array} 
\right), 
\eeq 
where $\eps$, $\tilde\eps$ are model-dependent phases. For this  
vector to belong to the kernel of the general matrix in 
(\ref{B2}), we must impose, among others, the condition 
$m_\g^1 n_\g^2 n_\g^3 = 0$. Now, notice that $m_\g^1 \neq 0$, or 
else there will be no intersection with branes $\a$ and $\d$, 
so we are finally led to consider $n_\g^2 = 0$ or $n_\g^3 = 0$. 
Since this will imply that the brane $\g$ has a twist vector 
\beq 
v_\g = \left(\th_\g, \pm \frac \pi2, 0\right) \quad {\rm or} \quad 
v_\g = \left(\th_\g, 0, \pm \frac \pi2\right), 
\label{rtwist} 
\eeq 
then in order to belong to one of the types of branes in (\ref{vectors}), 
we must impose $\th_\g = \frac \pi2$, which in turn implies that 
$n_c^1 = 0$. As a result, the brane $\g$ will not couple to any 
of the $B_2$ fields in (\ref{caplillos}), as can be seen by direct  
substitution in (\ref{B}). The $\B$ matrix will be effectively 
reduced to the $3 \times 3$ minor $(1,2,4)$, and its rank will depend 
exclusively on $\eps_\d$. Indeed, when $\eps_\d = -1$ we find that 
there are two massless $U(1)$'s whose generators are 
\beq 
\begin{array}{c} 
Q_{\g}, \\ (N_\d\ n_\d^1 m_\d^2 n_\d^3)\ Q_\a -  
(N_\a\ n_\a^1 m_\a^2 n_\a^3)\ Q_\d, 
\end{array} 
\label{massless2} 
\eeq 
whereas in the case $\eps_\d = 1$ only $Q_\g$ remains massless. 
 

\end{document}